\documentstyle[epsf,multicol,12pt]{article}

\renewcommand{\theequation}{\thesection.\arabic{equation}}

\def\dag{{\dagger}}
\def\sgn{{\rm sgn}}
\def\bx{{\bf x}}
\def\bs{{\bf S}}
\def\bsx{{{\bf S}x}}
\def\bsy{{{\bf S}y}}
\def\bt{{\bf T}}
\def\bq{{\bf q}}
\def\bQ{{\bf Q}}

\def\hbn{\hat{\bf n}}
\def\bn{{\bf n}}
\def\bk{{\bf k}}
\def\ba{{\bf a}}
\def\bA{{\bf A}}
\def\by{{\bf y}}
\def\wp{\widetilde{\phi}}
\def\ve{v_F^*}
\def\ves{v_F^{*2}}
\def\bp{{\bf p}}
\def\bu{{\bf U}}
\def\bv{{\bf V}}
\def\half{{{1}\over{2}}}

\begin{document}


\title{Multidimensional Bosonization}
\author{A. Houghton$^1$, H.-J. Kwon$^2$, and J. B. Marston$^1$}
\date{\today}
\bigskip
\maketitle

\hskip 0.2in $^1$Department of Physics, Box 1843, Brown University, 
Providence RI 02912-1843 USA

\hskip 0.2in $^2$Department of Physics, University of Florida, Gainesville, FL
32611-8440 USA

\begin{abstract}
Bosonization of degenerate fermions yields insight both into Landau Fermi 
liquids, and into non-Fermi liquids.  We begin our review with a pedagogical
introduction to bosonization, emphasizing its applicability in spatial 
dimensions greater 
than one.  After a brief historical overview, we present the essentials of
the method.  Well known results of Landau theory are recovered, demonstrating
that this new tool of many-body theory is robust.  Limits of multidimensional
bosonization are tested by considering several examples of non-Fermi liquids,
in particular the composite fermion theory of the half-filled Landau level.
Nested Fermi surfaces present a different challenge, and these may be 
relevant in the cuprate superconductors.  We conclude by discussing the
future of multidimensional bosonization.
[Published in {\it Advances in Physics} {\bf 49}, 141 - 228 (2000).]

\end{abstract}

\tableofcontents

\bigskip
\section{Introduction}
\label{Intro}

Bosonization, a concept which sounds quite mysterious, is in fact both easy
to understand and ubiquitous.  In this review we outline an 
approach via bosonization to the problem of many interacting fermions in
spatial dimensions greater than one.  As the virtue of multidimensional 
bosonization lies in the reduction of complicated four-fermion
interactions to a Gaussian problem, we stress the simplicity of the approach. 

Multidimensional bosonization is more delicate than bosonization in 
one dimension.  In particular the technique is not useful for certain strongly 
correlated systems, specifically those in which the Fermi surface is 
obliterated by singular or unscreened interactions, as it relies 
heavily on the existence of an abrupt change in the quasiparticle occupancy
near the Fermi wavevector.  It is also difficult to implement systematic 
perturbative expansions using bosonization.  In these instances, 
it is more useful to work with the
original fermionic variables.  Nevertheless bosonization provides a
straightforward approach to many interesting problems
and we begin this review with a brief introduction to the method in its
several forms.  

Historically bosonization has been associated closely with the quantum physics 
of particles moving in one spatial dimension or with statistical mechanics
in two dimensions.  (For up-to-date reviews of one-dimensional bosonization,
which contain material not covered here, see the articles by 
Schulz\cite{Schulz}, 
Voit\cite{Voit}, and von Delft and Schoeller\cite{vonDelft}.)  
But any fermion system which exhibits collective 
excitations has some sort of bosonic description.  Spin waves in a ferromagnet,
plasmons in a metal, and superconducting order are 
well known examples.  In these instances, some if not all of the low energy
degrees of freedom have a bosonic description.  We focus here on
cases when {\it all} of the low-energy fermion degrees of freedom may be 
replaced by
bosons.  Mathematically the replacement can be carried out, at least formally, 
by means of a Hubbard-Stratonovich transformation\cite{Strat}: 
auxiliary bosons are 
coupled to the fermion degrees of freedom, and then the fermions are integrated
out yielding a purely bosonic effective theory.  In practice, however, the
transformation is useful only if there are additional constraints.  The 
following two examples illustrate the power of bosonization when the fermions
are subject to constraints in the form of infinite symmetries.

Consider first the problem of representing a spin-1/2 degree of freedom in terms
of the underlying fermions.  The angular momentum algebra is reproduced 
faithfully if we introduce fermion creation and annihilation operators 
$c^{\dag \alpha}$ and $c_\alpha$ 
which obey the canonical anticommutation relations
$\{ c^{\dag \alpha},~ c_{\beta} \} = \delta^\alpha_\beta$.  Then we may write:
\begin{equation}
\vec{S} = \half c^{\dag \alpha} \vec{\sigma}_\alpha^\beta c_\beta 
\end{equation}
subject to the constraint
\begin{equation}
n \equiv c^{\dag \alpha} c_\alpha = 1 \ .
\end{equation}
Here an implicit summation over repeated raised and lower Greek spin indices
is assumed.  More generally, to describe a quantum magnet for instance, 
we attach a site index $\bx$ to the operators:
\begin{equation}
\vec{S}_\bx = \half c_\bx^{\dag \alpha}~ \vec{\sigma}_\alpha^\beta~ 
c_{\bx \beta} \ .
\label{fermi-spin}
\end{equation}
If, for example, we consider now the nearest-neighbor Heisenberg magnet
\begin{equation}
H = J~ \sum_{<\bx,\by>} \vec{S}_\bx \cdot \vec{S}_\by
\label{HAF}
\end{equation}
it is clear that its representation in terms of the underlying fermions
is not as economical as its representation in terms of the spins themselves, 
since the spin operators of equation \ref{fermi-spin}, 
and hence the Hamiltonian equation \ref{HAF}, are invariant under 
$U(1)$ gauge transformations at each lattice point\cite{Baskaran,AM}: 
\begin{eqnarray}
c_{\bx \alpha} &\rightarrow& e^{i \theta_\bx}~ c_{\bx \alpha}\ .
\nonumber \\
c^{\dag \alpha}_\bx &\rightarrow& e^{-i \theta_\bx}~ c^{\dag \alpha}_\bx 
\label{xU(1)}
\end{eqnarray}
because the local phase rotation $\theta_\bx$ cancels out.
We denote this infinite gauge symmetry $U(1)^\infty_x$.
The physical meaning of this local symmetry is simply that the charge
degrees of freedom are frozen out and play no role in the insulating magnet.
The underlying fermions have both charge and spin degrees of freedom, but
as the number of fermions on each site is fixed to be one, only the spin
degree of freedom is active.  

Spin operators $S^+_\bx$ and $S^-_\bx$ create particle-hole excitations, 
localized on a single site $\bx$.  
That is, $S^+_\bx$ creates a spin-$\uparrow$ fermion and 
destroys a spin-$\downarrow$ fermion at site $\bx$.  These local particle-hole 
excitations, taken together, have a bosonic character, as two fermions bound
together obey boson statistics.  As the charge degree of freedom is frozen
in place, the remaining spin degree of freedom
can be represented instead in terms of bosons which obey
the usual canonical commutation relations
$[ b_\alpha,~ b^{\dag \beta} ] = \delta_\alpha^\beta$.  Now it is permissible
to write:
\begin{equation}
\vec{S}_\bx = \half b_\bx^{\dag \alpha}~ \vec{\sigma}_\alpha^\beta~ 
b_{\bx \beta} \ .
\end{equation}
These spins obey precisely the same algebra, and if we now constrain the
number of bosons to be one per site,
\begin{equation}
n_{\bx} \equiv b^{\dag \alpha}_\bx b_{\bx \alpha} = 1 \ , 
\end{equation}
the operators are spin-$\half$.  Note that such a replacement of fermions
with bosons -- our first concrete example of bosonization -- 
would not have been 
permissible for a model which contains both charge and spin degrees of freedom,
such as the Hubbard model.  For instance, in the non-interacting 
$U \rightarrow 0$ limit, the Hubbard model reduces to the tight-binding
model  
\begin{equation}
H = -t~ \sum_{<\bx,\by>} c_\bx^{\dag \alpha} c_{\by \alpha} + H.c. 
\end{equation}
which is clearly not invariant under the $U(1)^\infty_x$ transformations,
equation \ref{xU(1)}.  Distinct $U(1)$ rotation angles at neighboring sites,
$\theta_x$ and $\theta_y$ do not cancel in general.  Physically this means
that the full Fermi character of the electrons is manifested, as they 
propagate from lattice site to lattice site.  Of course 
the ground state in the $U \rightarrow 0$ limit is a degenerate Fermi liquid.
For $U > 0$ at half-filling, and for a bipartite lattice, on the other
hand, the charge degrees of freedom are frozen out.  Then
the low energy effective theory of the resulting 
antiferromagnetic insulator does exhibit the $U(1)^\infty_x$ symmetry.

At this point it might be asked what {\it practical} 
benefit can be gained by the replacement
of the fermions with bosons.  Our model provides a nice illustration: with
bosons, semiclassical approximations may be contemplated.  In the case of 
the antiferromagnet with $J > 0$, for instance, 
an expansion in powers of $1/S$ about
the limit of large on-site spin S is now possible \cite{Auerbach},
as the on-site spin can be increased simply by
populating more bosons at each site: $S = \half n_b$.  (In the fermion 
representation this can only be done by attaching an additional orbital
index to the fermions, and then enforcing Hund's rule to align the fermions
into a maximal spin state.)  In particular, the
N\'eel ordered state of the unfrustrated antiferromagnet on a square lattice
can be accessed in this way even at the mean-field level. In contrast 
mean-field theories with fermions are disordered\cite{Auerbach}.

There is another system which possesses infinite $U(1)$ symmetry and 
which can be bosonized easily:  the free, degenerate, Fermi gas or 
liquid\cite{Haldane}.  We begin with the usual Hamiltonian:
\begin{equation}
H = \sum_{\bk} \epsilon_{\bk}~ c^{\dag \alpha}_{\bk} c_{\bk \alpha}\ .
\end{equation}
The Hamiltonian is invariant under an infinity of $U(1)$ rotations, but now
at each point in momentum space, rather than position space:
\begin{eqnarray}
c_{\bk \alpha} &\rightarrow& e^{i \theta_\bk}~ c_{\bk \alpha}\ .
\nonumber \\
c^{\dag \alpha}_{\bk} &\rightarrow& e^{-i \theta_\bk}~ c^{\dag \alpha}_{\bk}\ .
\label{kU(1)}
\end{eqnarray}
This $U(1)^\infty_k$ symmetry, like the $U(1)^\infty_x$ symmetry explored
above, has a clear physical interpretation:
in the absence of interactions and disorder, momentum is a good quantum 
number for the single-particle states, as these are infinitely long-lived.  
Thus we may make the identification:
\begin{eqnarray}
U(1)^\infty_x &\leftrightarrow& {\rm insulating~ solid} 
\nonumber \\ 
U(1)^\infty_k &\leftrightarrow& {\rm conducting~ liquid.} 
\end{eqnarray}

Again the particle-hole pairs can be bosonized, and this bosonization  
furnishes us with another example of how some properties can be seen 
more clearly
in the bosonic basis.  Recall the classic calculation of the specific
heat at constant volume $V$, $C_V$, of a degenerate 
Fermi liquid\cite{Ashcroft}.  Two integrals, each
a function of the temperature $T$ and the chemical potential $\mu$, must
be computed: the total energy of the system $U(T, \mu)$ and the total number of 
fermions $N(T, \mu)$.  
To extract $C_V$, the chemical potential must be determined first, as it 
depends on temperature.  Taking this variation into account, the correct 
answer at low temperatures is:
\begin{equation}
C_V = \bigg{(}\frac{\partial U}{\partial T}\bigg{)}_N = 
\frac{\pi^2}{3} k_B^2 T N(0) \times {\rm V}\ .
\label{freeC_V}
\end{equation}
Here $N(0)$ is the density of states per unit volume at the Fermi surface.
The same result may be recovered more readily by simply treating the 
particle-hole excitations as free bosons.  
More readily, because there is no chemical potential for the 
bosons as they are not conserved in number, but rather vanish in the 
zero-temperature limit.  Now the total energy of a system of free bosons
with single-particle energies $\epsilon_i$ is given by:
\begin{equation}
U = \sum_i  {{\epsilon_i}\over{e^{\epsilon_i / k_B T} - 1}}
\end{equation}
and therefore
\begin{equation}
C_V = {{1}\over{4 k_B T^2}}~ \sum_i {{\epsilon_i^2}
\over{\sinh^2(\frac{\epsilon_i}{2 k_B T})}}\ .
\label{C_V}
\end{equation} 

To apply this formula we need a sensible interpretation of 
$\epsilon_i$.  As we are interested in the low-temperature properties, we
first linearize the spectrum of particle-hole excitations:
\begin{eqnarray}
\epsilon_{\bk + \bq} - \epsilon_{\bk} &=& \frac{(\bk + \bq)^2 - \bk^2}{2 m}
\nonumber \\
&\approx& \frac{\bk \cdot \bq}{m} 
\end{eqnarray}
where the second line holds in the limit $|\bq| \ll |\bk|$.  
This restriction means that we are considering only particle-hole excitations 
where particle and hole have nearly the same momentum.  We show below
that this constraint, which is unphysical, can in fact be discarded. 
At low temperatures the particle-hole pairs lie near the Fermi surface, and
we may set $|\bk| = k_F$, the Fermi momentum.  Then the dispersion reads
\begin{equation}
\epsilon_{\bk + \bq} - \epsilon_{\bk} = v_F q_\parallel 
\label{ph-dispersion}
\end{equation}
where $q_\parallel$ is the component of the momentum parallel to the Fermi 
wavevector $\bk$ at the Fermi surface, 
$q_\parallel \equiv \hat{\bk} \cdot \bq$.  Of particular significance is the
fact that the energy of a pair has this simple form when $|\bq| \ll k_F$, 
not depending on $\bk$, but rather only on the relative momentum $q_\parallel$.
The energy of a particle-hole pair, like that of a single free particle, is
determined uniquely by a single momentum.      
Substituting the dispersion into the specific heat formula, 
equation \ref{C_V}, we obtain:
\begin{eqnarray}
C_V &=& \frac{\rm No.~ of~ states~ on~ the~ Fermi~ Surface}{4 k_B T^2} \times
\sum_{q_\parallel > 0} {{(v_F q_\parallel)^2}\over{
\sinh^2(\frac{v_F q_\parallel}{2 k_B T})}}
\nonumber \\
&=& {{8 \pi k_F^2 (L/2 \pi)^3}\over{4 k_B T^2}} \times 
\int_0^\infty dq_\parallel~ {{v_F^2 q_\parallel^2}
\over{\sinh^2(\frac{v_F q_\parallel}{2 k_B T})}}\ .
\end{eqnarray}
Upon making a change of variables, and using the fact that
\begin{equation}
\int_0^\infty {{x^2~ dx}\over{\sinh^2(x)}} = \frac{\pi^2}{6}
\end{equation}
the exact result, equation \ref{freeC_V}, is recovered with the correct
prefactor\cite{Haldane}.

Why is it permissible to restrict $|\bq| \ll k_F$?  At any non-zero temperature
there are a macroscopic number of particle-hole excitations, some with 
large $\bq$ and some with small $\bq$.  However, in the vicinity of any
point on the Fermi surface there are statistically equal numbers of 
particles and holes.  Whether or not these particles and holes originate from
small or large momentum scattering processes, they have the same statistical
mechanical description, and they can always be re-assigned
to small-$\bq$ pairs which are described accurately by the above bosonic 
calculation.  

With this introduction to multidimensional bosonization, we turn now
to a brief history of efforts to use bosonization to describe interacting
fermion liquids.
For more than forty years Landau Fermi liquid theory provided the framework to
discuss the physics of strongly interacting fermion systems\cite{Landau}.  
In the
phenomenological theory the coupling between electrons is given in terms of a
few parameters which are determined by comparison with experiment\cite{Baym}.  
Landau Fermi liquid theory, focusing on the collective excitations of the Fermi
surface, is a type of bosonization.  Zero sound modes, ordinary sound, 
and paramagnetic spin waves all have a bosonic character, 
despite their origin in terms of fermions.

\begin{figure}
\center
\epsfxsize = 10.0cm \epsfbox{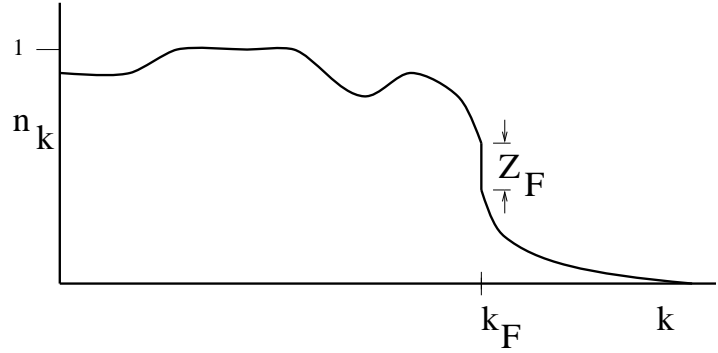}
\caption{
\label{discontinuity}
Particle occupancy of the interacting Fermi liquid.  The discontinuity at the 
Fermi surface has magnitude $Z < 1$; remaining spectral weight shows up
as incoherent structure at higher energies away from the Fermi surface.}
\end{figure}

\begin{figure}
\center
\epsfxsize = 10.0cm \epsfbox{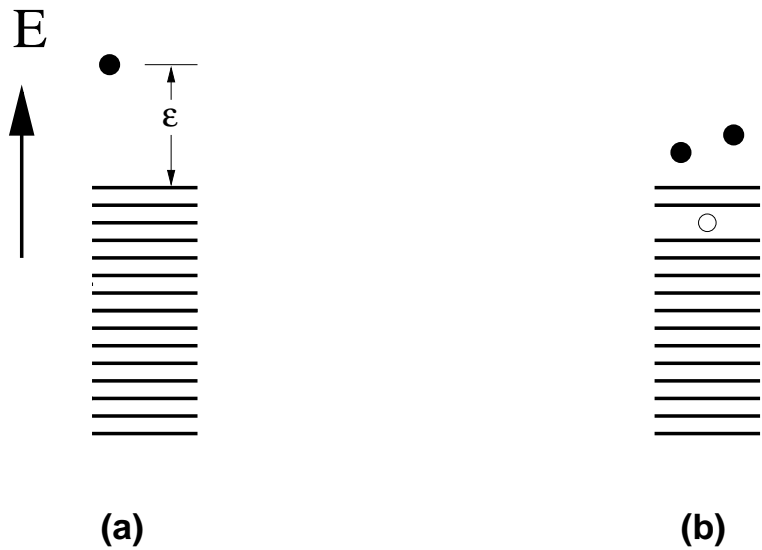}
\caption{
\label{phase_space}
A quasiparticle (a) of energy $\epsilon$ can lower its energy
via the formation of a particle-hole pair (b), but the phase-space for these
processes diminishes as $\epsilon^2$ in the $\epsilon \rightarrow 0$ limit.
This is known as Pauli blocking.}
\end{figure}

A microscopic interpretation of the Landau theory was provided in
the late 1950's and early 1960's via many body perturbation 
theory\cite{Landau-micro,AGD}.
Although the main idea behind Fermi liquid theory is deceptively simple, the
concept of a quasiparticle, its realization in the microscopic theory is 
far from straightforward. 
Qualitatively when an electron interacts with all of the other
electrons it polarizes its immediate vicinity due to the electron-electron
repulsion, and this complex entity is the Landau quasiparticle.  In this picture
the interaction between electrons does not affect the quantum numbers of
the electron, only its dynamics.  For example the effective mass
$m^*$ of the quasiparticle generally differs from the bare mass of the 
electron.  In the microscopic
approach, the existence of the quasiparticle can be ascertained by identifying
a simple pole in the single-particle Green's function.  At zero temperature 
this pole yields a $\delta$-function peak in the single-particle
spectral function at the Fermi surface.  In the non-interacting case all of
the weight is in this peak and there is no incoherent contribution to the
spectral function.  For free electrons the single-particle occupancy $n_\bk$ has
a discontinuity of magnitude $Z_F = 1$ at $|\bk| = k_F$. When interactions
are turned on, according to the Landau hypothesis, 
the discontinuity remains, but is reduced in magnitude to $0 < Z_F < 1$, 
as shown in figure \ref{discontinuity} with the remaining
spectral weight appearing as incoherent background.  
This idea is made 
explicit by considering a power-series expansion of the self-energy of the 
quasiparticle about the low-energy limit:
\begin{equation}
\Sigma(\omega, \bk) = a~ \omega + b~ \epsilon_\bk 
+ i c~ \omega^2 \sgn(\omega) + O(\omega^3)\ .
\label{self}
\end{equation} 
In the absence of singular interactions, Pauli blocking constrains
the phase space for scattering, as shown in figure \ref{phase_space}.  
The scattering rate vanishes as $\omega^2$ in the low-energy 
limit\cite{Nozieres,Negele},
and thus the $U(1)^\infty_k$ symmetry is recovered in this limit (see below). 
The mapping of the bare fermions $c_\bk$ onto the Landau quasiparticles 
$\psi_\bk$ may also be made explicit by writing:
\begin{equation}
\psi_\bk = Z_F^{-\half} c_\bk 
\end{equation}
where in terms of the self-energy, equation \ref{self}, we have
$Z_F^{-1} = 1 + a$ and $m^* = m \times (1 - a)/(1 + b)$. 
With this renormalization, the single-particle occupancy of the quasiparticles, 
$\langle \psi^\dag_\bk \psi_\bk \rangle$, like that of free fermions, has 
a unit step at the Fermi surface.  According to 
Luttinger's theorem\cite{Luttinger}, 
the discontinuity in the occupancy encloses the
same volume in momentum space.  In the special case of the
isotropic Fermi gas with a spherical Fermi surface, the surface
remains fixed at exactly the same position.  

The existence
of Landau quasiparticles implies the existence of a Fermi surface, but the
reverse does not necessarily hold.
In one spatial dimension it is well known that the Landau picture breaks down, 
perturbation
theory diverges, and the quasiparticle is destroyed by the interactions.
In this case a different paradigm, the ``Luttinger liquid,'' (the term having
been introduced in 1981 by Haldane\cite{old-Haldane}) describes a  
wide range of interacting fermion systems.  The foundation for this picture
was provided in early work by Tomonaga\cite{Tomonaga} who showed how to
bosonize a one-dimensional interacting fermion system with a band cutoff.
The foundations of one-dimensional bosonization were further developed
by Luther and Peschel\cite{Luther-Peschel}, Coleman\cite{Coleman}, and
Mandelstam\cite{Mandelstam}. 
Mattis and Lieb\cite{Mattis} showed that a model of interacting 
fermions proposed by Luttinger\cite{Luttinger-model} (a model which allowed
only forward scattering and in which the cutoff was taken to infinity) 
could be solved exactly. This solution demonstrated explicitly that the
low energy degrees of freedom of the interacting one-dimensional electron
gas are collective modes of charge density and spin density.  Nonetheless a
Fermi surface survives in the sense that the single-particle occupancy is 
non-analytic at exactly the same locations in momentum space as the free
Fermi gas, 
\begin{equation}
n_k \sim n_{k_F} - {\rm const}~ \sgn(k - k_F)~ |k - k_F|^{2 \zeta}\ ,
\end{equation}
where the exponent $\zeta$ depends on the interaction strength\cite{Voit}.    
The Luttinger liquid fixed point has $U(1)_L \otimes U(1)_R$ symmetry which
reflects the conservation of the number of quasiparticles at each of the two
Fermi points.  Due to this enhanced symmetry, the Hamiltonian can be expressed
in terms of left and right current operators, 
$J_L(x)~ =~ :\psi_L^\dag(x) \psi_L(x):$
and
$J_R(x)~ =~ :\psi_R^\dag(x) \psi_R(x):$,
and the bare two-body interaction can be replaced in the low-energy limit
by the effective interaction 
\begin{equation}
H_{int} = \int dx~ J_L(x) J_R(x)\ .
\end{equation}
As the current operators obey a simple type of harmonic-oscillator algebra, the
problem reduces to that of diagonalizing a Gaussian Hamiltonian. 

In spatial dimension $D > 1$, Haldane pointed out that the 
$U(1)_L \otimes U(1)_R$ symmetry enlarges, in the case of a Landau Fermi
liquid, to the infinite $U(1)^\infty_k$ symmetry\cite{Haldane}.  
Although the bare interacting fermion
Hamiltonian has only global $U(1)$ symmetry reflecting global conservation
of charge, the Landau semiclassical energy functional 
\begin{equation}
\epsilon_\bk = \epsilon^0_\bk + \sum_{\bk^\prime} f_{\bk \bk^\prime}~ 
\delta n_{\bk^\prime}\ ,
\end{equation}
where $\delta n_\bk$ is the
quasiparticle occupancy and $\bk$ and $\bk^\prime$ denote momenta on the Fermi
surface, has the same $U(1)^\infty_k$ symmetry as the free system because 
the energy $\epsilon_\bk$ depends only on the occupancies 
$n_\bk$, each of which separately is invariant under the $U(1)^\infty_k$ 
transformation equation \ref{kU(1)}.  Haldane
proposed that a fully quantum mechanical description of the interacting 
fermion system might be obtained by bosonization, at least in cases where
there is a well defined Fermi surface.  In this viewpoint the formation of the
Fermi surface is a zero temperature quantum critical phenomenon, an idea
anticipated by Anderson in his book {\it Basic Notions of Condensed Matter
Physics}\cite{PWAbook}.  
Of course the bare Hamiltonian does not exhibit the enlarged $U(1)^\infty_k$
symmetry; rather only the effective low-energy theory has this symmetry.  
Symmetry-breaking terms
are irrelevant perturbations of the fixed point effective Hamiltonian.  Indeed,
for fermions with short-range interactions, this was shown explicitly
in renormalization group (RG) calculations in a mathematical-physics style
by Feldman and Trubowitz\cite{Feldman} and 
by Benfatto and Gallavotti\cite{Benfatto}.  Shankar's particularly clear
exposition of the RG approach to interacting fermions is the one we draw upon
most heavily in the following\cite{Shankar}.   

A pioneering attempt to generalize the bosonization approach to spatial
dimension greater than one was made by Luther\cite{Luther}.  Luther studied
free fermions in general dimension by bosonizing the fermion fields 
at each point on the
Fermi surface.  The central idea was that the spectrum of the free fermion
Hamiltonian can be reproduced by the particle hole excited states in the radial
direction in momentum space, 
\begin{equation}
H_0 = v_F~ \sum_\bk (|\bk| - k_F)~ \psi^{\dag \alpha}_\bk \psi_{\bk \alpha}\ , 
\end{equation}
and therefore it is possible to take the fermions to be one dimensional in each 
radial direction $\bk$.  In this construction, however, conservation of 
momentum makes it impossible for particle-hole pairs in different radial 
sectors to interact.  This is where the subject remained until 
Haldane\cite{Haldane}
showed that coupling between different points of the Fermi surface could be
incorporated by adopting a different geometry: he coarse grained the degrees
of freedom near the Fermi surface into patches of finite extent rather than
along radial rays.  The approach was developed as a calculational tool by
the authors and their collaborators who emphasized the role of the current 
algebras\cite{Tony,HKM,HKMS,KHM,HKM2}.  Independently, Castro-Neto and Fradkin 
used a coherent-state formalism which emphasized fluctuations in the geometry 
of the Fermi surface to derive similar results\cite{Castro,Castro2}. 
A program to include systematically corrections to the Gaussian
bosonic fixed point has been initiated by Fr\"ohlich, G\"otschmann, and 
Marchetti\cite{Frohlich} and also 
by Kopietz and co-workers\cite{Kopietz,Kopietz_book}.  One of the goals of
this approach, which employs a Hubbard-Stratonovich transformation to 
replace the fermions with bosons, 
is to include non-linear terms in the fermion dispersion, as these
terms are especially important in gauge theories of interacting fermions.  
See also work by Khveshchenko\cite{geometrical}.
Some of these later developments are summarized in a review article
by Metzner, Castellani and DiCastro\cite{Metzner} who discuss the close
connection between multidimensional bosonization and the 
Ward identity approach\cite{Dzyaloshinsky,Castellani}.

It has been demonstrated how the main results of Fermi liquid theory can be
recovered within this multidimensional bosonization 
approach\cite{Haldane,Tony,HKM,HKMS,KHM,HKM2,Castro,Castro2,Kopietz}.  
Furthermore,
the stability of the Fermi liquid fixed point can be tested\cite{HKM,HKMS}. 
More important, bosonization provides a 
non-perturbative approach to the study of non-Fermi liquid fixed points that 
arise for example in the study of single layer quantum Hall fluids at 
even-denominator fillings $\nu = 1/2, 1/4, \ldots$, and which may possibly 
be relevant in the cuprate superconductors, as we discuss below in 
Section \ref{nest}.  In this instance, bosonization is a particularly
appropriate tool, as within limitations described below it does not presuppose
a Fermi liquid form for the quasiparticle propagator, and therefore non-trivial
zero-temperature fixed points can be accessed.   

The outline of the rest of our review is as follows.  In 
Section \ref{essentials} we 
introduce the essentials of multidimensional bosonization.  
In Section \ref{stability} 
we employ the renormalization group to examine the stability of the Fermi 
liquid fixed point.  The thermodynamics of interacting fermions is examined in 
Section \ref{thermodynamics}.  Collective modes are exhibited easily in the
bosonized picture, more easily perhaps than in the traditional fermion 
approach, and an illustrative calculation of the spectrum of the charge 
mode is presented in Section \ref{collective}.  
The difficult problem of calculating the fermion propagator
is addressed in Section \ref{propagator}.  Two-particle properties such as 
the linear response to external density perturbations are examined in 
Section \ref{two_particle}.  Although bosonization is limited to a description
of low energy phenomena, high momentum processes are contained accurately.  
In particular, we show that the Kohn anomaly at wavevector $|\bq| \approx 
2 k_F$ is recovered within bosonization. 
In Section \ref{hfll} we turn our attention to the 
non-Fermi liquid fixed points exemplified by composite fermion theories of
the half-filled Landau level in the fractional quantum Hall effect.  
The gauge-invariant density response is calculated in Section \ref{response}.
We discuss nested Fermi surfaces and their possible relevance to cuprate
superconductivity in Section \ref{nest}.
Finally we conclude in Section \ref{conclusion} with some further thoughts 
about the usefulness of multidimensional bosonization.  

\section{Basics of Bosonization}
\label{essentials}
\setcounter{equation}{0}

\begin{figure}
\center
\epsfxsize = 10.0cm \epsfbox{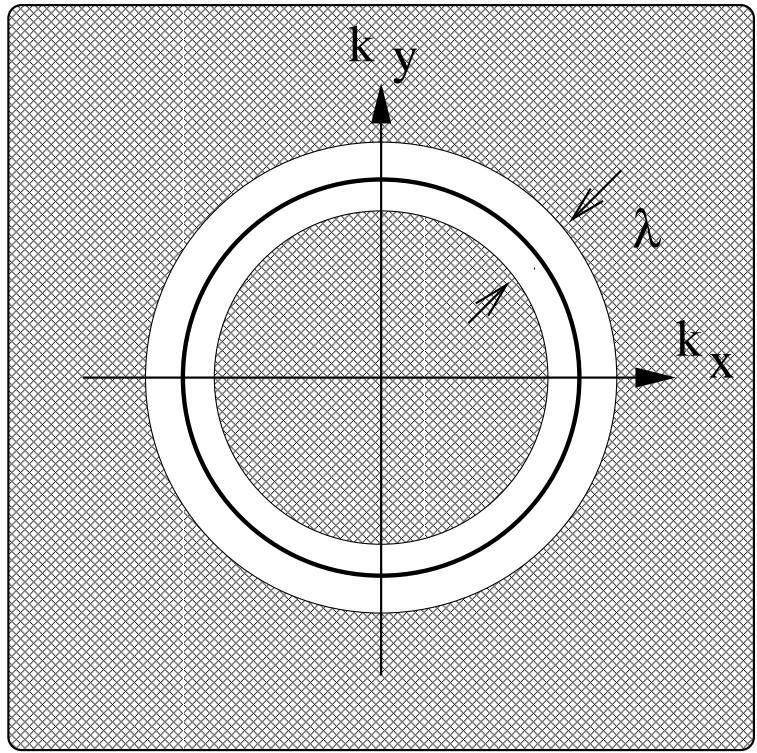}
\caption{
\label{integrate}
High-energy degrees of freedom, indicated by the hatched region, are
integrated out yielding an effective theory for the remaining low-energy 
degrees of freedom near the Fermi surface (circle).  These are to
be bosonized.} 
\end{figure}

\begin{figure}
\center
\epsfxsize = 10.0cm \epsfbox{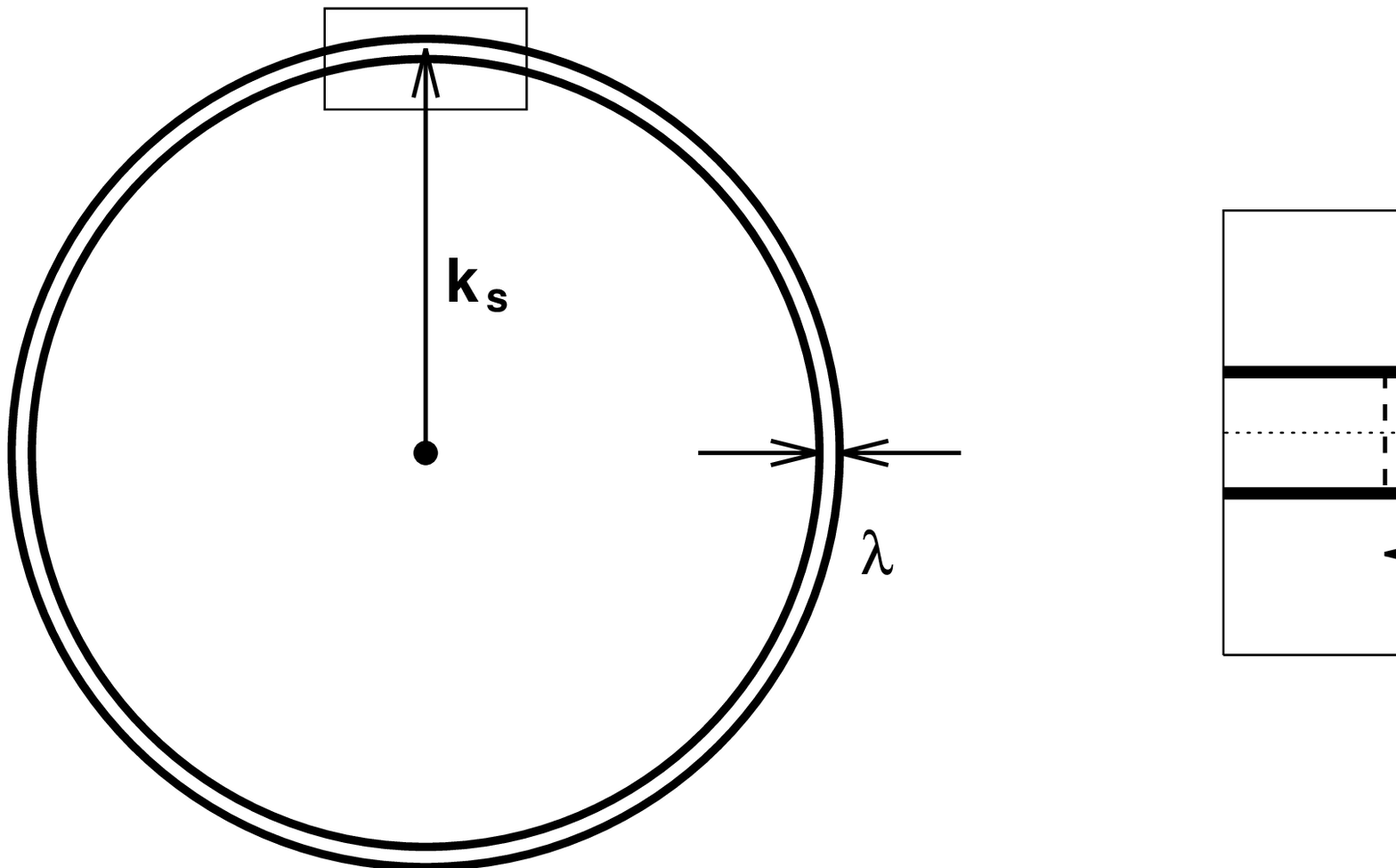}
\caption{
\label{shell}
The narrow shell of thickness $\lambda$ which contains the Fermi surface. 
The surface is tiled with squat boxes; the blowup shows one of the boxes at 
position $\bs$ which contains a particle-hole pair of momentum $\bq$. 
We require $|\bq| < \lambda \ll \Lambda \ll k_F$ for 
multidimensional bosonization to work (see text).}
\end{figure}

In this section we formulate the problem of an interacting Fermi liquid using a
bosonization scheme which is applicable in any spatial dimension. We start from
the bare Hamiltonian of fermions interacting via two-body interactions.
Initially, for simplicity, we consider a spherical Fermi surface. The bare
Hamiltonian is given by
\begin{equation}
H = \int d^Dx~ c^{\dag \alpha} ({\bf x}) \left(\frac{-\nabla^{2}}{2
m}\right) c_\alpha ({\bf x})+\frac{1}{2} \int d^Dx d^Dy~ V({\bf x}-{\bf y}) 
c^{\dag \alpha} ({\bf x}) c^{\dag \beta}({\bf y}) c_\beta({\bf y}) 
c_\alpha ({\bf x}) \ .
\label{bare-H}
\end{equation}
First we make use of the renormalization group to integrate out the high-energy
degrees of freedom\cite{Shankar}.  The region of integration is depicted in 
figure \ref{integrate}.  
The result is a low energy effective Hamiltonian
describing quasiparticles with an effective mass $m^*$ 
interacting via two-body and higher-order
interactions which, in general, include the long range Coulomb interaction
as well as short range Fermi liquid interactions.  Field operators
$\psi_{{\bf k}}^\dag$ and $\psi _{{\bf k}}$ create and destroy these
quasiparticles and are related to the bare fermion operators by
\begin{equation}
\psi_\bk = Z_\bk^{-1/2} c_\bk 
\label{Zk}
\end{equation}
for momenta ${\bf k}$ which are restricted to a narrow shell of thickness
$\lambda$ around the Fermi surface $k_F - \lambda/2 < |\bk| < k_F + \lambda/2$
depicted in figure \ref{shell}.  Note that an attempt is not being made to 
calculate the values of $m^*$, $Z_\bk$, or the other effective parameters
from first principles; there is no generally reliable way to do this.  Rather
we focus on the remaining low-energy degrees of freedom and the effective
theories which control them.  

\begin{figure}
\center
\epsfxsize = 10.0cm \epsfbox{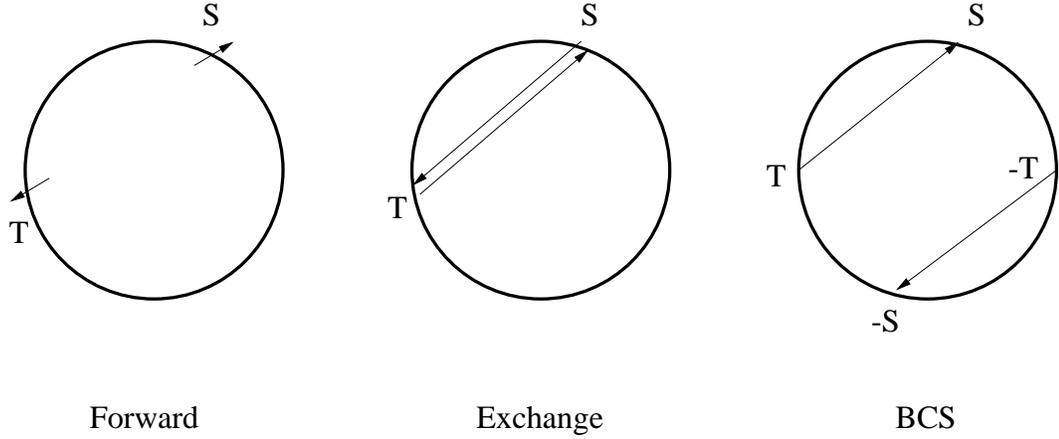}
\caption{
\label{channels}
The three marginally relevant scattering channels in $D = 2$ which are 
confined to the narrow shell of width $\lambda$ about the Fermi surface and
which conserve momentum.} 
\end{figure}

\begin{figure}
\center
\epsfxsize = 10.0cm \epsfbox{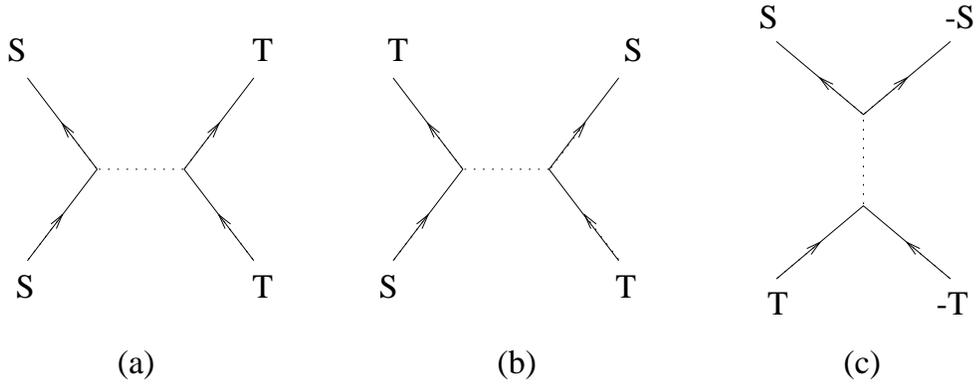}
\caption{
\label{diagrams}
Feynman diagrams corresponding to the three marginally relevant scattering
channels. The interactions between fermions (solid lines) are mediated by 
gauge fields (dotted line). (a) Forward scattering with amplitude $f_c(\bs, \bt)
- \frac{1}{4}~ f_s(\bs, \bt)$.  
(b) Exchange scattering with amplitude $\frac{1}{2}~ f_s(\bs, \bt)$.  
(c) BCS Cooper pair scattering with amplitude $V_{BCS}(\bs - \bt)$.}
\end{figure}

As the next step, the Fermi surface is coarse grained by tiling it
with squat boxes of height $\lambda$ perpendicular to the Fermi surface and
dimension $\Lambda^{D-1}$ along the surface.  The boxes are small and have a 
squat aspect ratio in the relevant limits $\lambda \ll \Lambda \ll k_F$ and 
$\Lambda^2 / k_F \ll \lambda$.  The reason for these limits, which are
of crucial import, is discussed below. 
We introduce operators 
$\psi_\alpha({\bf S}; {\bf x})$ and $\psi^{\dag \alpha}({\bf S}; {\bf x})$ 
which respectively 
annihilate and create fermion quasiparticles inside each patch
\begin{equation}
\psi_\alpha({\bf S}; \bx) = e^{i{\bf k}_{\bs} \cdot {\bf x}} 
\sum_\bp \theta({\bf S}; \bp) e^{i (\bp - \bk_\bs) \cdot {\bf x}}~ 
\psi_{\bp \alpha}\ , 
\end{equation}
where $k_{\bf S}$ is the Fermi momentum located at the center of the squat box
labeled by the coordinate ${\bf S}$.  Also, we introduce 
$\theta({\bf S}; {\bf p})$ which equals one if
$\bp$ lies inside the box and is zero otherwise.  In terms of these operators
the effective interaction is an instantaneous two-body interaction of the form 
\begin{equation}
\psi^{\dag \alpha}({\bf S}; {\bf x}) \psi^{\dag \beta}({\bf T}; {\bf y}) 
\psi_\beta({\bf U}, {\bf y}) \psi_\alpha({\bf V}, {\bf x})\ . 
\end{equation}
This is the only marginal interaction as all terms with more 
operators or derivatives
are irrelevant in the usual renormalization group sense\cite{Shankar};  
retardation may be neglected for the same reason.  The bare Hamiltonian,
equation \ref{bare-H}, is rotationally invariant, and
barring spontaneous magnetization
we can restrict our attention to rotationally invariant effective 
interactions.  The
interaction is heavily constrained by momentum conservation which demands that
\begin{equation}
|{\bf k}_{\bf U} + {\bf k}_{\bf V} - {\bf k}_{\bf S} - {\bf k}_{\bf T} | <
\lambda \ll \Lambda\ .
\end{equation} 
In two dimensions there are only three possibilities: ${\bf V}
= {\bf S}$ and ${\bf U} = {\bf T}$ (forward scattering); 
${\bf U} = {\bf S}, {\bf V} = {\bf T}$ (exchange scattering); 
and ${\bf S}= -{\bf T}, \ {\bf U} = -{\bf V}$ (BCS Cooper pair scattering, 
where the (-) sign connotes the antipodal point).
The three scattering processes are depicted in figure \ref{channels} and the
corresponding Feynman diagrams appear in figure \ref{diagrams}. 
In three dimensions there is the additional complication of out-of-plane
scattering\cite{Shankar}.  For the sake of simplicity we ignore these 
azimuthal processes here.

It is clear that the interactions in the forward and
exchange channels possess the large $U(1)^\infty_k$ symmetry:  
a $U(1)$ phase rotation of both spin species of fermions in
each squat box leaves the interaction in these two channels invariant as
patches $\bs$ and $\bt$ each contain a creation and an annihilation operator. 
(There is no corresponding $SU(2)^\infty_k$ spin symmetry, as the exchange
of spin between two separate patches on the Fermi surface breaks this
symmetry.) The BCS interaction which couples quasiparticles in four separate 
patches breaks the $U(1)^\infty_k$ symmetry down to the ordinary global
$U(1)$ symmetry which just reflects overall conservation of charge.  
We defer analysis of the BCS term until the next section,
where it is shown to be marginally irrelevant or relevant depending
on the sign of the coupling, and consider the remainder of the
effective Hamiltonian:
\begin{eqnarray}
H[\psi^\dagger, \psi] &=& 
\int d^Dx \sum_\bs \bigg{\{} \psi^{\dag \alpha}({\bf S}; \bx) 
\bigg{(} \frac{\bk_\bs}{m^*} \cdot \frac{{\bf \nabla}}{i} \bigg{)} 
\psi_\alpha({\bf S}; {\bf x}) \bigg{\}}  
\nonumber \\
&+& \int d^Dx d^Dy~ \sum_{\bs \bt} \bigg{\{} 
f_c({\bf S}, {\bf T}; {\bf x}-{\bf y}) 
[\psi^{\dag\alpha}({\bf S}; {\bf x}) \psi_\alpha({\bf S}; {\bf x})]
[\psi^{\dag \beta}({\bf T}; {\bf y}) \psi_\beta({\bf T}; {\bf y})] 
\nonumber \\
&+& \frac{1}{4}~ f_s({\bf S}, {\bf T}; {\bf x}-{\bf y}) 
[\psi^{\dag \alpha}({\bf S}; {\bf x})~ 
\vec{\sigma}^\beta_\alpha~ \psi_\beta({\bf S}; {\bf x})] \cdot 
[\psi^{\dag \gamma}({\bf T}; {\bf y})~ \vec{\sigma}_\gamma^\delta 
\psi_\delta({\bf T}; {\bf y})] \bigg{\}}\ .
\label{effective}
\end{eqnarray}
The quasiparticle spectrum has been linearized about the Fermi surface,
$\epsilon({\bf k}) = v_F^* (|{\bf k}|-k_F)$, and the quasiparticle velocity
defines the effective mass through the relation $v_F^* = k_F/m^*$.
Functions $f_c(\bs, \bt; \bx - \by)$ and $f_s(\bs, \bt; \bx - \by)$ 
parameterize the effective interactions in the forward and exchange channels.
Upon making use of the Pauli matrix identity
\begin{equation}
\vec{\sigma}^\beta_\alpha \cdot \vec{\sigma}_\gamma^\delta 
= 2 \delta^\beta_\gamma \delta^\delta_\alpha - 
\delta^\beta_\alpha \delta^\delta_\gamma
\end{equation}
it is straightforward to show that the forward scattering channel has
amplitude $f_c - f_s/4$ while the exchange channel has amplitude $f_s/2$.
Again no attempt is made here to calculate the short-range part of the 
parameters
$f_c$ and $f_s$ as this is of comparable difficulty to a first-principles
calculation of, say, the boiling point of water.  Long-range density-density
interactions such as the Coulomb repulsion, on the other hand, 
determine\cite{Silin} the form of $f_c(\bs, \bt; \bx - \by)$ in the limit of
large separation, $|\bx - \by| \rightarrow \infty$.  The remaining low-energy 
degrees of freedom within the shell are responsible for the screening of this 
long-range interaction, if screening occurs (see Section \ref{propagator}).  

To bosonize the effective Hamiltonian, equation \ref{effective}, we 
introduce the coarse grained currents
\begin{equation}
J_\alpha({\bf S}; {\bf q}) = \sum_{\bf k} \theta({\bf S}; {\bf k} - {\bf q}) 
\theta({\bf S}; {\bf k}) \left\{ \psi_{\bk - \bq \alpha}^\dag 
\psi_{\bk \alpha} - \delta^D_{{\bf q}, {\bf 0}} n_{\bk \alpha} \right\} 
\label{current}
\end{equation}
here 
$n_{\bk \alpha} = \langle \psi_{\bk \alpha}^\dag \psi_{\bk \alpha} \rangle$
and the spin index $\alpha$ is not summed over, so all of the spin indices
appear as subscripts.  
The index ${\bf S}$ labels a patch, with ${\bf S} \equiv (\theta, \phi)$ in
three dimensions, on the Fermi surface at momentum ${\bf k}_\bs$.  Recall that
$\theta({\bf S}; \bk) = 1$ if ${\bf k}$ lies inside a squat box centered on
${\bf S}$ with height $\lambda$ in the radial (energy) direction and area
$\Lambda^{D-1}$ along the Fermi surface; $\theta({\bf S}; {\bf k}) = 0$
otherwise.  These two scales are made small in the sense that 
$k_F \gg \Lambda \gg \lambda > |\bq|$ by setting $\lambda = k_F / N$, and
$\Lambda = k_F /N^\alpha$ where $\alpha < 1$ and $N \rightarrow \infty$. We also
require that $\alpha > 1/2$ as otherwise the circular Fermi surface curves out 
of the squat box in the $N \rightarrow \infty$ limit.  
The relationship between these currents and the usual {\it total} density 
operator $\rho(\bq)$ and total current operator ${\bf J}(\bq)$ 
familiar to condensed matter physicists is, for small-$\bq$,
\begin{equation}
\rho(\bq) = \sum_\bs \sum_{\alpha = \uparrow, \downarrow} J_\alpha(\bs; \bq)
\end{equation}
and
\begin{equation}
{\bf J}(\bq) = \sum_\bs \sum_{\alpha = \uparrow, \downarrow} 
{\bf v}_F(\bs)~ J_\alpha(\bs; \bq) 
\end{equation}
where ${\bf v}_F(\bs) = \bk_\bs / m^*$ is the fermion velocity at patch $\bs$.

The coarse-grained current $J_\alpha(\bs; -\bq)$ creates a particle-hole pair of
relative momentum $\bq$ and spin $\alpha$ in patch $\bs$ of the Fermi surface.
As discussed in the Introduction, we expect $J_\alpha(\bs; \bq)$ to have a 
bosonic character.  Indeed, the currents satisfy the $U(1)$ current algebra
\begin{equation}
\left[J_\alpha({\bf S}; {\bf q}), J_\beta({\bf T}; {\bf p})\right] =
\delta_{\alpha, \beta}~ \delta_{{\bf S}, {\bf T}}^{D-1}~
\delta_{\bq + \bp, 0}^D~ \Omega~ \hbn_\bs \cdot \bq \ .
\label{current_algebra}
\end{equation}
Here the quantity 
\begin{equation}
\Omega \equiv \Lambda^{D-1}~ \left( \frac{L}{2\pi} \right)^D,
\end{equation}
which equals the number of states in the squat box divided by the cutoff
$\lambda$, appears for the first of many times.
Also $\hbn_\bs$ is the unit normal to the 
Fermi surface at patch ${\bf S}$.  This result can be made plausible by the 
following argument.  First, the current operators $J_\alpha(\bs; \bq)$ and
$J_\beta(\bt; \bp)$ clearly commute if $\alpha \neq \beta$.  They also
commute if $\bs \neq \bt$ as the creation and
annihilation operators are in separate boxes.  The right hand side of 
equation \ref{current_algebra}, if it is a $c$-number and not an operator, must
be non-zero only when $\bq = -\bp$ as otherwise momentum is not 
conserved.  As the anomaly must change sign upon interchanging the two
current operators, it must contain odd powers of $\bq$.  Finally, as the 
current operators are scalars, a dot product of $\bq$ with some other
vector must appear.  The only other vector is $\hbn_\bs$.   

The right hand side
of equation \ref{current_algebra} can be determined rigorously by
direct computation on making use of the canonical
anticommutation relations $\{\psi_\bq^{\dag \alpha}, \psi_{\bp \beta}\}
= \delta^\alpha_\beta \delta_{\bq, \bp}^D$.
\begin{eqnarray}
&& \left[ J_\alpha({\bf S}; {\bf q}), J_\beta({\bf T}; {\bf p}) \right]
\nonumber \\
&=& \delta_{\alpha,\beta}~ \delta_{\bs,\bt}^{D-1}~ 
\bigg{[} \sum_\bk \theta({\bf S}; {\bf k} - {\bf q} - {\bf p}) 
\theta ({\bf S}; {\bf k}) \left[ \theta({\bf S}; {\bf k}
- {\bf q}) - \theta({\bf S}; {\bf k} - {\bf p}) \right]~ 
\delta_{{\bf q} + {\bf p}, 0}^D~ n_{\bk \alpha} 
\nonumber \\
&& + \sum_{\bf k} \theta ({\bf S; k-q-p}) \theta({\bf S}; {\bf k}) \left[
\theta({\bf S}; {\bf k} - {\bf q}) - \theta({\bf S}; {\bf k}-{\bf p}) \right]
\nonumber \\
&& \times 
\left\{ \psi_{{\bf k} - {\bf p} - {\bf q} \alpha}^\dag \psi_{\bk \alpha}
- \delta_{{\bf q+p}, 0}^D~ n_{\bk \alpha} \right\} \bigg{]} 
\nonumber \\
&=& \delta_{\alpha, \beta} \delta_{\bf S, T}^{D-1} \delta_{{\bf q} + {\bf p},
0}^{D}~ \Omega~ \hbn_\bs \cdot \bq + {\rm Error}.
\label{algebra_error}
\end{eqnarray}
The first term on the right hand side of equation \ref{algebra_error}, 
called the ``quantum anomaly'' or simply ``anomaly,'' is of order
$\lambda \Lambda^{D-1}(L/2\pi)^D$. The second term, the error, 
can be bounded by replacing the operator with a $c$-number:
\begin{equation}
(\psi^\dag_{{\bf k}-{\bf q}-{\bf p} \alpha}~ \psi_{\bk \alpha}
- \delta_{{\bf q}+{\bp}, 0}^D n_{\bf k})
\rightarrow (1 - \delta_{\bq+\bp, 0}^D) 
\end{equation}
and computing the volume of the geometrical intersection of the two $\theta$
functions 
$[\theta({\bf S}; {\bf k} - {\bf q}) - \theta({\bf S}; {\bf k} - {\bf p})]$. 
Here a crucial assumption is introduced: the quasiparticle occupancy 
$n_{\bk \alpha} = 1$ deep below the Fermi surface on the lower face of the
squat box, $|\bk| = k_F - \lambda/2$.  Likewise, $n_{\bk \alpha} = 0$ 
on the upper face.  Hence, matrix elements of the error term vanish on the
upper and lower faces of the squat box, and a simple computation 
then shows that 
\begin{equation}
{\rm Error} <  
\max \{|\bq|, |\bp| \} \times \lambda \Lambda^{D-2} (L/2\pi)^D 
< \lambda^2\Lambda^{D-2}(L/2\pi)^D 
\end{equation}
where $L$ is the length of one side of the system.  The error
therefore is negligible compared to the anomaly as long as the limit 
$\lambda / \Lambda \rightarrow 0$ (or equivalently $N \rightarrow \infty$), 
is chosen.  This establishes the key $U(1)$ current algebra, 
equation \ref{current_algebra}.
The calculation of the error term makes it clear why the squat aspect ratio is 
so important: Scattering of quasiparticles between adjacent boxes 
is minimized by the construction\cite{Tony}.  Unlike bosonization in one
spatial dimension, which is accurate even when the cutoff $\lambda$ is an
sizable fraction of 
the Fermi momentum, in higher spatial dimensions bosonization is more delicate.
Because the Fermi surface curves, we require $\Lambda \ll k_F$, and to 
minimize interpatch scattering we require $\lambda \ll \Lambda$; hence
$\lambda \ll k_F$ and the quasiparticle occupancy must change from $1$ to $0$
over the small momentum range $\lambda$.  
This point has not received sufficient attention in the literature.

The charge currents also have an ``Abelian'' bosonic representation.  If we set
\begin{equation}
J_\alpha({\bf S}; {\bf x}) = 
\sqrt{4\pi}~ \hbn_\bs \cdot {\bf \nabla} \phi_\alpha({\bf S}; {\bf x}),
\end{equation}
then we must demand that the bosonic fields 
$\phi_\alpha({\bf S}; {\bf x})$ satisfy the commutation relations:
\begin{equation}
\left[\phi_\alpha ({\bf S}; {\bf x}),
\phi_\beta({\bf T}; {\bf y})\right] = \left\{ \begin{array}{ll}
i \frac{\Omega^2}{4}~ \delta_{\alpha, \beta}~ \delta_{{\bf S}, {\bf T}}^{D-1} 
\epsilon[\hbn_\bs \cdot (\bx -\by)]; 
& | \hbn_\bs \times ({\bf x}-{\bf y})| \Lambda \ll 1 \\
\\ 
0; & | \hbn_\bs \times ({\bf x} - {\bf y})| \Lambda \gg 1\ .
\end{array}\right.
\label{commutation}
\end{equation}
Here $\epsilon(x) = 1$ if $x > 0$ and $-1$ if $x < 0$.
This can be accomplished by decomposing the boson fields into canonical 
boson annihilation and creation operators
\begin{equation}
\phi_\alpha({\bf S}; {\bf x}) =~ i \sum_{\bq, \hbn_\bs \cdot \bq > 0}
\frac{a_\alpha^\dag (\bs; -\bq) e^{-i{\bf q} \cdot {\bf x}} 
- a_\alpha({\bf S}; {\bf q}) 
e^{i{\bf q} \cdot {\bf x}}}{\sqrt{\hbn_\bs \cdot \bq}}
\label{represent}
\end{equation}
where
\begin{equation}
\left[a_\alpha (\bs; \bq), a_\beta^\dag(\bt; \bp) \right]
=\delta_{\bs, \bt}^{D-1}~ \delta_{\bq, \bp}^D~ \delta_{\alpha, \beta}
\end{equation}
then
\begin{equation}
J_\alpha({\bf S}; {\bf q}) = \sqrt{\Omega |\hbn_\bs \cdot \bq|} \left[
a_\alpha({\bf S}; {\bf q}) 
\theta(\hbn_\bs \cdot {\bf q}) + a_\alpha^\dag({\bf S}; {\bf -q})
\theta(-\hbn_\bs \cdot \bq) \right]
\label{current-rep}
\end{equation}
and equation \ref{current_algebra} follows immediately.  
We have thus replaced the fermion bilinear appearing in equation \ref{current} 
with a single boson operator.
The boson operator acts on the quiescent Fermi liquid to produce a single
particle-hole pair, which as noted above and in the Introduction, 
has bosonic character.

\begin{figure}
\center
\epsfxsize = 10.0cm \epsfbox{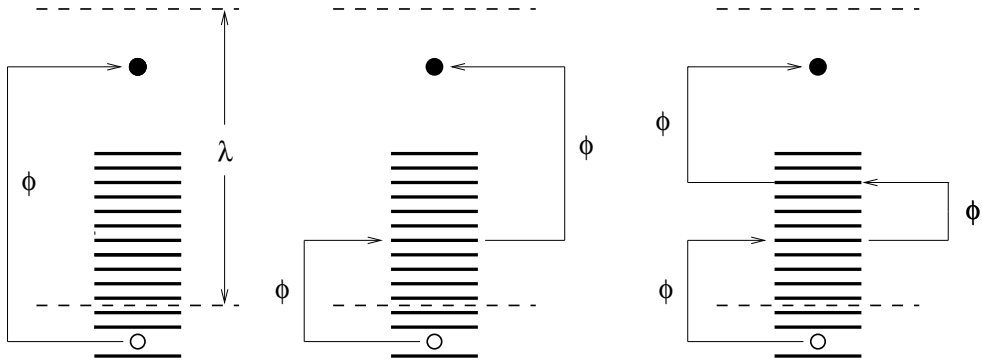}
\caption{
\label{ramp}
The Fermi quasi-particle $\psi^\dag$ formed by repeated action of the operator
$\phi$ which produces particle-hole pairs.  One hole is left behind beneath
the cutoff, so there are an odd number of fermions within the low-energy
region specified by the ultraviolet cutoff $\lambda$.}
\end{figure}

We now introduce the key formula of the bosonization procedure 
which expresses the fermion quasiparticle field operator $\psi$ as an 
exponential\cite{Schotte} of the boson fields:  
\begin{equation}
\psi_\alpha({\bf S}; {\bf x}) = \frac{1}{\sqrt{V}} \sqrt{\frac{\Omega}{a}}
e^{i{\bf k}_{\bf S} \cdot {\bf x}} \exp \left\{ i\frac{\sqrt{4\pi}}{\Omega}
\phi_\alpha({\bf S}; {\bf x}) \right\} \hat{O}({\bf S})\ .
\label{bosonization}
\end{equation}
Here $V$ is the system volume, $a \equiv 1/\lambda$ is the ultraviolet cutoff, 
and $\hat{O}({\bf S})$ is an
ordering operator introduced to maintain Fermi statistics in the angular
direction along the Fermi surface
\begin{equation}
\hat{O}(\bs) \equiv \exp \left[i \pi \sum_{\bt = 1}^{\bs - 1} 
\sum_{\alpha = \uparrow, \downarrow}~ J_\alpha(\bt; \bq = {\bf 0}) \right]
\label{order}
\end{equation}
where the patch labels ${\bf T}$ have been arranged consecutively. For example,
the order can be chosen to start at point $\bt = 1$ on the Fermi surface, 
spiral outward in a clockwise fashion, and finally converge at the 
corresponding antipode\cite{Luther}.  The operator 
$J_\alpha(\bs; {\bf 0})$ appearing in equation \ref{order} simply 
counts the number of quasiparticles in box $\bt$, so $\hat{O}(\bs)$ supplies
the extra minus signs, much like the Jordan-Wigner transformation\cite{Jordan},
which 
preserves anticommuting statistics.  In practice it is easy to identify these
minus signs by hand and the explicit form of $\hat{O}(\bs)$ is not
needed usually.  Anticommutation rules are satisfied
automatically in the direction perpendicular to the Fermi surface, as may
be verified easily either by using the commutation relations, 
equation \ref{commutation}, or by examination of the fermion propagator (see
Appendix \ref{fermion_details}).
How can the fermion operator $\psi$ possibly be equated to
the exponential of the  
bosonic operator $\phi$?  To understand this remarkable relationship it is 
important to appreciate the crucial role played by the cutoff $\lambda$.  As
depicted in figure \ref{ramp}, we can think of $\psi^\dag$ as a sort of coherent
state formed out of multiple particle-hole pairs\cite{vonDelft}.  
A hole is left beneath the
cutoff, and is therefore beyond the horizon of the low-energy degrees of 
freedom.

It is important to test for closure using this bosonized representation 
of the fermion fields.  We seek to verify that 
\begin{eqnarray}
J_\alpha({\bf S}; {\bf x}) &=&  V \lim_{\epsilon \rightarrow 0} 
\left\{ \psi_\alpha^\dag({\bf S}; {\bf x} + \epsilon \hbn_\bs) 
\psi_\alpha({\bf S}; \bx) 
- \langle \psi_\alpha^\dag({\bf S}; \bx + \epsilon \hbn_\bs)
\psi_\alpha(\bs; \bx) \rangle \right\} 
\nonumber \\
&=& \sqrt{4\pi}~ \hbn_\bs \cdot {\bf \nabla} \phi_\alpha({\bf S}; \bx)\ .
\label{closure}
\end{eqnarray}
Here, in the first line, the current operator has been normal-ordered by 
point-splitting.  That is, the expectation value of the fermion bilinear
with fields 
at slightly different positions $\bx$ and $\bx + \epsilon \hbn_\bs$ has been
subtracted, to accord with the definition in momentum space, 
equation \ref{current}.  To obtain the second line, 
first use the bosonization formula, equation \ref{bosonization}, to write
\begin{equation}
\psi_\alpha^\dag(\bs; \bx + \epsilon \hbn_\bs) \psi_\alpha(\bs; \bx) 
= \frac{1}{V} \frac{\Omega}{a} \exp \left[ -i\frac{\sqrt{4\pi}}{\Omega} 
\phi_\alpha(\bs; \bx + \epsilon \hbn_\bs) \right] 
\times \exp \left[i\frac{\sqrt{4\pi}}{\Omega} \phi_\alpha(\bs; \bx)\right]\ .
\end{equation}
Then apply the following useful operator identity\cite{Katmandu}: 
\begin{equation}
e^A e^B~ =~ :e^{A+B}:~ e^{\langle AB + \frac{A^2 + B^2}{2} \rangle} 
\label{key_identity}
\end{equation}
which holds if $A$ and $B$ are linear combinations of simple harmonic oscillator
creation and annihilation operators.
Here the notation $:A:$ indicates normal ordering, with 
all annihilation operators on the right, and all creation operators on the left.
This useful relationship, which we use frequently, can be derived using the
more familiar Baker-Hausdorff relation:
\begin{equation}
e^{A+B}= e^A e^B e^{-\frac{1}{2}[A,B]} = e^B e^A e^{\frac{1}{2}[A, B]}\ . 
\label{Baker-Hausdorff}
\end{equation}
Now we can write
\begin{eqnarray}
\psi_\alpha^\dag(\bs; \bx + \epsilon \hbn_\bs) \psi_\alpha(\bs; \bx)
&=& \frac{\Omega}{V a}~ 
\lim_{\epsilon \rightarrow 0} :\exp -i \frac{\sqrt{4\pi}}{\Omega} \left[
\phi_\alpha(\bs; \bx + \epsilon \hbn_\bs) - \phi(\bs; \bx) \right]: 
\nonumber \\
&\times& \exp \frac{4\pi}{\Omega^2} \langle \phi_\alpha(\bs; \bx + \epsilon
\hbn_\bs) \phi_\alpha(\bs; \bx) - \phi_\alpha^2(\bs; \bx) \rangle \ . 
\label{point-split}
\end{eqnarray}
The equal-time correlation function of the $\phi$ fields, derived in 
Appendix \ref{boson_correlation}, is given by
\begin{eqnarray}
G_\phi(\bs; \bx) = \langle \phi_\alpha(\bs; \bx) \phi_\alpha(\bs; {\bf 0}) 
- \phi_\alpha^2(\bs; {\bf 0}) \rangle 
&=& \left\{ \begin{array}{ll}
\frac{\Omega^2}{4\pi} \ln \frac{ia}{\hbn_\bs \cdot \bx + ia}; 
& | \hbn_\bs \times \bx | \Lambda \ll 1 \\
\\
-\infty; 
& | \hbn_\bs \times \bx | \Lambda \gg 1
\end{array}\right.
\end{eqnarray}
Inserting this result into equation \ref{point-split} yields:
\begin{equation}
\psi_\alpha^\dag(\bs; \bx + \epsilon \hbn_\bs) \psi_\alpha(\bs; \bx) 
= \frac{\Omega}{Va}~ : [1 - i \epsilon \frac{\sqrt{4\pi}}{\Omega} 
\hbn_\bs \cdot {\bf \nabla} \phi_\alpha({\bs}, \bx) + O(\epsilon^2)]:~ 
\bigg{(} \frac{i a}{\epsilon + i a} \bigg{)}.
\end{equation}
The first, constant, term in this expansion about $\epsilon = 0$ is subtracted
off in the point-splitting convention of equation \ref{closure} and the
continuum limit may be taken.  The correct order of limits is
$\lim_{\epsilon\rightarrow 0} \lim_{a \rightarrow 0}$ because the separation
between the fields $\epsilon$
must always be much greater than the ultraviolet lattice cutoff $a$. As required
by consistency $J_\alpha({\bf S}; {\bf x}) = \sqrt{4\pi} \hbn_\bs \cdot 
{\bf \nabla} \phi_\alpha(\bs; \bx)$.  Similarly, by careful 
use of point-splitting it can be shown that the free part of the Hamiltonian is 
quadratic in the currents,
\begin{eqnarray}
H_0 &=& \ve \sum_\bs \int d^Dx~ \psi^{\dag \alpha}(\bs; \bx)
\left(\hbn_\bs \cdot \frac{\bf \nabla}{i} \right) \psi_\alpha(\bs; \bx) 
\nonumber \\
&=& \frac{\ve}{2 \Omega V} \sum_{\bs; \alpha} \int d^Dx [J_\alpha(\bs; \bx)]^2
\nonumber \\
&=& \frac{2\pi \ve}{\Omega V} \sum_{\bs; \alpha} \int d^Dx 
[\hbn_\bs \cdot {\bf \nabla} \phi_\alpha(\bs; \bx)]^2
\end{eqnarray}
quadratic because it has been assumed that the spectrum can be linearized near
the Fermi surface.  Non-linear corrections to the dispersion, 
when significant, lead to cubic
and higher order terms in the boson fields which can only be treated
perturbatively\cite{old-Haldane}.  
As a final check on the bosonization formula we compute the equal-time
free fermion Green's function by again making use of 
equation \ref{bosonization}.
\begin{eqnarray}
G_{F\alpha}^0({\bf S}; {\bf x}) = \langle \psi_\alpha({\bf S}; {\bf x})
\psi^\dag_\alpha(\bt; {\bf 0}) \rangle 
&=& \frac{\Omega}{Va}~ \delta_{\bs, \bt}^{D-1}~ e^{i \bk_\bs \cdot \bx}~
\exp \left[\frac{4 \pi}{\Omega^2}~ G_\phi({\bf S}; {\bf x}) \right]  
\nonumber \\
&=& \left\{ \begin{array}{ll}
i~ \frac{\Lambda^{D-1}}{(2 \pi)^D}~ 
\frac{e^{i \bk_\bs \cdot \bx}}{\bx \cdot \hbn_\bs + ia}; & 
|\hbn_\bs \times \bx| \Lambda \ll 1 \\
\\
0; & {\rm otherwise}
\end{array}\right.
\end{eqnarray}
as expected.  The low-energy fermion quasiparticle is related to the 
patch fermions by a sum over the patches
\begin{equation}
\psi_\alpha(\bx) = \sum_\bs \psi_\alpha(\bs; \bx)~ ,
\end{equation} 
so the usual Green's function $G_F({\bf x})$ can be
obtained by summing over the patches, 
\begin{equation}
G_F({\bf x}) = {\sum_\bs}^\prime G_F(\bs; \bx),
\end{equation}
where the prime indicates that the sum is only over patches for which $|{\bf x}
\times \hbn_\bs| \Lambda < 1$.  Physically this means that fermions 
in each patch, $\psi(\bs; \bx)$, propagate along nearly straight rays,
reminiscent of semiclassical ballistic trajectories. 

The $U(1)^\infty_k$ transformation rotates the phase of the fermions by
a different amount $\theta_{\bs \alpha}$ in each patch:
\begin{equation}
\psi_\alpha(\bs; \bx) \rightarrow e^{i \theta_{\bs \alpha}} 
\psi_\alpha(\bs; \bx)~ .
\end{equation}
The transformation leaves the current operators $J_\alpha(\bs; \bx)$
invariant, as the phase factors $\pm \theta_{\bs \alpha}$ cancel out in the
first line of equation \ref{closure}.  Therefore Hamiltonians or actions
which can be expressed solely in terms of the patch currents possess the
large $U(1)^\infty_k$ symmetry.  In terms of the boson fields, the
$U(1)^\infty_k$ transformations amount to a translation of the fields by
a constant:
\begin{equation}
\phi_\alpha(\bs; \bx) \rightarrow \phi_\alpha(\bs; \bx) + 
\frac{\Omega}{\sqrt{4 \pi}}~ \theta_{\bs \alpha}~ .
\end{equation}
Any Hamiltonian or action which conserves total charge is invariant under
the global $U(1)$ symmetry $\theta_{\bs \alpha} = \theta$.  The BCS channel
breaks the $U(1)^\infty_k$ symmetry down to this global symmetry, as the
phases in the four separate patches $(\bs, -\bs)$ and $(\bt, -\bt)$ do not
cancel out in general.  The exchange channel, on the other hand, 
possesses a restricted
$U(1)^\infty_k$ symmetry: both spin species must rotate through the same
angle, $\theta_{\bs \alpha} = \theta_\bs$.  The restriction reflects the
fact that while the exchange channel transports spin between two different
patches of the Fermi surface, the total number of quasiparticles in each
patch remains unchanged.  Effective actions with only forward scattering
possess the full $U(1)^\infty_k$ symmetry, however, 
and these are particularly amenable to bosonization, as shown below. 

Having shown how fermion quasiparticles are bosonized in each patch, we can now
bosonize the effective Hamiltonian, equation \ref{effective}. 
If we choose $\hat{z}$ as the spin quantization axis, 
introduce charge and spin currents
\begin{equation}
\begin{array}{ll}
J_c({\bf S}; {\bf x}) = J_\uparrow({\bf S}; {\bf x}) 
+ J_\downarrow({\bf S}; {\bf x})
\nonumber \\
J_s({\bf S}; {\bf x}) = J_\uparrow({\bf S}; {\bf x}) 
- J_\downarrow({\bf S}; {\bf x})
\end{array}
\end{equation}
and the corresponding charge and spin bosons
\begin{equation}
\begin{array}{ll}
\phi_c({\bf S}; {\bf x}) = 
\phi_\uparrow ({\bf S}; {\bf x}) + \phi_\downarrow({\bf S}; {\bf x}) 
\nonumber \\
\phi_s({\bf S}; {\bf x}) = 
\phi_\uparrow({\bf S}; {\bf x}) - \phi_\downarrow({\bf S}; {\bf x})
\end{array}
\end{equation}
the effective Hamiltonian can be written in terms of the currents as
\begin{eqnarray}
H[\psi^\dagger, \psi] 
&=& \int d^Dx~ \sum_\bs~ \frac{v_F^*}{4\Omega} [J_c^2({\bf S}; \bx) +
J_s^2({\bf S}; {\bf x})]
\nonumber \\
&+& \int d^Dx d^Dy~ \sum_{\bs, \bt} \bigg{\{} 
\frac{f_c(\bs, \bt; \bx - \by)}{V^2} J_c(\bs; \bx) J_c(\bt; \by) 
+ \frac{f_s(\bs, \bt; \bx - \by)}{V^2} J_s(\bs; \bx) J_s(\bt; \by) 
\nonumber \\
&+& f_s(\bs, \bt; \bx - \by)~ 
[\psi_\uparrow^\dag(\bs; \bx) \psi_\downarrow(\bs; \bx) 
\psi_\downarrow^\dag(\bt; \by) \psi_\uparrow(\bt; \by)
+ \uparrow \leftrightarrow \downarrow \bigg{\}}
\end{eqnarray}
In terms of the $\phi$ fields it becomes
\begin{eqnarray}
H[\phi_c, \phi_s] &=& \int d^Dx~ \sum_\bs \frac{\pi v_F^*}{\Omega} \left\{ 
[\hbn_\bs \cdot {\bf \nabla} \phi_c({\bf S}; {\bf x})]^2 + 
[\hbn_\bs \cdot \nabla \phi_s({\bf S}; {\bf x})]^2\right\}
\nonumber \\
&+& \frac{1}{V^2}~ \int d^Dx d^Dy~ \sum_{\bs, \bt}~ \bigg{\{}
4 \pi f_c(\bs, \bt; \bx - \by) 
[\hbn_\bs \cdot {\bf \nabla} \phi_c({\bf S}; {\bf x})]
[\hbn_\bt \cdot {\bf \nabla} \phi_c({\bf T}; {\bf y})] 
\nonumber \\
&+& 4 \pi f_s(\bs, \bt; \bx - \by) \bigg{(}
[(\hbn_\bs \cdot {\bf \nabla} \phi_s({\bf S}; {\bf x})]
[\hbn_\bt \cdot {\bf \nabla} \phi_s({\bf T}; \by)] 
\nonumber \\
&+& 2 \left(\frac{\Omega}{a}\right)^2 
\cos \frac{\sqrt{4\pi}}{\Omega} [\phi_s({\bf S}; \bx) - \phi_s({\bf T}; \by)]
\bigg{)} \bigg{\}}
\label{hamspincharge}
\end{eqnarray}
Although the spin and charge degrees of freedom separate in the Hamiltonian,
equation \ref{hamspincharge}, it does not follow necessarily that the spin and
charge excitations propagate at different velocities.  

It is remarkable that the charge part of the Hamiltonian is completely 
bilinear in the currents, or equivalently, the $\phi_c$ fields, and 
therefore it is invariant under the full $U(1)^\infty_k$ symmetry and
can be diagonalized exactly.  In this instance the multidimensional
bosonization scheme has transformed the four point fermion interaction into
a problem of Gaussian bosons.  Diagonalization is non-trivial, but as
we show in subsequent sections, we can calculate the multi-point correlation 
functions exactly.  The spin sector of the Hamiltonian is more complicated
since the last, x-y, term in equation \ref{hamspincharge}, the coupling between 
the x-components and the y-components of the spin current, 
is a nonlinear cosine function
of the spin boson fields. Consequently it breaks the $U(1)^\infty_k$ symmetry
and the equations of motion are nonlinear.  
Nevertheless the low-energy behavior of the theory can be analyzed by the
RG method we discuss in the next section.

\section{Stability of Fermi Liquids}
\label{stability}
\setcounter{equation}{0}

In this section we examine the stability of the Fermi liquid fixed point 
against BCS pairing in a weak-coupling RG calculation. 
It is well known that an attractive interaction leads to a
superconducting instability, and here we show how this occurs in the bosonic
representation.  In contrast to mean-field theories, bosonization when combined
with the renormalization group provides an unbiased means for analyzing 
instabilities, as it treats all of the interaction channels on an equal footing.

For simplicity we consider the case of spinless fermions in two
dimensions as the extension to three dimensions is straightforward. 
Consider also a
circular Fermi surface to eliminate the possibility of nesting which could lead
to charge or spin density wave instabilities in the forward and exchange
channels.  These possibilities are discussed later in Section \ref{nest}. 
To further simplify the problem, initially we turn off the
interactions in the forward and exchange channels, as we shall see 
that these channels do not alter the renormalization of the BCS channel
at leading order, at least for the case of the circular Fermi surface.

In the BCS channel a fermion in patch ${\bf S}$ on the Fermi surface is
paired with the fermion in patch, $-{\bf S}$, directly opposite, 
see figure \ref{channels}.  
In cylindrical coordinates patches ${\bf S}$ and $-{\bf S}$ correspond to 
$\theta$ and $\theta + \pi$.  To carry out the RG calculation
we use the path integral framework, as we wish to rescale both
space and (imaginary) time.  The BCS action expressed in terms of the four 
Fermi fields is:
\begin{equation}
S_{int}[\psi^*, \psi] = \int d\tau~ \int d^2x~ \sum_{\bs, \bt}
\frac{V_{BCS}({\bf S}-{\bf T})}{N^*(0)}~ 
\psi^*(-{\bf T}; x) \psi^*({\bf T}; x) 
\psi({\bf S}; x) \psi(-{\bf S}; x)
\label{BCS-action}
\end{equation}
here $N^*(0) = m^*/(2 \pi)$ is the density of states at the Fermi energy
and ${\bf S}$ and ${\bf T}$ are restricted to 
range over only half of the Fermi surface. As the
fermions are spinless $V_{BCS}$ must change sign under inversion so
$V_{BCS}(\theta) = -V_{BCS} (\theta +\pi)$. Also the Hamiltonian is
Hermitian and therefore $V_{BCS}(\bs - \bt) = V_{BCS}(\bt - \bs)$. 
The interaction
can be written in bosonic representation using the bosonization formula,
equation \ref{bosonization}. 
In doing so it is important to keep track of the order of the
fermion operators. Formally the correct sign is set by the ordering operator
$\hat{O}({\bf S})$ but as noted in Section \ref{essentials} 
in practice it is easier to determine the sign by
inspection of the fermion operators.  With these remarks in mind the action is
written as
\begin{equation}
S_{int}[\phi] = \left[\frac{\Lambda \lambda}{(2\pi)^2}\right]^2
\sum_{\bs \bt} \int d\tau \int d^2 x \frac{V_{BCS}
(\bs-\bt)}{N^*(0)} \exp \left[ i \frac{\sqrt{4\pi}}{\Omega} [\phi_\bs(x)
-\phi_\bt(x) ]\right] 
\label{BCS-boson-action}
\end{equation}
where we have introduced the combination 
$\phi_\bs (x) \equiv \phi(\bs; x) + \phi(-\bs; x)$ and have defined the
usual $2+1$ dimensional space-time coordinate $x \equiv ({\bf x}, \tau)$.

Because an exponential of the boson field appears in the action, 
equation \ref{BCS-boson-action},
it is most convenient to implement the RG transformation in real space.
The RG transformation rescales only the component of length
parallel to the Fermi surface normal, $x_\parallel \equiv \hbn_\bs \cdot \bx$.
The perpendicular component, $x_\bot \equiv \hbn_\bs \times \bx$, remains
invariant so that the real space volume element $d^2x = dx_{\parallel} dx_\bot$
becomes $(s dx_{\parallel}) dx_\bot $ and $d\tau \rightarrow s d\tau$.  
The quadratic
part of the action is invariant under this transformation provided that the
boson field is invariant, $\phi \rightarrow \phi$. 
Then it follows that the BCS interaction is marginal
as expected.
The usual renormalization group procedure is to integrate out the fast
parts of the fields $\phi(\bs; \bx)$.  Here ``fast'' means high-energy. 
To be precise modes with momenta
\begin{equation}
\frac{\lambda}{2s} < | \hbn_\bs \cdot \bq | < \frac{\lambda}{2} 
\label{integrated-modes}
\end{equation}
are eliminated which in real space
translates into integrating out the short distance pieces of the fields
with $2a < | \hbn_\bs \cdot \bx | < 2 s a$, where we recall that 
$a \equiv \lambda^{-1}$ is the ultraviolet cutoff. 
To complete the transformation $x_\parallel \rightarrow s x_\parallel$, and 
$\tau \rightarrow s \tau$.  
Integration over the fast modes is accomplished as usual within the 
functional integral
\begin{equation}
e^{-S[\phi(\bs; x)]} = \prod_\bt \int_{2 a < |\hbn_\bt \cdot \bx |< 2as} 
D\phi(\bt; x)~ e^{-S_0[\phi] - S_{int}[\phi]} 
\label{integrate-fast}
\end{equation}
As only the fast parts of the fields are integrated out it is convenient to
split the $\phi$ fields into slow $\phi^\prime$ and fast $h$ fields
by writing $\phi = \phi^\prime + h$.  Then we can
write the right hand side of equation \ref{integrate-fast} 
as $\exp (-S_0[\phi])~ \times~ 
\langle \exp (-S_{int}[\phi^\prime + h]) \rangle$ 
where $\langle \rangle$ means an average over the fast $h(\bt; x)$ 
fields only.  The renormalized interaction is found perturbatively by 
expanding in powers of $V_{BCS}$ and at leading order 
\begin{eqnarray}
\langle S_{int}[\phi^\prime] \rangle &=&  
\left[\frac{\Lambda\lambda}{2\pi}\right]^2 \sum_{\bs, \bt}
\int d\tau  \int d^2x \frac{V_{BCS}({\bf S}-{\bf T})}{N^*(0)} \exp \left\{
i\frac{\sqrt{4\pi}}{\Omega}[\phi_\bs^\prime(x) -\phi_\bt^\prime(x)]\right\} 
\nonumber \\
&\times&  \left\langle \exp i \frac{\sqrt{4\pi}}{\Omega} 
[h_\bs(x) - h_\bt(x)]\right\rangle
\label{first-order}
\end{eqnarray}
To evaluate the restricted average $\langle \rangle$ we make use of 
the propagator of the $h$ fields (the derivation of the propagator is given in 
Appendix \ref{fast}): 
\begin{equation}
\langle h(\bs; x) h(\bs; 0) \rangle 
= \left\{ \begin{array}{ll} 
\frac{\Omega^2}{4\pi} \ln \left[ \frac{x_{\parallel} + iv_F^*\tau 
+ (2 i s a) \sgn(\tau)}
{x_{\parallel} +iv_F^* \tau + (2 i a) \sgn(\tau)}\right], 
&  \Lambda |x_\bot| \ll 1 \\ 
\\
0 ;  & {\rm otherwise.}
\end{array}\right.
\label{Nhh}
\end{equation}
In particular
\begin{equation}
\langle h^2(\bs; 0) \rangle = \frac{\Omega^2}{4\pi} \ln(s) \ .  
\label{Nhh0}
\end{equation}
Now use the fact that, for a Gaussian measure, 
\begin{equation}
\langle e^{i h} \rangle = e^{-\frac{1}{2} \langle h^2 \rangle}
\end{equation} 
and equation \ref{first-order} becomes
\begin{equation}
\langle S_{int}[\phi^\prime] \rangle = \left[ \frac{\Lambda \lambda
/s}{2\pi}\right]^2 \sum_{\bs, \bt} \int^\infty_{-\infty} d\tau \int^\prime d^2x
\frac{V_{BCS}(\bs - \bt)}{N^*(0)} \exp \left\{ \frac{i\sqrt{4\pi}}{\Omega}
[\phi_\bs^\prime(\bx)-\phi_\bt^\prime(\bx)]\right\}, 
\end{equation}
where the prime on the integral indicates that the region of spatial 
integration covers only small distances.  Then, upon  
rescaling space and time, we recover $S_{int}[\phi]$.  

\begin{figure}
\center
\epsfxsize = 10.0cm \epsfbox{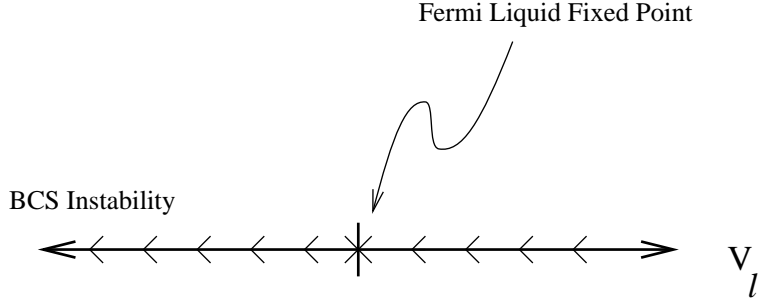}
\caption{
\label{flow}
RG flows decouple, channel by channel.  The Fermi liquid fixed point is 
at $V_\ell = 0$.}
\end{figure}

The first nontrivial
contribution to the renormalized interaction is at second order in the
expansion in powers of the BCS interaction, $V_{BCS}$. 
This term is evaluated by making use of the variant of
the Baker-Hausdorff formula, equation \ref{key_identity}, and the propagator 
of the $h$ fields:
\begin{eqnarray}
\langle S_{int}^2[\phi^\prime] \rangle &=& 
\left[\frac{\Lambda \lambda}{(2\pi)^2 s}\right]^4 \sum_{\bs, \bt}
\int d\tau \int d^2x 
\frac{V_{BCS}({\bf S} - {\bf U}) V_{BCS}({\bf U} - {\bf T})}{[N^*(0)]^2}  
\exp \left\{ i\frac{\sqrt{4\pi}}{\Omega}
[\phi_\bs^\prime(x) - \phi_\bt^\prime(x)] \right\}
\nonumber \\
&\times& \frac{1}{\Lambda} \int^\infty_{-\infty} d\sigma \int^{2 a s}_{2a} 
dy_\parallel \frac{[v_F^* \sigma + 2 a s~ \sgn(\sigma)]^2 + y_\parallel^2}
{[v_F^* \sigma + 2 a~ \sgn(\sigma)]^2 + y_\parallel^2}
\end{eqnarray}
where $y_\parallel \equiv \hbn_\bu \cdot \by$ is the spatial integration
variable and the factor of $1/\Lambda$ in the second line is the result
of integrating over the perpendicular coordinate, $y_\perp$.
As $s \rightarrow \infty$ the highest power of $s$ (in this case $\ln(s)$)
gives the leading
contribution to the renormalization.  The time integral over $\sigma$ may be
evaluated by contour integration. The structure is such that of the two
poles, one lies in the upper half plane and one in the lower, which is a direct
result of the coupling in the BCS channel between patches ${\bf U}$ and
$-{\bf U}$. The remaining spatial integral is proportional to $\ln(s)$ and we
find that the flow of the coupling function in the BCS channel is given by:
\begin{equation}
\frac{dV_{BCS}({\bf S} - {\bf T})}{d \ln(s)} = - \frac{\Lambda}{(2\pi)^2 k_F}
\sum_{\bf U} V_{BCS}({\bf S}-{\bf U}) V_{BCS}({\bf U}-{\bf T})~ . 
\label{BCS-beta}
\end{equation}
For the circular Fermi surface considered here 
\begin{equation}
\Lambda \sum_\bu = \frac{k_F}{2} \int^{2 \pi}_0 d\theta
\end{equation}
and the flow equation can be diagonalized by a Fourier decomposition of 
the BCS interaction into angular momentum channels: 
\begin{equation}
V_\ell \equiv \int_0^{2\pi} \frac{d\theta}{2\pi}~ e^{i\ell \theta}~ 
V_{BCS}(\theta)\ .
\label{Nangl}
\end{equation}
The flow equations decouple channel by channel, and are simplified 
further as $V_\ell$ is dimensionless: 
\begin{equation}
\frac{dV_\ell}{d \ln(s)} = -\frac{V_\ell^2}{4 \pi}\ . 
\label{BCS-flow}
\end{equation}
This result, which is identical to that derived by Shankar in the fermion 
basis\cite{Shankar}, 
tells us that the couplings $V_\ell$, which are marginal at tree level,
are in fact marginally relevant if negative, and marginally irrelevant 
if positive.  The
first case leads to the BCS instability, in the second case the Fermi liquid
fixed point is stable, as indicated by figure \ref{flow}.

We should ask now, what would happen were we to include at the outset 
the other marginal interactions, for example
the coupling between the charge-currents:
\begin{equation}
S_{FL} = \frac{1}{V^2}~ \sum_{\bs \bt} \int d\tau~ d^2x~ d^2y~ 
f_c(\bs, \bt; \bx - \by)~ J_c(\bs; \bx, \tau) J_c(\bt; \by, \tau)\ .
\end{equation}
Since the interactions are quadratic in the bose fields, they
parameterize infinitely many 
different Gaussian fixed points which do not 
flow under the RG transformations.  Incorporated into the RG 
calculation they contribute only subleading corrections to scaling and do not
influence the leading RG flows of the BCS interaction. For example, although
bosons in different patches are correlated via the interaction, 
this correlation is only a
fraction, $\Lambda / k_F$, of the correlations within the same patch,
as we prove below in Section \ref{propagator}.  The leading RG flows, 
equation \ref{BCS-flow}, are unchanged.

It is not difficult to see that the system develops an energy gap if the BCS
interaction is attractive in any channel.  The RG flows in attractive channels
scale to large negative values as $s \rightarrow \infty$, and the size of the
gap may be estimated by calculating the length scale at which the renormalized
coupling becomes of order one.  Alternatively, a mean-field approximation shows
how the gap develops.  Suppose for simplicity that 
$V_{BCS}(\theta) = V_{\ell = 1} \cos(\theta)$, conducive to p-wave ordering. 
As in the classic BCS calculation we may 
decouple the Cooper pairs via a Hubbard-Stratonovich
transformation upon introducing a field $\Psi(\bs) = \Delta(\bs) e^{i\Phi}$. 
Then the bosonized Hamiltonian for the pair $(\bs, -\bs)$ is given by
\begin{eqnarray}
H[\phi_\bs] &=& \int d^2 x  \left\{
\frac{2\pi v_F^*}{\Omega V} \left( [(\hbn_\bs \cdot {\bf \nabla})
\phi({\bf S}; {\bf x})]^2 + [(\hbn_\bs \cdot {\bf \nabla}) 
\phi(-{\bf S}; {\bf x})]^2\right)\right.    
\nonumber \\
&-& \left. \frac{\Delta(\bs) \Lambda}{2\pi k_F a} \cos
[\frac{\sqrt{4\pi}}{\Omega} (\phi_\bs(x) - \Phi)\right\} \ .
\end{eqnarray}
At the mean field level, $\Delta(\bs)~ e^{i \Phi}$ can be assumed to be constant
with no fluctuations, and
the problem is equivalent to a sine-Gordon model in each
direction $\hbn_\bs$.  The continuous $U(1)^\infty_k$ gauge symmetry 
of the action under translations of the boson field by an additive 
constant, $\phi_\bs \rightarrow \phi_\bs + \frac{\Omega}{\sqrt{4 \pi}}
(\theta_\bs + \theta_{-\bs})$, breaks 
spontaneously giving rise to the usual superconducting energy gap. 

There is in fact the possibility of a BCS instability even when the
interactions are repulsive: the Kohn-Luttinger effect\cite{KL}.  
This possibility only
becomes apparent beyond leading order.  The bare couplings, $V_\ell$, due to a
short-range repulsive interaction are all positive but tend rapidly to zero at
large $\ell$. Fermi liquid interactions as discussed above generate
irrelevant contributions in the BCS channel down by a positive power of
$\Lambda/k_F$, nevertheless this is sufficient to make some of the $V_\ell$
slightly negative, at least at
sufficiently large $\ell$, leading to an instability.
Because of the small magnitude of these coefficients, the effect can only be
important at extremely low temperatures and the essential physics is
controlled by the Fermi liquid fixed point.  However, nesting instabilities
can enlarge dramatically this effect, and we examine this situation in 
Section \ref{nest}.

Finally, it might be thought that in general an instability must 
occur in the spin sector since the interaction is a
cosine function of the spin boson.  Indeed this does occur in one spatial
dimension: the coupling between spin currents at the left and right Fermi
points flows under the RG transformation\cite{Emery,Solyom,Ian,Katmandu}.  
If this were to be the case in $D > 1$
it would spell disaster for the multidimensional bosonization approach to 
Landau Fermi liquids since the Landau parameters in
the spin sector should be constants parameterizing the Landau fixed point. 
However, on closer inspection it is clear that the spin couplings are indeed
fixed.  Repeating the weak-coupling RG calculations to second order for the
x-y part of the interaction in equation \ref{hamspincharge} gives
\begin{eqnarray}
\frac{1}{2} \langle S_{xy}^2[\phi_s^\prime] \rangle &=& \lambda^2 \left(
\frac{\Lambda}{(2\pi)^2}\right)^4 \int d\tau~ \int d^2x 
\sum_{\bs, \bt} \cos \frac{\sqrt{8\pi}}{\Omega} 
\left[ \phi_s^\prime(\bs; x) - \phi_s^\prime(\bt; x)\right]
\nonumber \\
&\times & \sum_\bu f_s({\bf S}, {\bf U}) f_s({\bf U}, {\bf T}) \frac{1}{\Lambda}
\int_{-\infty}^\infty d\sigma~ \int_{2 a < |y_\parallel| < 2 s a} dy_\parallel
\frac{1}{(y_\parallel + iv_F^* \sigma)^2} 
\end{eqnarray}
In this case, unlike the case of the BCS interaction, 
the correction vanishes since the integral over $(y_\parallel, \sigma)$
does not encompass a simple pole, 
and the spin interaction is strictly marginal, at least within this 
second-order perturbative calculation.  
A deeper, and non-perturbative, reason for the
exact marginal nature of the spin current coupling in $D > 1$
is given below in Section \ref{collective}.

\section{Thermodynamics of Fermi Liquids}
\label{thermodynamics}
\setcounter{equation}{0}

In this section we use the bosonization formalism to calculate the specific heat
of the interacting Landau Fermi liquid in three dimensions.  We show that a
well-known nonanalyticity in the specific heat, a term proportional to
$T^3 \ln T$, is reproduced by the method.  The existence of such a term
is consistent with careful measurements\cite{Greywall} of the specific heat
of helium-3.  It has also been seen in heavy fermion systems. 
For the sake of 
simplicity we turn off the spin-spin interaction and eliminate the spin
index.  Then the bosonic Hamiltonian of the interacting spinless fermion system
is given by
\begin{equation}
H = \frac{1}{2}~ 
\sum_{\bs \bt} \sum_\bq V(\bs, \bt; \bq) J(\bs; -\bq) J(\bt; \bq)
\end{equation}
where
\begin{equation}
V(\bs, \bt; \bq) = \frac{\ve}{\Omega} \delta_{\bs, \bt}^{D-1} 
+ \frac{f(\bs, \bt; \bq)}{V}. 
\label{coupling}
\end{equation}
As the nonanalytic behavior of the specific heat derives from small momentum
processes it is sufficient to take the Fermi surface to be locally flat. If we
again let $q_\parallel \equiv \hbn_\bs \cdot \bq$ denote the component of the
momentum $\bq$ parallel to the surface normal or Fermi momentum in patch $\bs$, 
the charge currents $J(\bs; \bq)$ can be represented in terms of canonical
boson creation and annihilation operators as follows 
\begin{equation}
J(\bs; \bq) = \left\{ \begin{array}{ll}
\sqrt{-\Omega q_\parallel}~ a^\dag(\bs; -\bq), & q_\parallel \leq 0 \\
\\
\sqrt{\Omega q_\parallel}~ a(\bs; \bq), & q_\parallel > 0
\end{array}\right.
\label{current-rep2}
\end{equation}
With the replacement of the current operators with the canonical bosons
the Hamiltonian becomes
\begin{equation}
H = \sum_{\bs, \bt}~ \sum_{\bq, q_\parallel > 0} V (\bs, \bt; \bq)~
\Omega q_\parallel~ a^\dag(\bs; \bq) a(\bt; \bq) 
\label{hamcv}
\end{equation}
which we can diagonalize directly to find the spectrum. 
The specific heat is then computed using the standard formula for bosons,
equation \ref{C_V}.

First we review the case of noninteracting fermions, $f(\bs, \bt; \bq) = 0$,
discussed in the Introduction.  The eigenenergies of the bosonized
Hamiltonian in this case are simply
\begin{equation}
\epsilon(\bs; \bq) = v_F^*~ \hbn_\bs \cdot \bq = \ve q_\parallel
\end{equation}
which agrees with equation \ref{ph-dispersion}, the energy of a particle-hole
pair. 
The sum over the patch index $\bs$ and the components of $\bq$ perpendicular
to the Fermi wavevector counts the number of states at the Fermi surface 
which we call $A$. The sum over the
components of $\bq$ parallel to the Fermi wavevector can be converted to an
integral.  Then assuming that the temperature is sufficiently low such that the
thermally excited particle-hole pairs lie within the cutoff, 
$k_B T \ll v_F^* \lambda$, we obtain
\begin{eqnarray}
C_V &=& \frac{A}{4 k_B T^2} \left(\frac{L}{2\pi}\right) 
\int_0^\infty \frac{\ves q_\parallel^2}
{\sinh^2(\frac{\ve q_\parallel}{2 k_B T})}~ dq_\parallel
\nonumber \\
&=& k_B^2 T \left[\frac{m^* k_F}{2\pi^2}\right] \frac{\pi^2}{3} L^3.
\end{eqnarray}
This is the correct result for spinless fermions in $D=3$ which,
of course, should
be multiplied by a factor of $2$ to account for spin.  It is remarkable that
the boson formula, equation \ref{C_V}, yields the full specific heat. 
Multidimensional bosonization reproduces the entire fermion Hilbert space.

\begin{figure}
\center
\epsfxsize = 10.0cm \epsfbox{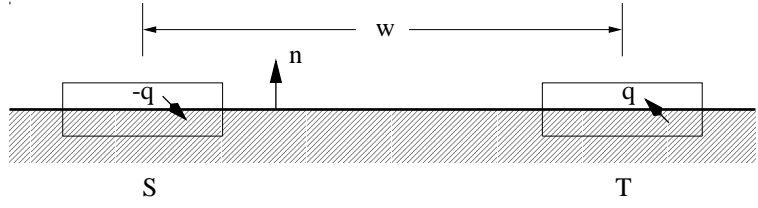}
\caption{
The Fermi surface treated as locally flat, with nearby particle-hole
pairs of momentum $\bq$ and $-\bq$ coupled together.}
\label{t3lnt}
\end{figure}

Now we follow Pethick and Carneiro\cite{Pethick} 
and focus on particle-hole pairs separated
only by a small momentum ${\bf w} = \bk_\bs - \bk_\bt$ (with $|{\bf w}|\ll k_F)$
since a consideration of
these processes is sufficient to demonstrate the existence of nonanalytic
contributions to the specific heat.  In this limit we may continue to 
approximate the Fermi surface as locally flat as shown in figure \ref{t3lnt}; 
this simplifies considerably the
problem of diagonalizing the quadratic boson Hamiltonian (see Section 
\ref{collective}).  Now the current-current coupling $f(\bs, \bt; \bq)$
may be expanded in powers of the dimensionless, and rotationally-invariant,  
expansion parameter $\bq^2 / ({\bf w}^2 + {\bq}^2)$:  
\begin{eqnarray}
f(\bs, \bt; \bq) &=& a + b~ \frac{\bq^2}{\bq^2 + {\bf w}^2} + \cdots
\nonumber \\
&=& a + b~ \frac{q_\parallel^2}{q_\parallel^2 + {\bf w}^2} + \cdots
\label{expansion}
\end{eqnarray}
In the second line we have dropped the perpendicular
$\bq_\perp \equiv \hbn_\bs \times \bq$ 
terms in the numerator since they do not contribute to the non-analyticity. 
The same approximation can be
made in the denominator, as the squat aspect-ratio of the patches enforces 
$|{\bf w}| \gg |\bq|$ in the $N \rightarrow \infty$ limit.  
The $q_\parallel^2$ term
in the denominator must be retained to eliminate the artificial divergence 
which would otherwise occur in the ${\bf w}^2 \rightarrow 0$ limit.
The expansion is controlled in low-temperature limit which ensures that 
$|q_\parallel|$,
the particle-hole momentum perpendicular to the Fermi surface, is small.  
The current-current interaction coefficient $b$, which therefore is irrelevant 
in the RG sense, gives subleading corrections to the specific heat. 
In actual fact, the interaction considered here differs from that studied by
Pethick and Carneiro in that it couples particle-hole pairs at point $\bs$ and
$\bt$ whereas the Pethick-Carneiro interaction couples the quasiparticle
occupancies $n_{\bf p}$ and $n_{{\bf u} +\bp}$. 
To be precise, the Pethick-Carneiro interaction has the
form $b^\prime~ \sum_{{\bf u}, \bp} ({\bf \hat{u}} \cdot {\bf \hat{p}})^2 
\delta n_\bp~ \delta n_{{\bf u} + \bp}$.  
This interaction cannot be expressed directly in
terms of the currents since it involves the products of quasiparticle 
{\it occupancies} above
and below the Fermi surface, whereas the current operator evaluated at zero
momentum $J(\bs; {\bf 0})$ averages the occupancy operator over the 
interior of a
squat box.  Therefore a direct connection with the earlier calculation 
of Pethick and Carneiro cannot be made.  

To proceed, we diagonalize the Hamiltonian with the aid of a Fourier transform
from patch index space $\bs$ to ${\bf X}$ space. Let
\begin{equation}
a^\dag (\bs; \bq) = \frac{\Lambda}{\sqrt{4\pi k_F^2}} \int d^2X~ 
e^{-i {\bf X} \cdot \bs}~ a^\dag ({\bf X}, \bq)
\end{equation}
where $\int d^2 X~ 1 = 4\pi k_F^2/ \Lambda^2$ is equal to the number of patches
which cover the Fermi surface.  Then
\begin{equation}
H = L^3~ \int d^2X~ \int_{-\Lambda/2}^{\Lambda/2} \frac{dq_{x}}{2\pi}~
\int_{-\Lambda/2}^{\Lambda/2}\frac{dq_{y}}{2\pi} \int_0^\lambda
\frac{dq_{z}}{2\pi}~
\epsilon({\bf X}; \bq) a^\dag({\bf X}; \bq) a({\bf X}; \bq). 
\end{equation}
Using equation \ref{coupling} and the representation of the current 
given in equation \ref{current-rep2} we obtain the eigenenergies
\begin{equation}
\epsilon({\bf X}; \bq) = \ve q_\parallel 
+ \frac{b \Lambda^2 q_\parallel^3}{(2\pi)^3} 
\sum_{\bf w} \frac{e^{-i{\bf X} \cdot {\bf w}/\Lambda}}
{q_\parallel^2 + {\bf w}^2} 
\end{equation}
The sum can be converted to a Riemann integral by the substitution
$\Lambda^2\sum_{\bf w} \rightarrow \int d^2 w$ and we find
\begin{equation}
\epsilon({\bf X}; \bq) = \ve q_\parallel - \frac{b q_\parallel^3}{(2\pi)^3} \pi 
\ln (q_\parallel^2 |{\bf X}|^2).
\end{equation}
In this equation we have discarded those terms proportional to $b$ 
that make analytic
contributions to the specific heat and have retained only the logarithmic
correction.  We treat this term as a perturbation and calculate the specific
heat to $O(b)$. The change in the specific heat $\delta C_V$ due to this 
perturbation is
\begin{eqnarray}
\delta C_V &=& -b
\frac{A}{k_B T^2} \frac{\ve \pi}{(2\pi)^2} \frac{L}{2\pi}
\int_0^\infty dq_\parallel \frac{q_\parallel^4 \ln q_\parallel}
{\sinh^2 \left[ \frac{\ve q_\parallel}{2k_B T}\right]}
\nonumber \\
&=&-b \frac{2 AL k_B^4}{ \pi^3 (\ve)^4}~ T^3 \ln T~ \int_0^\infty
\frac{x^4}{\sinh^2 x} dx
\end{eqnarray}
In the second line of this equation we have retained only the term containing
$T^3 \ln T$ temperature dependence. The integral in the second line equals
$\pi^4/30$ so the final result is 
\begin{equation}
\frac{\delta C_V}{L^3} = b \frac{k_B^4 (m^*)^4}{15 k_F^2} T^3 \ln T\ . 
\end{equation}
Not surprisingly this result has the same
form as that found by Pethick and Carneiro\cite{Pethick} as dimensional analysis
makes this likely. A direct comparison of the coefficient is meaningless, 
however, as the
interactions are not the same.  Nevertheless, the same point is made: 
subleading corrections to the linear specific heat have many different origins.
Irrelevant interactions such as equation \ref{expansion} or the 
Pethick-Carneiro form both make nonanalytic contributions to the specific
heat, as do other more
complicated irrelevant terms\cite{Bedell}.  Repeating the calculation in 
$D = 2$ we find that the interaction equation \ref{expansion} 
leads to a subleading correction
$\delta C_V \sim b T^2$, also in agreement with fermionic 
calculations\cite{Coffey}.

\section{Collective Modes}
\label{collective}
\setcounter{equation}{0}

The curvature of the Fermi surface did not play an important role in the
determination of the specific heat. In fact we took the Fermi surface to be
flat and consequently the Hamiltonian could be rewritten as the sum of products
of a single creation operator and a single annihilation operator. Collective
excitations of the Fermi surface, on the other hand, are possible because of
the curvature.
The Fermi surface itself may be viewed as dynamical, sloshing back and forth
relative to its equilibrium position in momentum space.
It is important to reproduce the collective mode spectrum within the
bosonic representation.  For a curved Fermi surface the Hamiltonian contains the
products of two annihilation operators and the products of two creation
operators in addition to the bilinear combination of annihilation and creation
operators found for a flat Fermi surface.  Therefore a straightforward 
diagonalization of the bosonic Hamiltonian is no longer possible, 
see however\cite{Castro,Castro2}.

In order to discuss the spectrum of collective modes it is necessary to
elaborate further on the Hamiltonian.  In the absence of an attractive
BCS interaction the excitations of the interacting Fermi liquid
are in the particle-hole channel and carry either charge or spin. To exhibit
this factorization into charge and spin sectors explicitly the bosonized
Hamiltonian can be written as $H = H_c + H_s$. The Hamiltonian 
for the charge sector in 
$D$ dimensions, as before, is bilinear in the charge current operators
$J(\bs; \bq) = \sum_\alpha  J_\alpha (\bs; \bq)$.  We drop the ``c'' subscript
on the charge currents as no ambiguity results.   
\begin{equation}
H_c = \frac{1}{2} \sum_{\bs \bt} \sum_{\bq} V_c(\bs, \bt; \bq)~ 
J (\bs; -\bq)~ J(\bt; \bq)
\end{equation}
The prefactor of $1/2$ compensates for the double-counting 
$\bq \leftrightarrow -\bq$ in the sum over the momentum. 
For the present we consider only the case of short-range Fermi liquid 
interactions $f_c(\bs, \bt)$:
\begin{equation}
V_c(\bs, \bt; \bq) = \frac{1}{2} \Omega^{-1} \ve \delta_{\bs, \bt}^{D-1} 
+ \frac{1}{V} f_c(\bs, \bt)\ .
\label{short-range}
\end{equation}
The charge currents satisfy the equal-time $U(1)$ current algebra, equation
\ref{current_algebra}, but with twice the anomaly as the spin index is summed
over. The factor of $\frac{1}{2}$ in the first term of equation
\ref{short-range} compensates for this doubling.  

For our analysis of the spin collective modes it is convenient to 
introduce spin currents which are manifestly $SU(2)$ covariant instead of 
continuing to employ the spin quantization axis as in Section \ref{essentials}.
These so-called ``non-Abelian'' currents\cite{Witten} are defined to be:  
\begin{equation}
\vec{J}(\bs; \bq) = \frac{1}{2} \sum_{\bk} \theta(\bs; \bk - {\bf q}) 
\theta (\bs; \bk)~ \psi_{\bk - \bq}^{\dag \alpha} \vec{\sigma}_\alpha^\beta 
\psi_{\beta \bk} 
\label{spin-current}
\end{equation}
where again $\vec{\sigma}$ are Pauli matrices.  Note that, unlike the case
of the charge currents, there is no need to subtract the vacuum expectation
value as the quiescent Fermi liquid is non-magnetic. 
Like the charge currents, 
spin currents in different patches on the Fermi surface commute.  They
also commute with the charge currents. Spin currents in the same box satisfy 
the more complicated $SU(2)$ current algebra: 
\begin{equation}
\left[J^a (\bs; \bq), J^b (\bt; \bp)\right] = 
\frac{1}{2}~ \Omega~ \hbn_\bs \cdot \bq~
\delta_{\bs, \bt}^{D-1}~ \delta^{ab} \delta_{\bq + \bp, {\bf 0}}^D 
+ i \delta_{\bs, \bt}^{D-1} \epsilon^{abc} J^c(\bs; \bq + \bp)\ ;
\label{non-Abelian}
\end{equation}
here $(a, b, c) = (x, y, z)$ label the three components of the spin, and
$\epsilon^{abc}$ is the completely antisymmetric tensor.  The coefficient
of the first term on the right hand side of the current algebra, 
equation \ref{non-Abelian}, is known as the ``central charge,'' $k$.
In gapless one-dimensional antiferromagnets\cite{Affleck-Haldane} 
with half-odd-integer spin-$S$, generically $k = 1$ except at special integrable
points where $k = 2 S$. Here, by contrast, 
$k = (\Lambda L / 2 \pi)^{D-1}  \gg 1$. 
The effective spin is large because the coarse-graining procedure sums
over a multitude of momentum points inside each squat box; consequently the
spin currents are semiclassical in nature.

The spin Hamiltonian is quadratic in the non-Abelian
spin currents (unlike the Abelian case,
there are no cosines) and Fermi liquid
spin-spin interactions $f_s$ which appear in the Hamiltonian couple
together spin currents in different patches:
\begin{equation}
H_s = \frac{1}{2} \sum_{\bs \bt} \sum_\bq V_s(\bs, \bt; \bq)~ 
\vec{J}(\bs; -\bq) \cdot \vec{J}(\bt; \bq)
\end{equation}
where
\begin{equation}
V_s(\bs, \bt; \bq) = \frac{2}{3} \ve \Omega^{-1} \delta_{\bs, \bt}^{D-1} +
\frac{1}{V} f_s({\bs, \bt})\ . 
\label{spin-params}
\end{equation}
Note the factor of $2/3$ appearing in the construction of the free Hamiltonian,
which we discuss below.

Instead of diagonalizing $H$ directly, we find instead the equations of motion
for the charge and spin currents which we shall show are equivalent to the 
well-known collective-mode equations.  In the charge sector, 
\begin{eqnarray}
i\frac{\partial}{\partial t} J(\bs; \bq) &=& [J(\bs; \bq), H_c]
\nonumber \\
&=& \ve~ \hbn_\bs \cdot \bq~ J(\bs; \bq) 
+ \hbn_\bs \cdot \bq \frac{2\Omega}{V} \sum_\bt f_c(\bs, \bt) J(\bt; \bq) \ .
\label{charge-collective-mode}
\end{eqnarray}
The first term on the right hand side of equation \ref{charge-collective-mode}
is due to the free dispersion relation for the particle-hole pairs of
momentum $\bq$ in the patch $\bs$. The second term couples currents in
different patches. Note that
\begin{eqnarray}
\frac{\Omega}{V} \sum_\bt &=& 
\int \frac{d^D k}{(2\pi)^D} \delta (|\bk| - k_F)
\nonumber \\
&=& \frac{k_F^2}{(2 \pi)^3} \int_{-1}^1 d(\cos \theta) \int_0^{2 \pi} d \phi
\quad (D = 3)\ ;
\end{eqnarray}
so the second term reduces to the usual integral over the Fermi surface in the
continuum limit of a large number of patches, $N\rightarrow \infty$. 
The equations of motion for the
spin currents contain, in addition to these two terms, a third
term which makes the spins precess in the local magnetic field, a phenomenon
which is especially important when the system is magnetically polarized.
\begin{eqnarray}
i \frac{\partial }{\partial t} J^a(\bs; \bq) &=& 
\ve \hbn_\bs \cdot \bq J^a(\bs; \bq) + \frac{\Omega}{2 V}~ \hbn_\bs \cdot \bq~
\sum_\bt f_s(\bs, \bt) J^a(\bt; \bq) 
\nonumber \\
&-& \frac{i}{V} \epsilon^{abc} \sum_\bk J^b(\bs; \bk) \sum_\bt f_s(\bs, \bt) 
J^c(\bt; \bq -\bk)\ .
\label{spin-collective-mode}
\end{eqnarray}
Note that the factor of $2/3$ multiplying the Fermi velocity $\ve$ 
in the free
part of equation \ref{spin-params} does not appear in equation 
\ref{spin-collective-mode}. The origin of the
factor of $2/3$ is easy to understand in the Sugawara construction 
of the free fermion Hamiltonian out of current 
bilinears\cite{Sugawara,Goddard}. It reflects the $SU(2)$
invariance of the spin currents which permits the replacement $\vec{J}(\bs; \bq)
\cdot \vec{J}(\bs; -\bq) \rightarrow 3 J_z(\bs; q) J_z(\bs; -\bq)$, at least
for the purpose
of computing the spectrum. The derivation given here, on the other hand, does
not rely on this argument as spin rotational invariance is respected
explicitly. Rather the factor of $2/3$ in equation \ref{spin-params} 
cancels contributions to the equations of motion that arise from both the 
$\delta^{ab}$ and $i \epsilon^{abc}$ terms in the spin current algebra.

Now we are in a position to 
understand the deeper reason for the absence of renormalization of the spin
Landau parameters alluded to in Section \ref{stability}.  Unlike the $D = 1$
case, the spin current operators defined by equation \ref{spin-current} 
involve a sum over a large number of momenta inside the squat box.  This
is apparent in the non-Abelian algebra equation \ref{non-Abelian} as the 
diagonal anomaly (the first term proportional to $\delta^{ab}$) 
is of order $\Omega |\hbn_\bs \cdot \bq| \gg 1$. In the absence of macroscopic
spin polarization the anomaly overwhelms the off-diagonal, 
$i \epsilon^{abc}$, term. 
Physically this means that, in zero applied magnetic field, 
the spin equation is identical in form to the charge equation,
and spin waves propagate freely as expected\cite{Baym}.  
In the opposite limit of a large external magnetic field, the last
term in equation \ref{spin-collective-mode} dominates. Spins precess locally 
in this limit.  Returning now to the question of the spin Landau parameters,
we see that they remain fixed, as only the non-linear $i \epsilon^{abc}$
term can drive renormalization flows and it can be neglected.  This proof,
which is non-perturbative, and therefore independent of the size of the spin
Landau parameters, is consistent with the absence of renormalization found
in the weak-coupling calculation presented in Section \ref{stability}.

As it stands these operator equations are exact, at least in the $N\rightarrow
\infty$ limit in which the current algebras, equations \ref{current_algebra}
and \ref{non-Abelian}, become exact.  Solving
the charge collective mode equation \ref{charge-collective-mode} is 
straightforward as it is linear.  The equation for the spin
currents is more difficult to solve due to the nonlinear term.  
Here for completeness we solve the charge equation for a spherical Fermi
surface in $D = 3$ with one Landau parameter, $f_c(\bs, \bt) = f_0$.  As usual
we search for a normal mode solution of the form: 
\begin{equation}
J(\bs; \bq, t)  = e^{-i \omega t} u(\bs; \bq)\ ;
\end{equation}
also set $\bq = q \hat{z}$ for definiteness.  Then equation 
\ref{charge-collective-mode} becomes:
\begin{equation}
(\ve q \cos \theta - \omega) u(\theta, \phi; q) + \ve q \cos \theta
\frac{f_0 k_F^2}{4 \pi^3 \ve} \int_{-1}^1 d(\cos \theta^\prime)
\int_0^{2 \pi} d \phi^\prime u(\theta^\prime, \phi^\prime; q) = 0 
\label{EOM}
\end{equation}  
and the solution then is given by 
\begin{equation}
u(\theta, \phi; q) = \frac{\cos \theta}{x - \cos \theta}
\label{simple-mode}
\end{equation}
where we have introduced the dimensionless quantity 
$x \equiv \frac{\omega}{\ve |\bq|}$.  Equation \ref{EOM} determines the
allowed values of $x$.  Substituting equation \ref{simple-mode} into 
equation \ref{EOM} we find that propagating modes exist for $x > 1$ with
$x$ given implicitly by:
\begin{eqnarray}
1 &=& \frac{k_F m^* f_0}{2 \pi^2} \left(x \ln \frac{x+1}{x-1} - 2 \right) 
\nonumber \\
&=& \frac{F_0}{2} \left( x \ln \frac{x+1}{x-1} - 2 \right) 
\label{implicit}
\end{eqnarray}
with the introduction of the conventional dimensionless form of the 
Landau parameters
\begin{equation}
F_\ell = N^*(0) f_\ell
\end{equation}
and recalling that in $D = 3$
\begin{equation}
N^*(0) = 2 \int_0^\infty \frac{4 \pi k^2 dk}{(2 \pi)^3} 
\delta(\epsilon(\bk) - \epsilon_F) = \frac{m^* k_F}{\pi^2}\ . 
\end{equation}
Analytic solutions of equation \ref{implicit} exist\cite{Negele} in the 
extreme limits $0 < F_0 \ll 1$ and $F_0 \gg 1$:
\begin{equation}
x = \left\{ \begin{array}{ll}
1 + 2 e^{-2/F_0}, & 0 < F_0 \ll 1 \\ 
\\
\sqrt{F_0/3}, & F_0 \gg 1\ .
\end{array}\right.
\end{equation}
It is remarkable that the charge sector can be solved exactly without taking
the semiclassical limit of a macroscopically occupied collective mode.  We
show in the next Section that the position of the poles of the 
boson correlation function agree with the dispersion of the
collective mode found here, as must be the case.

To complete the discussion, we show how to construct spin collective modes
in the semiclassical limit.  We seek to replace the quantum spin
current operator ${\bf J}^a$ with its semiclassical expectation value:
\begin{equation}
\langle J^a(\bs; \bq, t) \rangle = S^a(\bs; \bq, t) 
\label{expectation}
\end{equation}
Note that the collective mode amplitudes are real valued in $\bx$-space as 
$[\vec{J}(\bs; \bq)]^\dagger = \vec{J}(\bs; -\bq)$. 
The replacement of the current operators by $c$-numbers can be
accomplished formally by introducing a coherent state basis which spans the
space of geometric distortions of the Fermi surface\cite{Castro,Castro2}.
Spin collective mode coherent states are generated by exponentials of the spin 
current operator
\begin{equation}
| \Psi[\vec{S}] \rangle = \exp \left\{ \sum_\bt \sum_\bq 
2~ \frac{\theta(\hbn_\bt \cdot \bq)}{\Omega \hbn_\bt \cdot \bq}~ 
\vec{S}(\bt; \bq) \cdot \vec{J}(\bt; -\bq) \right\} |0\rangle
\end{equation}
where $|0 \rangle$ represents the quiescent Fermi liquid. A simple computation
using the spin current algebra equation \ref{non-Abelian} then shows that
\begin{equation}
\frac{\langle \Psi[\vec{S}] | J^a(\bt; \bp) | \Psi[\vec{S}] \rangle}
{\langle \Psi[\vec{S}] | \Psi[\vec{S}] \rangle} = S^a(\bt; \bp)
\end{equation}
consistent with our definition, equation \ref{expectation}. 
In the equation for the spin collective modes
it is also necessary to decouple the expectation value of the product
of two spin currents into the product of expectation values 
$\vec{S}(\bt; \bk) \times \vec{S}(\bt; \bq - \bk)$. 
The decoupling is exact only in the semiclassical limit of a
macroscopically occupied spin mode.  

In summary, multidimensional bosonization captures all of the well-known
physics of charge and spin collective modes.
The derivation of the equations of motion is straightforward and relies only 
on the existence of the Abelian and non-Abelian current-current
commutation relations and the quadratic Hamiltonian which describes the
$U(1)^\infty_k$ fixed point.

\section{Quasiparticle Properties}
\label{propagator}
\setcounter{equation}{0}

In this section we use the bosonization scheme to determine the fermion
propagator of a system of spinless fermions interacting via two body forces. The
question we address is:  What is the nature of the ground state and how
does the ground state depend on the spatial dimension $D$ and the range of the 
interactions?

Making use of the bosonization formula, equation \ref{bosonization},
the fermion propagator for a single patch on the Fermi surface can be expressed
in terms of the boson propagator
\begin{eqnarray}
G_F({\bf S}; {\bf x}, \tau) &=& 
- \langle \psi({\bf S}; {\bf x}, \tau) \psi^\dag(\bs; {\bf 0}, 0) 
\rangle
\nonumber \\
&=& -\frac{\Lambda^{D-1}}{(2\pi)^D} \frac{e^{i \bk_{\bf S} \cdot {\bf x}}}{a}
\langle \exp [ i\frac{\sqrt{4\pi}}{\Omega} \phi({\bf S}; {\bf x}, \tau)] 
\exp [-i \frac{\sqrt{4\pi}}{\Omega} \phi({\bf S}; {\bf 0}, 0)] \rangle 
\end{eqnarray}
which can be rewritten by means of the Baker-Hausdorff formula, 
equation \ref{Baker-Hausdorff} as
\begin{equation}
G_F({\bf S}; {\bf x}, \tau) = -\frac{\Lambda^{D-1}}{(2\pi)^D}
\frac{e^{i \bk_\bs \cdot {\bf x}}}{a} \exp\left[\frac{4\pi}{\Omega^2}
\langle \phi({\bf S}; {\bf x}, \tau) \phi(\bs; {\bf 0}, 0) 
- \phi^2(\bs; {\bf 0}, 0) \rangle \right] 
\label{fermion-propagator}
\end{equation}
It is convenient to express the fields, $\phi$ in terms of the canonical boson
fields, $a$ and $a^\dag$, equation \ref{represent}. 
Then the correlation function $G_\phi(\bs; \bx, \tau)$ can be written as
\begin{equation}
G_\phi({\bf S}; \bx, \tau) = \langle \phi(\bs; \bx, \tau) \
\phi(\bs; {\bf 0}, 0) - \phi^2(\bs; {\bf 0}, 0) \rangle =
\frac{\Omega}{4\pi} \sum_\bq \int \frac{d\omega}{2\pi} [e^{i(\bq \cdot
\bx - \omega \tau)} - 1] 
\frac{ \langle a({\bf S}; \bq) a^\dag(\bs; \bq) \rangle} {\hbn_\bs \cdot \bq} 
\label{back-transformed}
\end{equation}
which determines the fermion Green's function once the Fourier transform of
the boson Green's function
\begin{equation}
G_B({\bf S}; {\bf x}, \tau) = 
-\langle a(\bs; {\bf x}, \tau) a^\dag(\bs; {\bf 0}, 0) \rangle
\end{equation}
is known. This procedure although straightforward in principle can be difficult
to carry out in practice. As we show below, the Fourier transform of the boson
Green's function can be obtained exactly for any interaction and for any 
spatial dimension, but to determine the fermion Green's function this result 
must be back transformed, equation \ref{back-transformed},
and exponentiated, equation \ref{fermion-propagator}. 
Finally, to extract the nature of the
fermion propagator, this space-time result must be 
Fourier transformed back into momentum and frequency space.

To determine the boson propagator we introduce source fields $\xi(\bs; q)$ and
$\xi^*(\bs; q)$ where $q \equiv (\bq, \omega)$ is shorthand notation for
the $D+1$ dimensional vector and construct the generating functional:
\begin{equation}
Z[\xi, \xi^*] = \int Da(\bs; q) Da^*(\bs; q) \exp \left[ -S[a^*, a]
- \sum_\bq \sum_{\bs, \hbn_\bs \cdot \bq > 0} \int \frac{d\omega}{2\pi} \left\{ 
\xi^*(\bs; q) a(\bs; q) + \xi(\bs; q) a^*(\bs; q) \right\}\right]\ . 
\end{equation}
Here the action is given in terms of the Hamiltonian as:
\begin{equation}
S[a^*, a] = \int d\tau~ \left\{ \sum_\bs \int d^Dx~ 
a^*(\bs; x) \frac{\partial}{\partial \tau} a(\bs; x) + H[a^*, a] \right\}\ .
\end{equation}
Then the boson Green's function may be found by differentiating with respect
to the source fields:
\begin{equation}
G_B({\bf S}, {\bf T}; q) = -\langle a(\bs; q) a^*(\bt; q) \rangle = 
-\frac{\delta^2 \ln Z}{\delta \xi(\bt; q) \delta \xi^*(\bs; q)}
\bigg{|}_{\xi = \xi^* = 0} \ . 
\label{xi-derive}
\end{equation}
To illustrate this procedure consider a system of spinless fermions in two
dimensions interacting via short range Fermi liquid interactions.  The
action is 
\begin{equation}
S[a^*, a] = S_0[a^*, a] + S_{FL}[a^*, a]
\end{equation}
where the free fermion contribution to the action, written in terms of the 
canonical boson fields, is:
\begin{equation}
S_0[a^*, a] = 
-\sum_\bq \sum_{\hbn_\bs \cdot \bq > 0} \int \frac{d\omega}{2\pi}(i\omega
-v_F^* \hbn_\bs \cdot \bq) a^*(\bs; q) a(\bs; q)
\label{112}
\end{equation}
and the term due to Fermi liquid interactions can be expressed as:
\begin{equation}
S_{FL}[a^*, a] = \frac{1}{2V} \int \frac{d\omega}{2\pi} \sum_{\bs, \bt} 
\sum_\bq f({\bf S}, {\bf T}) J({\bf S}; -q) J({\bf T}; q)
\label{landau-action}
\end{equation}
where we recall that
\begin{equation}
J(\bs; q) = \sqrt{\Omega \hbn_\bs \cdot \bq} \left[a(\bs; q)
\theta(\bs; \bq) + a^*(\bs; -q) \theta(\bs; -\bq) \right] . 
\end{equation}
In equation \ref{112} and in what follows we abbreviate 
$\sum_\bq \sum_{\bs, \hbn_\bs \cdot \bq > 0}$ by simply 
$\sum_\bq \sum_{\hbn_\bs \cdot \bq > 0}$. 

For free fermions the generating function is
\begin{eqnarray}
Z[\xi, \xi^*] &=& \int Da Da^*  \exp \bigg{\{}
\sum_\bq \sum_{\hbn_\bs \cdot \bq > 0} \int \frac{d\omega}{2\pi}  
a^*(\bs; q) (i \omega - v_F^* \hbn_\bs \cdot \bq) a(\bs; q) 
\nonumber \\
&-& \xi^*({\bf S}; q) a({\bf S}; q) - \xi({\bf S}; q) a^*({\bf S}; q) \bigg{\}}
\ .
\end{eqnarray}
The action is quadratic in the bose fields which can be integrated out to give
\begin{equation}
Z[\xi, \xi^*] = \exp - \sum_\bq \sum_{\hbn_\bs \cdot \bq > 0} \int
\frac{d\omega}{2\pi} \xi^*(\bs; q) \frac{1}{(i\omega-v_F^* \hbn_\bs \cdot \bq)}
\xi(\bs; q) 
\end{equation}
and therefore
\begin{equation}
G_B^0(\bs, \bt; q) = -\langle a(\bs; q) a^*(\bt; q) \rangle 
= \frac{\delta^{D-1}_{\bs,\bt}}{i \omega - v_F^* \hbn_\bs \cdot \bq}
\end{equation}
On taking the modified Fourier transform, equation \ref{back-transformed},
the patch-diagonal piece of the $G_\phi$ correlation function is given by:
\begin{equation}
G_\phi^0(\bs; \bx, \tau) = -\frac{\Omega^2}{4\pi} \ln \left[\frac{\hbn_\bs
\cdot \bx + i v_F^* \tau}{ia}\right] \ .
\end{equation}
Exponentiating this equation, 
the free fermion Green's function is found to be given by
\begin{equation}
G_F^0({\bf S}; {\bf x}, \tau) = \frac{i\Lambda}{(2\pi)^2} \frac{e^{i \bk_\bs
\cdot \bx}}{\hbn_\bs \cdot \bx + i v_F^* \tau}\ . 
\end{equation}
Additional details of the calculation are presented 
in Appendix \ref{fermion_details}. 

The analysis can be extended now to include the interactions, 
equation \ref{landau-action}.
First Fourier decompose the coupling $f({\bf S}, {\bf T})$ and 
introduce Landau parameters $f_n$:
\begin{equation}
f({\bf S}, {\bf T}) = \sum_n f_n \cos [n \theta(\bk_\bs, \bk_\bt)]
\end{equation}
and recall the addition theorem in $D=2$
\begin{equation}
\cos [n \theta(\bk_\bs, \bk_\bt)]
= (-1)^n \bigg{[} \cos[n \theta(\bk_\bs, \bq)] \cos[n \theta(\bk_\bt, \bq)] 
+ \sin[n \theta(\bk_\bs, \bq)] \sin[n \theta(\bk_\bt, q)] \bigg{]}\ .
\end{equation}
Here $\theta(\bk_\bs, \bk_\bt)$ is the angle between $\bk_\bs$ and $\bk_\bt$.
To avoid the complicated Bogoliubov transformation needed to diagonalize the
Hamiltonian, 
we introduce instead auxiliary gauge fields $A_n^{\ell}$ and $A_n^t$, 
two fields for each $n$, and perform a Hubbard-Stratonovich transformation to 
decouple the interaction\cite{KHM}.  We call these auxiliary
fields ``gauge'' fields since in the case of the long-range Coulomb 
interaction which we consider below they correspond to 
the scalar component of the 
electromagnetic potential, $A_0$, which mediates the interaction 
(see figure \ref{diagrams}).  
Written in terms of the gauge fields the short-range interaction has the form
\begin{eqnarray}
S_{FL} &=& \frac{1}{2V} \int \frac{d\omega}{2\pi}
\sum_\bq \left\{ (-1)^n  \left[ A_n^{\ell}(-q) \frac{1}{f_n} A_n^{\ell}(q) 
+ A_n^t(-q) 
\left( \frac{1}{f_n} -\delta_{n, 0} \right) A_n^t (q) \right] \right. 
\nonumber \\
&+& i \sum_\bs \left[ A_n^{\ell} (-q) \cos [n \theta(\bk_\bs, \bq)]
+ A_n^t(-q) \sin [n \theta(\bk_\bs, \bq)] \right] J(\bs; q)
\nonumber \\
&+& \left. i\sum_\bs \left[A_n^{\ell}(q) \cos [n \theta(\bk_\bs, -\bq)] 
+ A_n^t(q) \sin [n \theta(\bk_\bs, -\bq)] \right] J(\bs; -q) \right\}
\end{eqnarray}
where it is understood that the repeated index $n$ is summed over.
When the gauge fields are integrated out in the generating functional the
original interaction is reproduced.  Now if we express the currents
$J(\bs; q)$ in terms of the canonical boson fields we see that
\begin{eqnarray}
&& \int \frac{d\omega}{2 \pi}~ 
\sum_{\bq, \bs} A_n^\ell(-q) \cos [n \theta(\bk_\bs, \bq)] J({\bf S}; q)
\nonumber \\
&=& \int \frac{d\omega}{2 \pi}~
\sum_{\bq, \bs} A_n^\ell(q) \cos [n \theta(\bk_\bs, -\bq)] J({\bf S}; -q)   
\nonumber \\
&=& \int \frac{d\omega}{2 \pi}~ \sum_{\bq, \hbn_\bs \cdot \bq > 0} 
\sqrt{\Omega \hbn_\bs \cdot \bq}
\left[ a({\bf S}; q) A_n^\ell(-q) \cos [n \theta(\bk_\bs, \bq)] \right. 
\nonumber \\
&+& \left. a^*(\bs; q) A_n^\ell(q) \cos [n \theta(\bk_\bs, -\bq)] \right] 
\end{eqnarray}
and therefore
\begin{eqnarray}
S_{FL} &=& \frac{1}{V} \int \frac{d\omega}{2\pi} \sum_\bq \bigg{\{}
\left[ A_n^\ell(-q) \frac{(-1)^n}{2f_n} A_n^\ell(q) 
+ A_n^t (-q)(-1)^n \frac{(1-\delta_{n,0})}{2f_n} A_n^t(q) \right] 
\nonumber \\
&+& i \sum_{\hbn_\bs \cdot \bq > 0} a({\bf S}; q) \left[ \sqrt{\Omega
\hbn_\bs \cdot \bq} [A_n^\ell(-q) \cos[n \theta(\bk_\bs, \bq)] +
A_n^t(-q) \sin [n \theta(\bk_\bs, \bq)] \right]  
\nonumber \\
&+& a^*(\bs; q) \left[ \sqrt{\Omega \hbn_\bs \cdot \bq} 
[A_n^\ell(q) \cos[n \theta(\bk_\bs, -\bq)] 
+ A_n^t(q) \sin [n \theta(\bk_\bs, -\bq)] \right] \bigg{\}} \ .
\end{eqnarray}
The action is now in a form which allows the canonical bose fields, $a$ and
$a^*$, to be integrated out and the resulting 
generating functional is expressed as
\begin{eqnarray}
Z[\xi, \xi^*] &=& \int DA_n^\ell DA_n^t \exp -S_G[A]
\nonumber \\
&\times&  \exp -\left\{ \int \frac{d\omega}{2\pi} 
\sum_\bq  \sum_{\hbn_\bs \cdot \bq > 0}
\frac{i}{V} \frac{\sqrt{\Omega \hbn_\bs \cdot \bq}}
{i \omega - \ve \hbn_\bs \cdot \bq} \left[\xi^*(\bs; q) b_n^m(\bs; -\bq) 
A_n^m(q) + \xi(\bs; q) b_n^m(\bs; \bq) A_n^m(-q)\right]\right.
\nonumber \\
&+& \left. \xi^*(\bs; q) \frac{1}{(i\omega
-v_F^* \hbn_\bs \cdot \bq)} \xi(\bs; q)\right\}
\end{eqnarray}
here the index $m$ ranges over $(\ell, t)$ and the 
associated 2-vectors $b_n^m(\bs; \bq)$ have components:
\begin{eqnarray}
b_n^\ell(\bs; \bq) &\equiv& \cos [n \theta(\bk_\bs, \bq)] 
\nonumber \\
b_n^t(\bs; \bq) &\equiv& \sin[n \theta(\bk_\bs, \bq)])\ .
\end{eqnarray} 
Using this notation the gauge action can be written as
\begin{equation}
S_G[A] = \frac{1}{V} \int \frac{d\omega}{2 \pi} \sum_{\bq;  
\hbn_\bs \cdot \bq > 0} A_n^m(-q)~ K_{nn^\prime}^m(q)~ A_{n^\prime}^m(q) 
\end{equation}
where the matrices $K_{nn^\prime}$ are given by
\begin{equation}
K_{nn^\prime}^\ell(q) = (-1)^n \left[\frac{\delta_{nn^\prime}}{f_n} -
\frac{\Lambda}{(2\pi)^2}\sum_\bs
\frac{\hbn_\bs \cdot \bq}{(i\omega-v_F^* \hbn_\bs \cdot \bq)}
\cos [n \theta (\bk_\bs, \bq)] \cos [n^\prime \theta(\bk_\bs, \bq)] \right] 
\label{lmatrix}
\end{equation}
and
\begin{equation}
K_{nn^\prime}^t(q) = 
(-1)^n \left[\frac{\delta_{nn^\prime} (1-\delta_{n,0})}{f_n}
- \frac{\Lambda}{(2 \pi)^2} \sum_\bs
\frac{\hbn_\bs \cdot \bq}{(i\omega-v_F^* \hbn_\bs \cdot \bq)}
\sin [n \theta(\bk_\bs, \bq)] \sin[n^\prime \theta(\bk_\bs, \bq)] \right].
\end{equation}
In these equations the $\bq$-sum is now restricted to $\hbn_\bs \cdot \bq > 0$, 
and the sum over $\bs$ (integral over
angles) runs free. Recognizing that $\Lambda \sum_{\bs} = 2 \pi k_F$, 
equation \ref{lmatrix} can be written in the more familiar form:
\begin{equation}
K_{nn^\prime}^\ell(q) = (-1)^n \left[ \frac{\delta_{nn^\prime}}{f_n} - N^*(0)
\int \frac{d\theta}{2\pi} \frac{\cos \theta}{x - \cos \theta} 
\cos(n \theta) \cos(n^\prime \theta) \right] 
\end{equation}
and a similar equation for $K_{nn^\prime}^t$, here as usual 
$x \equiv \omega/(\ve |\bq|)$ and
$N^*(0) = m^*/2\pi$ is the density of states of spinless fermions
in two dimensions.

The gauge fields can now be integrated out to provide an explicit expression
for the generating functional
\begin{eqnarray}
Z[\xi, \xi^*] &=& \exp-\left\{ \sum_{\bq, \hbn_\bs \cdot \bq > 0}
\int \frac{d\omega}{2\pi} \left[\sum_\bs \xi^*({\bf S}; q) \frac{1}{i\omega
-v_F^* \hbn_\bs \cdot \bq} \xi(\bs; q) \right.\right.
\nonumber \\
&+& \left.\left. \sum_{\bf ST} \xi^*(\bs; q) 
\frac{\sqrt{\Omega \hbn_\bs \cdot \bq}}{i\omega - v_F^* \hbn_\bs \cdot \bq}
\left( b_n^m({\bf S}; -\bq)~ D_{nn^\prime}^m(q)~ 
b_{n^\prime}^m({\bf T}; \bq) \right) \right. \right.
\nonumber \\
&\times& \left. \left.
\frac{\sqrt{\Omega \hbn_\bt \cdot \bq}}{i\omega -v_F^* \hbn_\bt \cdot \bq} 
\xi(\bt; q) \right] \right\}\ .
\end{eqnarray}
Here the gauge propagator is given by the matrix inverse of the $K$-matrix:
\begin{equation}
{\bf D}^m(q) = \left[ {\bf K}^m(q) \right]^{-1}\ .
\end{equation}
The boson propagator now can be determined directly by differentiating
the generating functional, equation \ref{xi-derive}: 
\begin{equation}
G_B({\bf S}, {\bf T}; q)= \frac{\delta_{\bs, \bt}^{D-1}}{i\omega -v_F^*
\hbn_\bs \cdot \bq} + \frac{\sqrt{\Omega \hbn_\bs \cdot \bq}}
{i \omega - v_F^* \hbn_\bs \cdot \bq} b_n^m(\bs; -\bq)~
D_{nn^\prime}^m(q)~  b_{n^\prime}^m(\bt; \bq)
\frac{\sqrt{\Omega \hbn_\bt \cdot \bq}}{i\omega -v_F^* \hbn_\bt \cdot \bq}
\label{direct-boson}
\end{equation}
and the fermion propagator follows immediately upon exponentiating: 
\begin{eqnarray}
G_F(\bs; \bx, \tau) &=& \frac{\Lambda}{(2 \pi)^2}~
\frac{e^{i \bk_{\bf S} \cdot {\bf x}}}
{\hbn_\bs \cdot \bx + iv_F^* \tau} \exp \bigg{[}
- \sum_\bq \int \frac{d\omega}{2\pi} 
(e^{i({\bf q} \cdot {\bf x} - \omega \tau)} - 1)
\frac{b_n^m(\bs; -\bq)~ D_{nn^\prime}^m(q) b_{n^\prime}^m(\bs; \bq)}
{(i\omega - v_F^* \hbn_\bs \cdot \bq)^2} \bigg{]}
\nonumber \\
&=& \frac{\Lambda}{(2 \pi)^2}~
\frac{e^{i \bk_\bs \cdot \bx}}{\hbn_\bs \cdot \bx + iv_F^*\tau} 
\exp [-\delta G_\phi(\bs; \bx, \tau)]; 
\quad \bx_\bot \equiv \bx \times \hbn_\bs = 0
\label{key}
\end{eqnarray}
This key result is exact within the bosonization scheme. The condition 
$\bx_\bot = 0$ can be replaced by $|\bx_\bot| \Lambda < 1$ 
provided we control the
logarithmic infrared divergence of the free boson correlation function by
placing the system in a large box.  The weak divergence, which is a consequence
of treating the Fermi surface as locally flat within each patch, presents no
real difficulties in practice and is ignored in the following. For $|\bx_\bot|
\Lambda > 1$ the fermion Green's function vanishes since bosons separated by
large perpendicular  distances, $\bx_\bot$, are uncorrelated. For convenience in
the second line of equation \ref{key} we have introduced the notation 
$\delta G_\phi({\bf S}; {\bf x}, \tau)$ 
to denote the modified Fourier transform of the
additive correction to the free $\phi$-boson propagator due to interactions
\begin{equation}
\delta G_\phi({\bf S}; {\bf x}, \tau) = \int_{-\lambda/2}^{\lambda/2} 
\frac{d^D q}{(2\pi)^D}\int
\frac{d\omega}{2\pi}  [e^{i({\bf q} \cdot \bx -\omega \tau)}-1] 
\delta G_\phi({\bf S}; {\bf q}, \omega)
\end{equation}
where we can identify $\delta G_\phi({\bf S}; {\bf q}, \omega) $ as
\begin{equation}
\delta G_\phi({\bf S}; {\bf q}, \omega) = \frac{b_n^m({\bf S}; \bq)~
D_{nn^\prime}^m(q) b_{n^\prime}^m({\bf S}; \bq)}
{(i\omega -v_F^* \hbn_\bs \cdot \bq)^2} 
\end{equation}
In equation \ref{key} 
the momentum integral ranges over $(-\lambda/2, \lambda/2)$ in
each direction. By contrast in the free part of the boson and fermion
propagators the perpendicular momentum ranges over the much larger interval
$(-\Lambda/2, \Lambda/2)$. The reason for this lies in the fact that the Fermi
surface normal points in a different direction in each patch on the Fermi
surface. As interactions couple together different patches, 
and as the patches are squat boxes with dimension 
$\Lambda^{D-1} \times \lambda$, only wavevectors 
$|\bq|< \lambda$ are permitted by the geometry of the construction. 

To complete the program of computing the fermion two point function, the
in-patch Green's function equation \ref{key} 
is summed over all patches on the Fermi surface,
\begin{equation}
G_F({\bf x}, \tau) = {\sum_\bs}^\prime G_F(\bs; \bx, \tau).
\label{total-fermion}
\end{equation}
As in section \ref{essentials}, 
the prime indicates that the sum is only over patches for which $|{\bf x}
\times \hbn_\bs| \Lambda < 1$. The expression for the fermion Green's function
given here, equations \ref{key} and \ref{total-fermion}, 
is identical to that found by Castellani et al.\cite{Castellani} 
who made use of Ward
identities derived in the fermion basis\cite{Dzyaloshinsky} 
to obtain equation \ref{key}.

In general the calculation of correlation functions is complicated as it
involves large matrices, but to illustrate the method, as is common practice, 
we can
truncate the Landau expansion and set all parameters except $f_0$ equal to
zero. First, to demonstrate the nonperturbative nature of the approach
explicitly, we consider the propagator of a fermion liquid in one spatial
dimension. In one dimension there are two Fermi points and fermion
quasiparticles can only propagate to the right or to the left. In this instance
the generating functional is given by
\begin{eqnarray}
Z[\xi, \xi^*] &=& \exp - \sum_q \int \frac{d\omega}{2 \pi} 
\left\{ \xi^*(R, q) \frac{1}{i\omega -v_F^* q}\xi(R, q) 
- \xi^*(L, q) \frac{1}{i\omega+v_F^*q} \xi(L, q)
\right.
\nonumber \\
&+& \left. \frac{q}{2\pi} \left( \frac{ \xi^*(R, q)}{i\omega-v_F^* q} -
\frac{\xi (L, -q)}{i\omega+v_F^*q}\right)
K_{00}^{-1}(q) \left( \frac{\xi(R, q)}{i\omega
-v_F^* q}-\frac{\xi^*(L, -q)}{i\omega+v_F^*q}\right)\right\}
\end{eqnarray}
where the single element of the matrix $K_{00}$ is given by
\begin{equation}
K_{00}(q) = 
\left[ \frac{1}{f_0} -\frac{q}{2\pi} \left( \frac{1}{i\omega -v_F^* q}-
\frac{1}{i\omega +v_F^* q}\right) \right]
\end{equation}
The boson propagator for quasiparticles propagating to the right, for example,
is given by
\begin{equation}
G_B(R; q) =\frac{1}{i\omega -v_F^* q} + 
\frac{q}{2\pi} \frac{K_{00}^{-1}(q)}{(i\omega -v_F^* q)^2}
\end{equation}
which after some algebra can be rewritten as
\begin{equation}
G_B(R; q)=\frac{a}{i\omega -qv_F^* (1+f_0/\pi v_F^*)^{1/2}}+
\frac{b}{i\omega +qv_F^* (1+f_0/\pi v_F^*)^{1/2}} 
\end{equation}
where
\begin{equation}
\left( \begin{array}{c}
a \\ b
\end{array} \right) = \frac{1}{2}\left[1\pm \frac{\left(1+f_0/2\pi v_F^*
\right)}{(1+f_0/\pi v_F^*)^{1/2}}\right].
\end{equation}
Now $v_F^\prime = v_F^* (1+ f_0/2\pi v_F^*)$ is the renormalized Fermi velocity
due to {\it intrapatch} scattering.  Introducing the dimensionless coupling $F_0
\equiv f_0 / (\pi v_F^\prime)$ 
the boson Green's function can be expressed as
\begin{equation}
G_B(R, q) = \frac{\cosh^2 \eta}{i\omega - q v_F^\prime
(1-(\frac{F_0}{2})^2)^{1/2}} + \frac{\sinh^2 \eta}
{i\omega - q v_F^\prime (1-(\frac{F_0}{2})^2)^{1/2}}
\end{equation}
where
\begin{equation}
\cosh^2\eta = \frac{1}{2} \left[1 + \frac{1}{\sqrt{1-(F_0/2)^2}}\right]
\end{equation}
reflecting the fact that {\it interpatch} interactions further renormalize
the speed of right and left moving quasiparticles, 
$\bar{v}_F = v_F^\prime \sqrt{1-(F_0/2)^2}$. The fermion propagator can now be
found analytically using equations \ref{fermion-propagator} and 
\ref{back-transformed} as
\begin{equation}
G_F(R; x, \tau) = i e^{i k_F x}~ 
\frac{(x^2 + \bar{v}_F^2 \tau^2)^{-\zeta}}{(x + i\bar{v}_F\tau)} 
\label{luttinger}
\end{equation}
The result is exact and identical to that found by diagonalizing the bosonic
Hamiltonian via a Bogoliubov transformation\cite{Voit}.  
It illustrates the well-known fact that in one spatial dimension 
even short-range
interactions destroy the Fermi liquid fixed point. The propagator is
non-analytic in the coupling constant $f_0$, the anomalous exponent $\zeta
= \sinh^2 \eta$, and the simple pole of the noninteracting Fermi gas at
$x = -iv_F^*\tau$ is replaced by a branch cut. The distribution function for
fermions near the Fermi surface is obtained by Fourier transform of the 
equal-time ($\tau = 0$) propagator:  
\begin{equation}
n_R(k)=n_R(k_F) -C \sgn(k-k_F)|k-k_F|^{2\zeta}
\end{equation}
where $C$ is a nonuniversal constant.  The occupancy no longer has a step
discontinuity at the Fermi surface, rather it is continuous, although
non-analytic, with a divergent
slope at the Fermi points, characteristic of a Luttinger liquid.

We have shown that bosonization transforms the problem of interacting fermions
into a problem of Gaussian bosons which in one dimension can be solved exactly.
In one dimension the conventional Fermi liquid picture breaks down and the
fixed point is a Luttinger liquid.  
In the remainder of this section we use the multidimensional bosonization
approach to study the properties of interacting fermions systems in
higher dimension, especially two spatial dimensions.
In $D=2$, the zero-zero elements of the inverse gauge propagator are given by
\begin{equation}
K_{00}^\ell(q) = \frac{1}{f_0} + \chi_0(q) 
\end{equation}
where 
\begin{equation}
\chi_0(\bq, \omega) = N^*(0) 
\int \frac{d\theta}{2\pi}\frac{\cos \theta}{\cos \theta 
- \frac{i \omega}{\ve |\bq|}}
\end{equation}
It follows from equation \ref{direct-boson} 
that the boson propagator is given by
\begin{equation}
G_B(\bs,\bt; q)= \frac{\delta_{\bs ,\bt}}{i\omega-v_F^* \hbn_\bs \cdot \bq} 
+ \frac{\Lambda}{(2\pi)^2} \frac{\sqrt{\hbn_\bs \cdot \bq}}{i\omega-v_F^*
\hbn_\bs \cdot {\bf q}} \left( \frac{f_0}{1+f_0 \chi_0(q)} \right)
\frac{\sqrt{\hbn_\bt \cdot \bq}}{i\omega-v_F^* \hbn_\bt \cdot \bq} 
\end{equation}
In particular the in-patch boson propagator is:
\begin{eqnarray}
G_B(\bs; q) &=& \frac{1}{i\omega-v_F^* \hbn_\bs \cdot {\bf q}} +
\frac{\Lambda}{(2\pi)^2}
\frac{\hbn_\bs \cdot \bq}{(i \omega - v_F^* \hbn_\bs \cdot \bq)^2}
\frac{f_0}{1+f_0 \chi_0(q)} 
\nonumber \\
&\approx&  \left[i\omega -v_F^* \hbn_\bs \cdot \bq
- \frac{\Lambda}{(2\pi)^2}(\hbn_\bs \cdot \bq)
\frac{f_0}{1 + f_0 \chi_0(q)}\right]^{-1}\ .
\label{long-prop}
\end{eqnarray}

Analytically continuing to real frequency, the longitudinal susceptibility
is given by the Lindhard function:
\begin{equation}
\chi_0(x) = N^*(0) \left[ 1- \frac{\theta(x^2-1)|x|}{\sqrt{x^2-1}} +
\frac{i\theta(1-x^2)|x|}{\sqrt{1-x^2}}\right]
\label{chi_0}
\end{equation}
where $x \equiv \omega/(v_F^* |\bq|)$. 
In the so-called $q$-limit (the limit $\omega \rightarrow 0$ is taken 
{\it before} ${\bf q} \rightarrow 0$ and hence the ratio $\omega/|\bq| 
\rightarrow 0$) the Lindhard function has an imaginary part: 
$\chi_0 \approx N^* (0) [1+ i\omega/(v_F^* |\bq|)]$.  The bosons have a finite
lifetime because they scatter between patches. 
In this limit the boson propagator is expressed most naturally in terms
of the scattering amplitude $A_0 \equiv F_0/(1+F_0)$ where as before 
the dimensionless coupling $F_0 \equiv N^*(0) f_0$.  
\begin{equation}
G_B({\bf S}; \bq, \omega) = \delta_{\bf S,T}^{D-1} \bigg{\{}
\omega-v_F^* \hbn_\bs \cdot \bq
\bigg{[} 1 - \left(\frac{\Lambda}{2 \pi k_F}\right) A_0 \bigg{]}
- i \hbn_\bs \cdot {\bf q} A_0^2
\frac{|\omega|}{|\bq|} \bigg{\}}^{-1}
\end{equation}
The bosons have a finite lifetime, as they scatter between patches, but
unlike in one dimension, the renormalization of the
boson velocity due to interactions amongst the bosonized shell of excitations 
with energy less than $v_F^*\lambda$ is insignificant, vanishing in the 
$N = 2 \pi k_F / \Lambda \rightarrow \infty$ limit.

To consider the effect of long range interactions, for example the Coulomb
interaction $V(\bq) = 2\pi e^2 / |\bq|$, 
we simply replace the Landau coupling $f_0$ in equation \ref{long-prop} by 
the momentum-dependent interaction\cite{Silin} $V(\bq)$. 
In the $q$-limit, the effective interaction
$V(\bq)/[1 + \chi_0 V(\bq)]$ is screened and the analysis given above holds. 
In the opposite limit, the $\omega$-limit, the collective excitations of the
interacting fermion system are given by the zeros of the boson propagator 
where the self-energy diverges,
${\rm det} {\bf K} = 0$. In particular, $\chi_0 \approx -1/(2 x^2)$ 
and therefore
\begin{equation}
{\rm det} {\bf K} = 1-\frac{\pi e^2}{|\bq|} N^*(0) 
\left(\frac{\ve |\bq|}{\omega}\right)^2
\end{equation}
which vanishes when $\omega^2 = 2 \pi e^2 n_f |\bq| / m^*$ where 
$n_f \equiv k_F^2/ 4 \pi$ is the density of spinless fermions.
However, it is well known 
that the frequency of the longitudinal collective mode, the plasma frequency, 
should only depend on the bare mass.  This requirement, known as Kohn's 
theorem\cite{Kohn}, is an immediate consequence of Galilean invariance, as
large numbers of electrons sloshing back and forth in response to an applied
oscillating electric field carry momentum corresponding to the product of
their speed multiplied by the bare electron mass.
To see that Kohn's theorem holds within the bosonization scheme, recall that the
effective Hamiltonian of a fermion system interacting via long range forces
describes a gas of low energy excitations coupled by both long range Coulomb
interactions and by short range Fermi liquid forces.  For our purposes
it is sufficient 
to set all of the Landau parameters except $f_1$ equal to zero. 
$K$ is now a three by three matrix whose determinant is given by
\begin{equation}
{\rm det} {\bf K} = \left[ \begin{array}{ccc}
V(\bq)^{-1} + \chi_0, & \chi_{\ell 0}, & 0 \\
\chi_{\ell 0}, & f_1^{-1} + \chi_{\ell \ell}, & 0 \\
 0, & 0,& f_1^{-1} + \chi_{tt}
\end{array}
\right]
\end{equation}
where the susceptibilities are given by
\begin{eqnarray} 
\chi_{\ell 0}(x) &=& N^*(0) \int
\frac{d\theta}{2\pi}\frac{\cos^2\theta}{\cos\theta -x} 
\approx - \frac{N^*(0)}{2x}
\nonumber \\
\chi_{\ell \ell}(x) &=& N^*(0) \int
\frac{d\theta}{2\pi}\frac{\cos^3\theta}{\cos\theta -x} 
\approx - \frac{3 N^*(0)}{8x^2}
\end{eqnarray}
and
\begin{equation}
\chi_{tt}(x) 
= N^*(0) \int \frac{d\theta}{2\pi}\frac{\cos\theta\sin^2 \theta}{\cos
\theta - x} \approx - \frac{N^*(0)}{8x^2}
\end{equation}
the approximate expressions are given in the $\omega$-limit. 
The determinant vanishes when
\begin{equation}
\frac{1}{V(\bq)} = -\chi_0(x) + f_1 \chi_{\ell 0}^2(x) \approx \frac{N^*(0)}{2}
\left(\frac{v_F^* |\bq|}{\omega}\right)^2 (1 + \frac{F_1}{2})~ ,
\end{equation}
but $(1+F_1/2)= m^* / m$ and therefore the zeros are located at
$\omega^2 = (2 \pi e^2 n_f / m) |\bq|$, 
which is the correct plasma frequency of a two dimensional Fermi liquid.

Now we can determine the quasiparticle propagator of fermions interacting
via the short-range Landau interaction $f_0$ using:
\begin{equation}
G_F({\bf S}; {\bf x}, \tau) = G_F^0({\bf S}; {\bf x}, \tau) 
\exp [-\delta G_\phi({\bf S}; {\bf x}, \tau)]\ . 
\end{equation}
In general this is a complicated procedure; but for our purposes here it is
sufficient to expand in powers of $\delta G_\phi$, as the expansion 
converges quickly for low fermion energies.  The Fourier transform can then
be taken directly,
\begin{equation}
G_F({\bf S; k}, \omega) = \sum_{\bq} \int \frac{d\nu}{2 \pi}~ 
G_F^0(\bs; \bk-\bq, \omega-\nu) [\delta_{{\bf q}, 0} \delta(\nu) 
+ \delta G_\phi({\bf S}; \bq, \nu) + \ldots] 
\label{f0}
\end{equation}
where
\begin{equation}
\delta G_\phi({\bf S}; {\bf q}, \nu) 
= -\frac{1}{(i\nu-v_F^* \hbn_\bs \cdot \bq)^2}~ 
\frac{f_0}{1+f_0 \chi_0({\bf q})} \ .
\label{correction}
\end{equation}
If we further 
expand in powers of $f_0$, the first order correction is purely real and
leads to only an infinitesimal velocity shift to the fermion propagator. The
leading contribution to the imaginary part of the fermion self-energy appears at
next order,
\begin{equation}
\delta G_\phi({\bf S; q}, \nu)
= \frac{f_0^2 \chi_0(\bq, \nu)}{(i\nu-v_F^* \hbn_\bs \cdot \bq)^2}\ . 
\end{equation}
Now the only dependence of $\delta G_\phi$ 
on the perpendicular momentum $q_\bot \equiv \hbn_\bs \times \bq$ 
is through the susceptibility $\chi_0(\bq, \nu)$:
\begin{equation}
\chi_0(\bq, \nu) = N^*(0) \left( 
1 - \frac{\nu~ \sgn(\nu)}{\sqrt{\nu^2 + (\ve q_\parallel)^2 
+ (\ve q_\bot)^2}} \right)\ ; 
\end{equation}
here, as usual, $q_\parallel \equiv \hbn_\bs \cdot \bq$.
Carrying out the integral over the perpendicular momentum in equation 
\ref{f0}, and retaining only the imaginary part we find:
\begin{equation}
{\rm Im} \int_{-\lambda/2}^{\lambda/2} dq_\bot~ \chi_0({\bf S}; {\bf q}, \nu) =
\frac{N^*(0)}{v_F^*} \nu~ \sgn(\nu)~
\ln \left( \frac{q_\parallel^2 + (\nu/v_F^*)^2}{\lambda^2} \right) 
\end{equation}
and therefore the contribution to the quasiparticle Green's function is
\begin{eqnarray}
\delta G_F(\bs; k_\parallel, \omega) &=& i \frac{f_0^2}{(2\pi)^2}
\frac{N^*(0)}{v_F^*} \int_0^\omega d\nu \int_{-\lambda/2}^{\lambda/2} 
\frac{dq_\parallel}{2 \pi}~ 
i \nu \sgn(\nu) \ln \left( (\nu/\ve)^2 + q_\parallel^2 \right)
\nonumber \\
&\times& \frac{1}{i(\nu-\omega)-v_F^*(q_\parallel - k_\parallel)}~
\frac{1}{(i\nu-v_F^* q_\parallel)^2}\ .
\end{eqnarray}
The integral over $q_\parallel$ can be carried out
by complex integration and the final
integral over $\nu$ evaluated exactly with the result that
\begin{eqnarray}
\delta G_F(\bs; k_\parallel,\omega) &=& i \left(\frac{f_0}{2\pi}\right)^2
\frac{N^*(0)}{2(v_F^*)^2} \frac{\sgn(\omega)}{(v_F^* k_\parallel -i\omega)^2}
\bigg{\{} \bigg{[}(i \omega)^2 
+ \left(\frac{i\omega-v_F^* k_\parallel}{2}\right)^2 \bigg{]} \ln
\left|\frac{i\omega -v_F^* k_\parallel}{\ve \lambda}\right|  
\nonumber \\
&+& \bigg{[} (i\omega)^2 - \left(\frac{i\omega-v_F^* k_\parallel}{2}\right)^2 
\bigg{]} \ln \left| \frac{i\omega +v_F^* k_\parallel}{\ve \lambda} \right| 
- \frac{i\omega}{2}(2i\omega -\ve k_\parallel)\bigg{\}}\ .
\end{eqnarray}
Analytically continuing to real frequencies we find that the imaginary part of
the fermion self-energy is given by
\begin{eqnarray}
{\rm Im} \Sigma_F (\bs; k_\parallel, \omega) &=& \left(\frac{f_0}{2\pi}\right)^2
\frac{N^*(0)}{2 (v_F^*)^2}
\sgn(\omega) \bigg{\{} \bigg{[} \omega^2 + \left(\frac{\omega-v_F^*
k_\parallel}{4}\right)^2 \bigg{]} 
\ln \left( \frac{\omega - v_F^* k_\parallel}{\ve \lambda}\right)
\nonumber \\
&+& \bigg{[} \omega^2-\left(\frac{\omega-v_F^* k_\parallel}{4}\right)^2
\bigg{]} \ln \left( \frac{\omega + v_F^* k_\parallel}{\ve \lambda}\right) 
- \frac{\omega}{2} (2\omega - \ve k_\parallel )\bigg{\}}\ . 
\end{eqnarray}
The appearance of the combination $\omega-v_F^* k_\parallel$ in the 
self-energy is the result of linearizing the fermion dispersion about the
Fermi energy.  This form is not expected
to survive in detail, either within the fermionic representation\cite{Halboth}, 
or perturbatively within bosonization\cite{old-Haldane,Kopietz_book}, when
the quadratic term in the fermion dispersion is included.  This is not 
surprising, as both the quadratic dispersion and the self-energy are
equally irrelevant in the RG sense.  A consistent calculation of the
self-energy must retain the quadratic term.

The imaginary part of the self-energy at the quasiparticle pole is the inverse
of the quasiparticle lifetime.  The location of the pole has been shifted from
its bare value to $\omega=v_F^\prime k_\parallel$ due to renormalization of the
Fermi velocity $v_F^\prime = v_F^* [1+\frac{\Lambda}{(2\pi k_F)}
\frac{F_0}{1+F_0}]$.  Note that $\Lambda$, not $\lambda$, appears in this 
expression because it is the patch-diagonal piece of the 
current-current interaction, $(f_0/V) J(\bs; \bq) J(\bs; -\bq)$, which 
is responsible for the renormalization of the velocity. 
As a result the imaginary part of the self-energy at the pole is given by
\begin{equation}
{\rm Im} \Sigma_F(\bs; k_\parallel, \omega) = \frac{f_0^2}{(2 \pi)^2}~
\frac{N^*(0)}{2 \ves}~ \sgn(\omega) \left\{ \omega^2 \ln \frac{\Lambda F_0
\omega^2}{\pi \ves \lambda^2 k_F} -\frac{\omega^2}{2} \right\} 
\end{equation}
the form of which we recognize immediately from previous work on
two-dimensional Fermi liquids\cite{Hodges,Bloom,Giuliani,Stamp}. 
This quantity is always negative since $\omega^2
\ll \lambda^2 k_F /\Lambda$. Despite the dependence of the self-energy on 
$\ln \omega$, the weight of the quasiparticle pole, $Z_\bk$, is unchanged
in the $|\bk| \rightarrow k_F$ limit.  
This is as expected, since as noted above short-range Fermi liquid interactions 
make only irrelevant contributions to the quasiparticle
self-energy and do not destroy the Fermi liquid
fixed point.  The result is not changed if we consider the more general
case of nonlocal short range interactions $f({\bf S, T}; \bq)$ as the additional
momentum dependence only gives rise to irrelevant operators which do not change
the behavior of the system.

The question remains as to whether or not singular long-range interactions can
destroy Fermi liquid behavior. In fact Bares and Wen\cite{Bares} 
have shown that for a
sufficiently long-ranged interaction, logarithmic in real space, Fermi liquid
behavior in two dimensions does break down.  Here we show how this comes
about within the bosonization framework and then discuss why Fermi liquid
theory remains intact for fermions interacting via Coulomb or shorter-ranged
forces. As
bosonization is a nonperturbative approach which does not presuppose the form
of the fermion propagator it provides a particularly appropriate framework
within which to address this question.

If we consider a gas of fermions interacting via Coulomb interactions 
$V(\bq) \propto e^2 / |\bq|^{D-1}$ 
then the correction to the boson Green's function is given by equation 
\ref{correction}, slightly generalized to include momentum dependence:
\begin{equation}
\delta G_\phi({\bf S; q}, \omega) = \frac{V(\bq)}{1+V(\bq) \chi_0 (\bq)}
\frac{1}{(i\omega -v_F^* \hbn_\bs \cdot \bq)^2} 
\end{equation}
In dimension greater than one, for small $\bq$ and small $\omega$, 
with $\omega \ll |\bq|$
screening is complete, $\delta G_\phi$ no longer depends on
$V(\bq)$, 
\begin{equation}
\delta G_\phi(\bs; \bq, \omega) \approx \frac{\chi_0^{-1}(q)}
{(i\omega -v_F^* \hbn_\bs \cdot \bq)^2}\ . 
\end{equation}
The imaginary part of the fermion self-energy can be obtained in the manner
we have described above and is given in two dimensions by
\begin{equation}
{\rm Im} \Sigma_F(\omega)_{\rm pole} \approx \frac{1}{(2\pi)^2 \ves N^*(0)} 
\sgn(\omega) \left( \omega^2 \ln \frac{2\pi \omega^2\Lambda}{\ves \lambda^2 k_F}
-\frac{\omega^2}{2} \right) 
\end{equation}
and in three dimensions by
\begin{equation}
{\rm Im} \Sigma_F(\omega)_{\rm pole}~ \propto~  
\frac{\lambda}{\ves k_F^2} \omega^2 \sgn(\omega) + O(\omega^2/k_F^3)  
\end{equation}
in agreement with the well known results of Fermi liquid theory\cite{Baym}. 
As in Fermi liquid theory the momentum cutoff appears in this result. In the 
random phase approximation (RPA) the cutoff $\lambda$ is replaced by an energy
of the order of the plasma frequency\cite{Quinn}.  Azimuthal non-planar 
scattering processes are included in RPA, accounting for the shorter lifetime.
These processes are neglected in our $D = 3$ calculation, though it should
be possible to treat them perturbatively. 

It is clear from this discussion that the breakdown of the Fermi liquid
behavior can only be brought about, if at all, by high-energy processes for
which screening is ineffective. To resolve this question we determine the
equal-time fermion propagator of a gas of
fermions in two dimensions interacting via the logarithmic potential $V(\bq) =
g/\bq^2$ introduced by Bares and Wen\cite{Bares}. Then
\begin{equation}
\delta G_\phi({\bf S}; {\bf x}, \tau = 0) = -\sum_{|\bq| < \lambda} 
(e^{i \bq \cdot \bx} - 1) \int_{v_F^* |\bq|}^{v_F^* \lambda/2} 
\frac{d\omega}{2\pi} \frac{K_{00}^{-1}(q)}{(i\omega- \ve \hbn_\bs \cdot \bq)^2} 
\end{equation}
where
\begin{equation}
K_{00}^{-1}(q) = \frac{g/\bq^2}{1+(g/\bq^2)\chi_0 ({\bf q}, \omega)}
\end{equation}
$\chi_0(x \gg 1) \approx N(0)/2x^2$ and therefore in the $\omega$-limit,
\begin{equation}
K_{00}^{-1}(q) = -\frac{g \omega^2/\bq^2}{(i\omega)^2-g n_f / m}. 
\end{equation}
For this super long-range potential, the collective excitations of the system,
plasmons, have a finite energy, given by the zeros of $K_{00}, \omega_p
=\sqrt{g\frac{n_f}{m}}$. Here we have adjusted the density of states and Fermi
velocity such that the plasma frequency depends on the bare mass as it
should by Kohn's theorem.  The cutoff
$\lambda$ has to be chosen such that the energy cutoff is much larger than the
plasmon energy $v_F^* \lambda \gg \omega_p$.
Obviously bosonization is less accurate for systems with a large plasmon gap,
nevertheless we  expect it to be qualitatively correct. Now since 
$\omega > v_F^* |\bq|$ we may replace 
$(i \omega - v_F^* \hbn_\bs \cdot \bq)^2$ by $\omega^2$ in the
frequency integral and carry out the integrations. The frequency integral gives
a constant due to the plasmon pole at finite frequency. It remains to evaluate
the two dimensional integral over the momentum $\bq$ which because of the factor
of $\bq^2 = q_\bot^2 + q_\parallel^2$ 
in the denominator gives rise to a logarithmic dependence on 
$x_\parallel \equiv \hbn_\bs \cdot \bx$. 
To obtain the fermion quasiparticle propagator $\delta G_\phi$ 
is exponentiated and we find
\begin{eqnarray}
G_F({\bf S}; x_\parallel, \tau = 0) = \left\{ \begin{array}{ll}
\frac{\Lambda}{(2 \pi)^2}~
\frac{e^{ik_Fx_\parallel}}{x_\parallel} |\lambda x_\parallel|^{-\zeta}\ , &
\quad \lambda x_\parallel \gg 1 \\ 
\\
\frac{\Lambda}{(2 \pi)^2}~
\frac{e^{ik_F x_\parallel}}{x_\parallel}, &
\quad \lambda x_\parallel \ll 1
\end{array}\right.
\end{eqnarray}
where the exponent $\zeta$ is given by:
\begin{equation}
\zeta \equiv \frac{1}{2} \sqrt{\frac{g}{\pi v_F^* k_F}}. 
\end{equation}
The nonanalytic dependence of $G_F$ on $x_\parallel$ reflects the
destruction of the quasiparticle pole by the emission and absorption of
plasmons. The power law dependence with exponent $\zeta$ is a consequence of
the nonperturbative treatment of the interaction within the bosonization
approach. The perturbative RPA approach in the fermion basis gave instead  a
logarithmic dependence on $x_\parallel$. Bares and Wen\cite{Bares} 
then used a variational
wavefunction to argue that the logarithm should be exponentiated into 
a power law with the same exponent as obtained above.  
Multidimensional bosonization obviously 
allows us to obtain the power law directly. 
As we have discussed, the appearance of an anomalous power law 
in the Green's function results in the elimination of the discontinuity in the 
quasiparticle occupancy at the Fermi surface, replacing it by a continuous but
nonanalytic crossover.  The quasiparticle occupancy decreases from one to zero
over a momentum scale of the order of $\lambda$. 
The derivation of the current algebra only requires that the occupancy
change over a momentum scale of the same order, providing a consistent picture.
It should be noted, on the other hand, that consistency does not necessarily
mean that the result is correct.  For example, it is possible that the super
long-ranged interaction actually induces Wigner crystallization\cite{crystal}.  
Either way, the destruction of the Fermi liquid is complete.

It is instructive to repeat the calculation of the equal-time Green's function
for the realistic case of a fermion liquid interacting via Coulomb forces in
both two and three dimensions\cite{HKMS}. 
In two dimensions the plasmon frequency is
proportional to $\sqrt{|\bq|}$. In this case the integral over $\omega$ is no
longer constant, as it was for the Bares and Wen potential, but rather it is
proportional to $\sqrt{|\bq|}$; as a result the momentum integral vanishes as
$\sqrt{\lambda/k_F}$. In three dimensions the plasmon frequency is non-zero but
the integral over $\bq$ is again suppressed, in this instance because of the
reduced phase space at small momentum, and as a result $\delta G_\phi$ 
vanishes as
$\lambda/k_F$. In each case the quasiparticle propagator has the standard
Fermi liquid form. The weight at the pole undergoes no additional
renormalization beyond the factor of $Z_k$ due to integrating out the
high-energy degrees of freedom, equation \ref{Zk}.

In this section we examined the effects of both short and long range
longitudinal interactions on fermion liquids. We found that if the
superconducting instability is suppressed then the Fermi liquid state is the
ground state of a degenerate gas of fermions interacting via short-range forces,
or by Coulomb forces, in either two or three dimensions. Bosonization provides a
means to go beyond an assumed Fermi liquid form for the quasiparticle
propagator. Indeed had we considered fermions with spin we would have found
that the bosonized Hamiltonian separates into two parts, charge and spin,
leading to the possibility that, as in one dimension, the quasiparticle
propagator would reflect spin charge separation. Spin charge separation would
destroy the Fermi liquid, as the key element, the existence of a pole in the
single particle Green's function with non-zero spectral weight, would be
replaced by a branch cut. In dimension greater than one 
this does not occur because 
in the $\Lambda \rightarrow 0$ limit the location of the
pole in the boson propagator is unchanged from its free value. Consequently
the spin and charge velocities are equal and both degrees of freedom propagate
together in the usual quasiparticle form. On the other hand, the Fermi liquid
form is destroyed both in one dimension, where the ground state is a Luttinger
liquid, and in the case of the superlong range interaction in two dimensions
studied by Bares and Wen. The non-Fermi liquid fixed points of these two
examples are accessible via bosonization.

\section{Two Particle Properties: Density Response}
\label{two_particle}
\setcounter{equation}{0}

In the previous section we discussed the single particle properties of an
interacting fermion system in some detail.  Here we turn to the consideration
of two-particle properties, specifically the density response.  We are 
concerned with two limits in particular, low-wavevectors, $|\bq| \approx 0$
and high-wavevectors, $|\bq| \approx 2 k_F$.  

In the low-wavevector limit we show that the classic RPA result for the
density response is exact within multidimensional bosonization.  In fact
in early work in the fermion basis this result was anticipated by Geldart and
Taylor\cite{Geldart}.  These authors showed that, in the calculation of the
density response, self-energy corrections to the fermion propagator were
canceled by vertex corrections at leading order in an expansion in the 
Coulomb interaction.  This question has been addressed most recently by 
Kopietz, Hermisson, and Sch\"onhammer\cite{Kopietz} who were able to show that,
provided the quasiparticle dispersion is linear, this result holds to all
orders in a loop expansion.  It is not immediately obvious {\it a priori} that
a bosonization scheme with the restriction $|\bq| < \lambda \ll \Lambda$ can
describe processes with large momentum transfer.  In fact, as in one dimension,
multidimensional bosonization can describe processes with either small or
large momentum transfer provided they are of sufficiently low energy, 
$|\omega| \ll \ve \lambda$.  To illustrate this point we determine the
density response of a Fermi gas at $|\bq| \approx 2 k_F$ which, due
to the sharp reduction in the phase space for low energy scattering, is
nonanalytic, as was first pointed out by Kohn\cite{Kohna}.

To begin we write the fermion density-density correlation function in the
bosonic representation.  The fermion density fluctuation is given by
\begin{equation}
\delta \rho({\bf x}, \tau)  =~ :\psi^\dag({\bf x}, \tau) \psi ({\bf x}, \tau):~
= \sum_{\bs \bt} 
:\psi^\dag({\bf S}; {\bf x}, \tau) \psi({\bf T}; {\bf x}, \tau):
\end{equation}
where the colons denote normal ordering which in this case is equivalent to
subtracting the uniform background density. Then the density-density
correlation function can be written as
\begin{eqnarray}
\langle \delta \rho({\bf x}, {\bf \tau}) \delta \rho({\bf 0},0) \rangle
&=& \frac{1}{V^2} \sum_{\bs \bt} \langle J(\bs; {\bf x}, \tau) 
J(\bt; {\bf 0}, 0) \rangle
\nonumber \\
&+& \left( \frac{\Omega}{Va}\right)^2 \sum_{\bs \neq \bt, \bu \neq \bv}
e^{i({\bf k}_\bs - \bk_\bt) \cdot {\bf x}} \bigg{\langle} 
\exp \left\{ i\frac{\sqrt{4\pi}}{\Omega} [\phi(\bs; \bx, \tau) 
- \phi(\bt; \bx, \tau)]\right\} 
\nonumber \\
&\times& 
\exp \left\{ i\frac{\sqrt{4\pi}}{\Omega} [\phi(\bu; {\bf 0},0) -\phi
(\bv; {\bf 0}, 0)]\right\} \bigg{\rangle}
\label{rhorho}
\end{eqnarray}
For small momentum transfer $|\bq| < \lambda \ll k_F$ only the first term is
relevant, whereas the second term is important at large momentum transfer.
We separate the analysis of these two regimes in the following subsections.

\subsection{Response at Small Wavevectors} 
Consider the density response of a free Fermi gas at long wavelength and
low frequency. For free fermions, the currents in different patches are
uncorrelated and the density response is given by
\begin{equation}
\Pi^0(\bq, \omega) = - \sum_\bs 
\langle J({\bf S}; {\bf q}, \omega) J ({\bf S}; -{\bf q}, -\omega) \rangle 
\end{equation}
where  $J(\bs; {\bf q}, \omega)$ is given by equation \ref{current-rep} 
and therefore
\begin{eqnarray}
\sum_\bs \langle J({\bf S; q}, \omega)~ J ({\bf S}; -{\bf q}, -\omega) \rangle 
&=&\Omega \sum_\bs (\bn_\bs \cdot \bq)~  
\langle a({\bf S; q}) a^*({\bf S;  q}) \rangle
\nonumber \\
&=& -\sum_\bs \frac{\hbn_\bs \cdot \bq}{i\omega-v_F^* \hbn_\bs \cdot \bq}
\end{eqnarray}
If we consider the specific case of two dimensions and recall that
$\frac{\Lambda}{v_F^* (2\pi)^2} \sum_\bs = N^*(0) \int \frac{d\theta}{2\pi}$ 
then 
\begin{equation}
\Pi^0(\bq, \omega) 
= N^*(0) \int \frac{d\theta}{2\pi}\frac{\cos\theta}{\cos\theta-x} 
\end{equation}
where again $x \equiv \omega/(v_F^* |\bq|)$ and therefore
\begin{equation}
\Pi^0({\bf q}, \omega) = N^*(0) \left( \frac{|x|}{\sqrt{x^2-1}} - 1 \right) 
= -\chi_0 ({\bf q}, \omega) 
\end{equation}
which is the well known result for the density response of the free electron
gas in the limit of small momentum.

Generalizing to include a long range Coulomb interaction, currents in different
patches are now correlated and the correlation function written out in full
reads
\begin{eqnarray}
\langle J({\bf S}; q) J({\bf T}; -q) \rangle
&=& \Omega \sqrt{(\hbn_\bs \cdot \bq) (\hbn_\bt \cdot \bq)} 
\times \bigg{[} \langle a({\bf S}; q) a({\bf T}; -q) \rangle 
\theta(\hbn_\bs \cdot \bq) \theta(-\hbn_\bt \cdot \bq) 
\nonumber \\
&+& \langle a(\bs; q) a^\dag(\bt; q) \rangle \theta(\hbn_\bs \cdot \bq)
\theta(\hbn_\bt \cdot \bq)
\nonumber \\
&+& \langle a^\dag(\bs; -q) a(\bt; -q) \rangle \theta(-\hbn_\bs \cdot \bq)
\theta(-\hbn_\bt \cdot \bq)
\nonumber \\
&+& \langle a^\dag(\bs; -q) a^\dag(\bt; q) \rangle 
\theta(-\hbn_\bs \cdot \bq) \theta(\hbn_\bt \cdot \bq) \bigg{]}
\label{mess}
\end{eqnarray}
and contains contributions from all possible two point boson Green's functions.
All of these functions can be obtained from the generating functional, which in
this instance is given by 
\begin{eqnarray}
Z[\xi, \xi^*] &=& \int DA
\exp\left[-\frac{1}{2} \sum_\bq \int \frac{d\omega}{2 \pi}
\bigg{\{} \frac{1}{V(\bq)} - 2 \sum_{\hbn_\bs \cdot \bq > 0} 
\frac{\hbn_\bs \cdot \bq}{i \omega - v_F^* \hbn_\bs \cdot \bq} \bigg{\}}
A(-q) A(q)\right] 
\nonumber \\
&\times& \exp \sum_\bq \int \frac{d \omega}{2 \pi} 
\sum_{\hbn_\bs \cdot \bq > 0} \bigg{[}
\frac{i\Omega \sqrt{\hbn_\bs \cdot \bq}}{i\omega - v_F^* \hbn_\bs \cdot \bq} 
[A(-q) \xi(\bs; q) + A(q) \xi^*(\bs; q)]
\nonumber \\
&-& \xi^*(\bs; q) \frac{1}{i\omega - v_F^* \hbn_\bs \cdot \bq} \xi(\bs; q) 
\bigg{]}\ .
\end{eqnarray}
here $A(q)$ is the longitudinal gauge field introduced to decouple the Coulomb
interaction. The correlation functions in equation \ref{mess} 
are found in the usual way, for example
\begin{eqnarray}
\langle a(\bs; q) a^\dag(\bt; q) \rangle &=& -\frac{\delta^2 \ln Z}
{\delta \xi^*(\bs; q) \delta \xi(\bs; q)} \left|_{\xi = \xi^* = 0}\right.
\nonumber \\
&=& \frac{\Omega \sqrt{(\hbn_\bs \cdot \bq)(\hbn_\bt \cdot \bq)}}
{(i\omega - v_F^* \hbn_\bs \cdot \bq)(i \omega - v_F^* \hbn_\bt \cdot \bq)} 
\langle A(q) A(-q) \rangle\ .
\end{eqnarray}
Likewise, the anomalous propagator is given by
\begin{equation}
\langle a(\bs; q) a(\bt; -q) \rangle = -\frac{\Omega \sqrt{(\hbn_\bs \cdot \bq) 
(\hbn_\bt \cdot \bq)}}{(i\omega-v_F^* \hbn_\bs \cdot \bq) 
(i\omega-v_F^* \hbn_\bt \cdot \bq)} \langle A(q) A(-q) \rangle.
\end{equation}
Combining all of the terms in equation \ref{mess} we obtain
\begin{eqnarray}
\langle J(\bs; \bq, \omega)~ J(\bt; -\bq, -\omega) \rangle 
&=& \frac{\Omega \hbn_\bs \cdot \bq}
{i\omega - \ve \hbn_\bs \cdot \bq}~ \delta^{D-1}_{\bs,\bt}
\nonumber \\
&+& \Omega \frac{\Lambda}{(2\pi)^2} \frac{(\hbn_\bs \cdot \bq)(\hbn_\bt \cdot
\bq)}{(i\omega -v_F^* \hbn_\bs \cdot \bq)(i\omega -v_F^* n_\bt \cdot \bq)}
\nonumber \\
&\times& \frac{V(\bq)}{1-V(\bq) \Pi^0(\bq)}\ . 
\label{JJ}
\end{eqnarray}
Then summing over $\bs$ and $\bt$ we find 
\begin{equation}
\Pi(\bq, \omega) = \frac{\Pi^0(\bq, \omega)}{1-V(\bq) \Pi^0(\bq,\omega)},
\end{equation}
which is 
identical to the random phase approximation result for the density response,
and is exact in the limit $\bq, \omega \rightarrow 0$.

\subsection{Response at $|\bq| \approx 2k_F$}
\begin{figure}
\center
\epsfxsize = 7.0cm \epsfbox{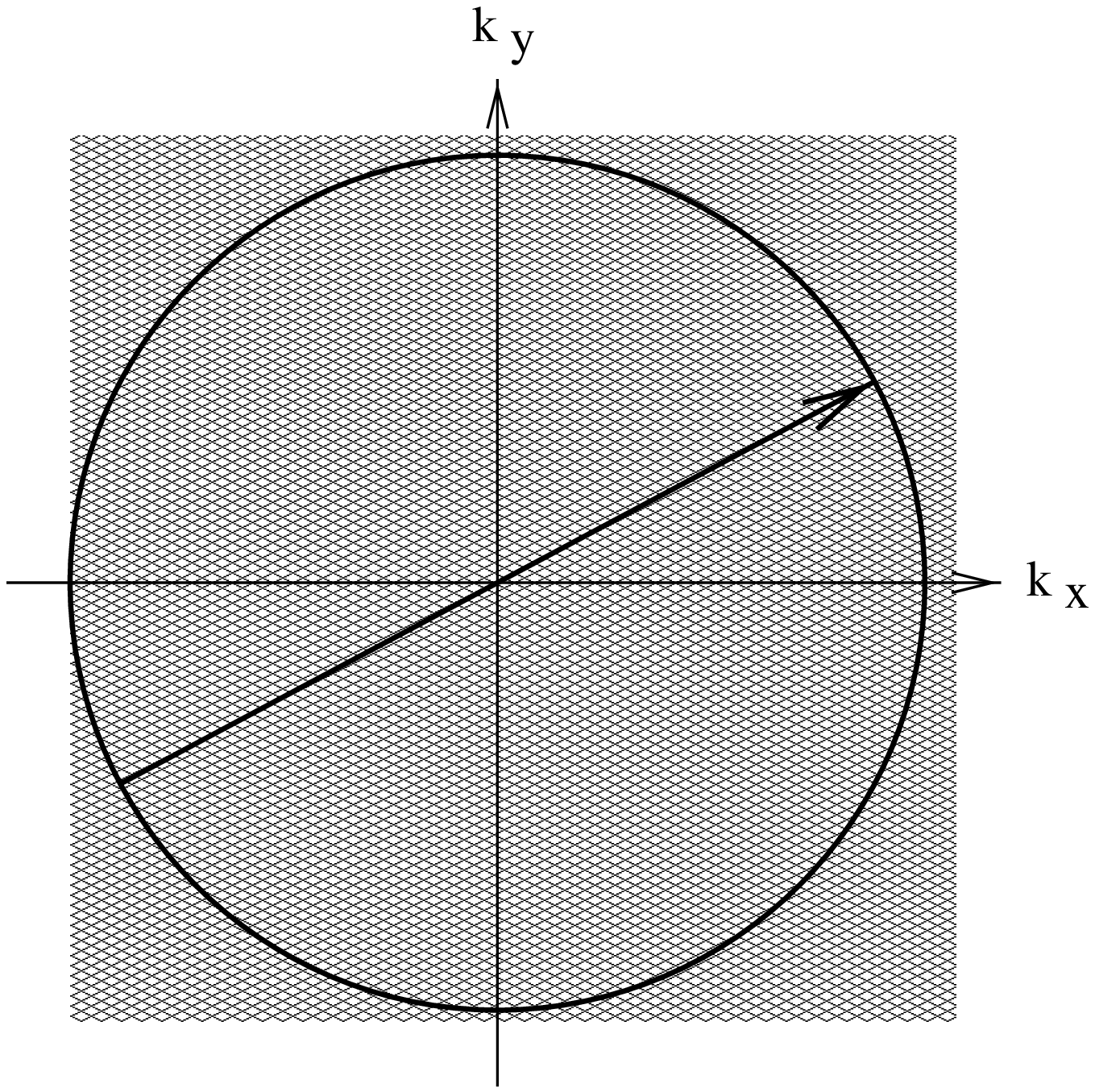}
\caption{
\label{2kf}
Wavevector of magnitude $2 k_F$ just spans the Fermi surface.}
\end{figure}

We have seen that bosonization successfully captures the physics of the long
wavelength response of an interacting fermion system.  As a further test of
the formalism we consider the density response at short wavelengths, in
particular around $|\bq| = 2k_F$. 
At wavevectors greater than the spanning vector of the
Fermi surface it is no longer possible to excite particle hole pairs at zero
frequency; see figure \ref{2kf}.  Consequently the density response is 
nonanalytic at $|\bq| = 2k_F$, as was first pointed out by 
Kohn\cite{Kohna}.

At first sight it is not obvious how large momentum transfer processes can be
described within a bosonization framework with the restriction 
$|\bq| < \lambda < k_F$. 
The reason why this is possible is made clear on noticing that scattering
between patches is allowed, and that in fact the second term in 
equation \ref{rhorho} 
contains these processes explicitly; recall that anomalies in $2k_F$ 
scattering are described correctly by standard one dimensional 
bosonization\cite{Emery}. Bosonization in one and
higher dimensions can describe processes with either small or large momentum
provided they are of low enough energy to lie inside the shell depicted
in figure \ref{shell}. 

Since for free fermions the bosons in different patches are independent, the
correlation function in the second term of equation \ref{rhorho}
is zero except when
${\bf V} = \bs$ and ${\bf U} = \bt$ and the contribution to the density response
function is given by
\begin{eqnarray}
\Pi^0 (\bx, \tau) &=& -\left(\frac{\Omega}{Va}\right)^2 \sum_{\bs \neq \bt}
e^{i (\bk_\bs - \bk_\bt ) \cdot \bx } \bigg{\langle} 
e^{i\frac{\sqrt{4\pi}}{\Omega} [\phi(\bs; \bx, \tau) -\phi(\bt; \bx, \tau )]}~
e^{i \frac{\sqrt{4\pi}}{\Omega} [\phi(\bt; {\bf 0}, 0) 
- \phi(\bs, {\bf 0}, 0)]} \bigg{\rangle} 
\nonumber \\
&=& \left[ \frac{\Lambda^{D-1}}{(2\pi)^D} \sum_\bs \frac{e^{i \bk_\bs
\cdot \bx}}{\hbn_\bs \cdot \bx +iv_F^* \tau}\right] \left[
\frac{\Lambda^{D-1}}{(2\pi)^D} \sum_\bt \frac{e^{-i \bk_\bt \cdot
\bx}}{(\hbn_\bt \cdot \bx +iv_F^*\tau )}\right]\ .
\label{2Pi}
\end{eqnarray}
Implicit in the last line is the restriction $|{\bf x}_\bot| \Lambda < 1$. 
As we are concerned with the response at $|\bq| \approx 2k_F$, 
only patches which are nearly opposite
to each other on either side of the Fermi surface participate 
in the scattering, $\hbn_\bs \approx -\hbn_\bt$. 
Therefore we may carry out the sum over patches in
each of the two terms in equation \ref{2Pi} keeping in mind that the spatial
directions perpendicular to the Fermi surface normal are the same for both
terms. To simplify the calculation we consider the static response, $\omega
=0$, only and specialize to the case of a circular Fermi surface in two
dimensions. Integrating the density response over time $\tau$ we obtain
\begin{equation}
\Pi^0(\bx, \omega = 0) = \pi v_F^* [N^*(0)]^2 \int \frac{d\theta_1}{2\pi} 
\int \frac{d\theta_2}{2\pi} \frac{e^{ik_F r(\cos\theta_1
-\cos\theta_2)}}{r(\cos\theta_1-\cos\theta_2)}~ 
[\sgn(\cos \theta_1) - \sgn(\cos \theta_2)] 
\label{180}
\end{equation}
where $\theta_{1, 2}$ are the angles between $\bx$ and $\bk_\bs$, and $\bx$
and $\bk_\bt$, respectively, and $r \equiv |\bx|$. 
Now we can focus on the case of $|\bq| \approx 2k_F$ by
limiting the range of integration over the angular variables $\theta_{1,2}$ to
angles near $\theta_1 =0 $ and $\theta_2 =\pi$. In this limited range the
cosines appearing in equation \ref{180} may be expanded in a Taylor series as
$\cos\theta_{1,2}=\pm (1-\theta_{1,2}^2/2)$ and the Gaussian integral over
angles can be carried out. Finally take the Fourier transform to $\bq$-space.
Then if $\theta$ is the angle between $\bq$ and $\bx$, the density response
function is given by
\begin{equation}
\Pi^0(\bq, \omega = 0) = i~ 
\frac{[N^*(0)]^2}{2m^*} \int_0^\infty rdr \int_0^{2\pi}
d\theta \frac{\exp i (2k_F - |\bq| \cos\theta) r}{r^2}\ .
\end{equation}
Now
\begin{equation}
\frac{1}{\pi} \int_0^\pi d\theta e^{-i |\bq| r \cos \theta} = J_0 (|\bq| r)
\approx \sqrt{\frac{2}{\pi |\bq| r}} \cos (|\bq| r -\frac{\pi}{4})\ , 
\quad |\bq| r \gg 1 \\
\end{equation}
and therefore
\begin{equation}
\Pi_{2k_F}(\bq, \omega=0) = i~ \frac{[N^*(0)]^2}{2m^*}
\sqrt{\frac{\pi}{k_F}} e^{\frac{i\pi}{4}} \int_a^\infty dr
\frac{\exp[(i(2k_F-|\bq|)r]}{r^{3/2}} 
\end{equation}
Here we have assumed $|\bq| \approx 2k_F$ and 
we have also introduced the ultraviolet cutoff
$a$ to eliminate the artificial divergence at small $r$. Finally,
asymptotic analysis of the radial integral gives the nonanalytic contribution
to the density response function
\begin{equation}
\int_a^\infty \frac{e^{i(2k_F - |\bq|)r}}{r^{3/2}}~ dr \approx 
-2 \sqrt{\pi i} \sqrt{|\bq| - 2k_F}
\end{equation}
and therefore
\begin{equation}
\Pi_{2k_F}(\bq, \omega = 0) = \left\{\begin{array}{ll}
\frac{\sqrt{2}m^*}{4\pi} \sqrt{\frac{|\bq|-2k_F}{2k_F}}+C, 
& |\bq| > 2k_F \\
C, & |\bq| < 2k_F\ .
\end{array}\right. 
\label{Kohn-anomaly}
\end{equation}
Here $C$ is a constant which can be determined by a more careful evaluation of
the integrals in the small $r$ region. Continuity across $|\bq| = 2k_F$ 
ensures that
the same constant appears in both limits of equation \ref{Kohn-anomaly}. 
The Kohn anomaly of equation \ref{Kohn-anomaly} 
agrees precisely with that found directly in the fermion basis\cite{FStern}.
Repeating the calculation in $D = 3$ we show easily that the static response is
given by
\begin{equation}
\Pi_{2k_F}(\bq, \omega = 0) = \frac{k_F m^*}{8 \pi^2} 
\frac{2k_F-|\bq|}{|\bq|}~ \ln \left| \frac{|\bq|-2k_F}{2k_F} \right| + \ldots 
\label{Kohn-anomaly3d}
\end{equation}
Which again agrees with the exact result\cite{Kohna} 
found in the fermion basis in the
limit $|\bq| \rightarrow 2k_F$. The Kohn anomaly persists even if longitudinal
interactions are included. This has already been anticipated in 
equations \ref{Kohn-anomaly} and \ref{Kohn-anomaly3d} 
where the effective mass appears rather than the bare mass.

\section{The Half-Filled Landau Level}
\label{hfll}
\setcounter{equation}{0}
The two-dimensional electron gas in a magnetic field with even denominator
Landau level filling fraction has been the subject of extensive study.  
For an overview see \cite{Halperin}.  Experiments at
even denominator filling point to a compressible metallic 
state\cite{Jiang,Du,Leadley,Willett,Goldman} rather than the
incompressible state found at odd denominator fractional filling.  
Furthermore, the remarkable similarity between phenomena observed at high
magnetic fields near filling fraction $\nu = 1/2$, and those found for 
electrons in weak magnetic fields can be explained within the 
composite fermion framework\cite{Jain}.  A composite fermion is formed
by attaching an even number of ``statistical'' flux quanta to an electron.  
Specifically, in the case of the half-filled Landau level, 
two such flux quanta are attached.  Then if the statistical magnetic field
due to each attached flux line is spatially averaged, the external magnetic
field is effectively canceled and the composite fermions travel in 
zero net field.  The mean-field ground state is, therefore,
a filled Fermi sea with a Fermi momentum $k_F$ larger 
than that of the free, unpolarized, 
Fermi electron gas by a factor\cite{HLR} of $\sqrt{2}$ consistent with
experimental observation\cite{Goldman}.  

This picture has been formalized as a Chern-Simons gauge theory by Lopez
and Fradkin\cite{Lopez}, by Kalmeyer and Zhang\cite{Kalmeyer}, and by
Halperin, Lee, and Read\cite{HLR} (HLR).  In this approach the 
electrons are subjected to a singular gauge transformation converting them 
into composite fermions which interact with each other
via a statistical transverse gauge field,
the Chern-Simons field, in addition to the usual Coulomb or short-range
interaction, and the mean-field theory is constructed.  However the effect of
dynamical
fluctuations of the gauge field (equivalent to fluctuations in the electron
density) must be addressed.  In this section and in the following section
we employ multidimensional bosonization in an attempt to answer this question.

The Hamiltonian for a two dimensional electron gas of density $n_f$ in a
perpendicular magnetic field $B$ is given by
\begin{equation}
H[c^\dag_e, c_e] 
= \frac{1}{2m} \int d^2x~ c_e^\dag (\bx) \left( \frac{\bf \nabla}{i} 
+ e{\bf A}(\bx) \right)^2 c_e(\bx) + H_{\rm Coulomb}[c_e^\dag, c_e]\ . 
\end{equation}
We have dropped the spin index as we assume that the spin of 
all of the electrons has been aligned parallel to the external magnetic field
by the Zeeman interaction.
The vector potential ${\bf A}(\bx)$ is due to the uniform external magnetic
field ${\bf B}$ which points in the $\hat{z}$ direction normal to the plane and
$m$ is the band mass of the electrons.  We use units where $c = 1$ and the
electron charge is $-e$. At a value $B=4\pi n_f/e$, half of the states in the
lowest Landau level are filled. The composite-fermion theory of the half
filled Landau level makes use of a local gauge transformation to describe
the system as a collection of quasiparticles that obey Fermi statistics:
\begin{equation}
c^\dag (\bx) = c_e^\dag (\bx) \exp \left[ 
-i\widetilde{\phi} \int d^2 y \arg (\bx
-\by) c_e^\dag ({\bf y}) c_e ({\bf y}) \right]\ . 
\end{equation}
When $\widetilde{\phi}$ is an even integer the quasiparticles are fermions.
Each quasiparticle is a composite object comprising the physical electron
together with a flux tube of integer $\widetilde{\phi}$ quanta. The transformed
Hamiltonian is given by
\begin{equation}
H[c^\dag, c] = 
\frac{1}{2m}~ \int d^2x~ c^\dag(\bx) [\frac{\bf \nabla}{i} + e{\bf A}(\bx)
-{\bf a}(\bx)]^2 c(\bx) 
\end{equation}
where
\begin{equation}
\ba(\bx) = \widetilde{\phi} \int d^2y~ 
\frac{{\bf \hat{z}} \times (\bx - \by)}{|\bx - \by|^2}~ c^\dag(\by) c(\by)
\end{equation}
and
\begin{equation}
{\bf \nabla} \times \ba (\bx) = 2\pi~ \widetilde{\phi}~ n_f(\bx)
\label{flux-constraint}
\end{equation}
The average field strength of the flux tubes can now be adjusted to cancel out
the external magnetic field.  At half filling, setting $\widetilde{\phi}=2$
\begin{equation}
{\bf \nabla} \times [e \bA(\bx) - \ba(\bx)] = e \left( B - \frac{2\pi n_f
\widetilde{\phi}}{e} \right) = 0.
\end{equation}
Consequently at the mean field level, the quasiparticles behave as if they are
free fermions in zero net field; the infinite Landau degeneracy is lifted and
presumably the state is stable against two body Coulomb interactions. The
important question to ask now is whether gauge fluctuations or, equivalently
quasiparticle density fluctuations (equation \ref{flux-constraint}) 
modify or destroy the mean field state. In the following we assume that 
despite strong gauge fluctuations, multidimensional bosonization is applicable.
Later we verify that this assumption is correct for the case of Coulomb or
longer-range interactions.

The bare action for composite fermions interacting via a Coulomb interaction
and subject to the constraint, equation \ref{flux-constraint}, can be written
\begin{eqnarray}
S[c^*, c, A_\mu, {\bf a}] &=& \int d^2x~ d\tau~
\bigg{\{} c^*(x) \left(\frac{\partial}{\partial \tau} - i A_0(x) \right) c(x)  
\nonumber \\
&+& \frac{1}{2 m} c^*(x) \left[\frac{{\bf \nabla}}{i} + e \bA(x) - \ba(x)
\right]^2 c(x) +\frac{iA_0(x)}{2\pi \hat{\phi}} [{\bf \nabla}\times \ba(x)]
\bigg{\}}
\nonumber \\
&+& \frac{1}{2} \frac{1}{(2\pi \hat{\phi})^2} 
\int d^2x~ d^2y~ d\tau~ [{\bf \nabla}_\bx \times \ba(\bx, \tau)] 
V(\bx -\by) [\nabla_{\by} \times \ba(\by, \tau)] 
\label{HFLL-action}
\end{eqnarray}
As usual we work in the Coulomb gauge ${\bf \nabla \cdot a} = 0$. 
The next to the last
term in equation \ref{HFLL-action} is the Chern-Simons term introduced by Lopez
and Fradkin\cite{Lopez}, by Kalmeyer and Zhang\cite{Kalmeyer}, and by
Halperin, Lee, and Read\cite{HLR} (HLR)
so that the integration over $A_0$ enforces the constraint, 
equation \ref{flux-constraint}. The
last term in the action is the longitudinal Coulomb interaction which has been
written in terms of the gauge field by making use of the constraint. It is
important to notice that in the Chern-Simons gauge theory the longitudinal
interaction plays an important role in suppressing fermion density
fluctuations. Since density fluctuations are equivalent to gauge fluctuations
in a Chern-Simons theory, repulsive long-range Coulomb interactions moderate
and control the gauge fluctuations.  As shown below, gauge fluctuations are
stronger for shorter-range interactions, and in fact completely obliterate the 
Fermi surface. In this case multidimensional bosonization is not useful.  

As before we integrate out the high-energy fermion degrees of
freedom using the renormalization group.  We also make the replacement
${\bf a} - e {\bf A} \rightarrow {\bf A}$, reflecting the fact that now we are
studying fluctuations about zero net mean magnetic field. 
The resultant effective action reads:
\begin{eqnarray}
S_{eff}[\psi^*, \psi, A_\mu] &=& \int d^2x~ d\tau~ \left\{ \psi^*(x)
\left(\frac{\partial}{\partial\tau} - i A_0(x) \right) \psi(x)\right. 
\nonumber \\
&+& \left. \frac{1}{2m^*} \psi^*(x) \left(\frac{\bf \nabla}{i} - \bA(x)
\right)^2 \psi(x) +i \frac{A_0(x)}{2\pi\hat{\phi}}\left({\bf \nabla}
\times \bA(x)\right) \right\} 
\nonumber \\
&+& \frac{1}{2} \frac{1}{(2\pi\widetilde{\phi})^2} \int d^2x~ d^2y~ d\tau~
[{\bf \nabla}_\bx \times \bA(\bx, \tau)]~ V(\bx - \by)~ 
[{\bf \nabla}_\by \times \bA(\by, \tau)]\ .
\label{HLR-effective}
\end{eqnarray}
At filling fraction $\nu = 1/2$ we set $\widetilde{\phi} = 2$ to cancel out the
external field.  As stated above, 
$\bA$ should be interpreted as the fluctuation 
about the mean field value. The effective action can now be expressed 
in terms of Abelian charge currents 
\begin{equation}
J(\bs; \bq, \omega) = \sqrt{\Omega | \hbn_\bs \cdot \bq |} 
[a(\bs; \bq, \omega) \theta(\hbn_\bs \cdot \bq)
+ a^*(\bs; -\bq, \omega) \theta(-\hbn_\bs \cdot \bq)]
\end{equation}
as
\begin{equation}
S_{eff}[A_\mu, a] = S_G^0[A_\mu] + S[A_\mu, a]
\end{equation}
where
\begin{eqnarray}
S[A_\mu, a] &=& \sum_\bq \sum_{\hbn_\bs \cdot \bq > 0} 
\int \frac{d\omega}{2\pi} 
(i\omega -v_F^* \hbn_\bs \cdot \bq) a^*(\bs; q) a(\bs; q)
\nonumber \\
&+& \frac{1}{V} \sum_\bq \int \frac{d\omega}{2\pi} \left\{ -\sum_\bs J(\bs; q)
\left[ i A_0(-q) + \frac{k_\bs}{m^*} \cdot \bA (-q)\right] \right.
\nonumber \\
&+& \left.\frac{n_f}{2m^*} \bA (q) \cdot \bA(-q)
\right\}~ ,
\end{eqnarray}
and the bare gauge action can be read off from equation \ref{HLR-effective},
\begin{eqnarray}
S_{G}^0[A_\mu] &=& \frac{1}{V} \sum_\bq \int
\frac{d\omega}{2\pi} \left\{ -\frac{1}{4\pi \widetilde{\phi}} \left[A_0(-q) 
[\bq \times \bA(q)] -A_0(q) [\bq \times \bA(-q)] \right] \right.
\nonumber \\
&+& \left. \frac{1}{2} \frac{\bq^2}{(2\pi\wp)^2} V(\bq) \bA(q) \cdot \bA(-q)
\right\}~ .
\end{eqnarray}
Fermi liquid interactions should also be included in the action\cite{Simon}: 
If all Landau coefficients except $f_1$ are set equal to zero 
they are described by
\begin{equation}
S_{FL} = -\frac{f_1}{2Vk_F^2} \sum_{\bs, \bt; \bq}
\int \frac{d\omega}{2\pi}~ J(\bs; q)~ (\bk_\bs \cdot \bk_\bt)~ J(\bt; -q) 
\label{FLact}
\end{equation}
which is made gauge covariant by the replacement 
\begin{equation}
\bk \rightarrow \frac{{\bf \nabla}}{i} - \bA\ .
\end{equation} 
As stated, for computational convenience we work
in the Coulomb gauge which eliminates the longitudinal component of the vector
gauge field, only the transverse component 
$A_T(q) \equiv \hat{\bq} \times \bA(q)$ remains.

Now we can determine the composite quasiparticle two
point Green's function.  It is important to note that the propagator for
physical electrons is not the same as the propagator for
the composite particles, as the 
physical electrons do not have attached flux\cite{HLR}.  The
first step as usual is to determine the boson
propagator; for the moment we set the short-ranged part of the 
interaction, the Landau parameters generated upon integrating out the
high-energy fields, to
zero. We construct the generating functional and integrate out the bose fields
\begin{eqnarray}
Z[\xi, \xi^*] &=& \int DA_\mu \exp -S_G[A_\mu]  
\nonumber \\
&\times& \exp \int \frac{d\omega}{2\pi} \sum_\bq \sum_{\hbn_\bs \cdot \bq > 0}
\bigg{\{}
\frac{\sqrt{\Omega\hbn_\bs \cdot \bq}}{V[i\omega-v_F^*\hbn_\bs \cdot \bq]}
\big{[} \xi(\bs; q) [iA_0(-q) - v_F^* \sin(\theta) A_T(-q)]  
\nonumber \\
&& + \xi^*(\bs; q) [iA_0(q) + v_F^* \sin(\theta) A_T(q) \big{]} \bigg{\}} 
\nonumber \\
&\times& \exp -\int \frac{d\omega}{2\pi} \sum_{\bq, \hbn_\bs \cdot \bq > 0} 
\xi^*(\bs; q) \frac{1}{i\omega - v_F^* \hbn_\bs \cdot \bq} \xi(\bs; q) 
\label{hfllgenfunc}
\end{eqnarray}
where the gauge action
\begin{eqnarray}
S_G[A_\mu] &=& S_{G}^0[A_\mu] + \int \frac{d\omega}{2\pi} 
\sum_\bq \sum_{\hbn_\bs \cdot \bq > 0}
\left[ [iA_0(-q) - v_F^* \sin(\theta) A_T(-q) ] \frac{\Omega \hbn_\bs \cdot \bq}
{i\omega-v_F^* \hbn_\bs \cdot \bq} \right.
\nonumber \\
&\times& \left. [iA_0(q) + v_F^* \sin(\theta) A_T (q)] 
- \frac{n_f}{2 m^*} A_T (-q) A_T (q) \right]\ .
\label{hfllact}
\end{eqnarray}
In writing equations \ref{hfllgenfunc} and \ref{hfllact} 
we have used the fact that 
\begin{eqnarray}
\frac{\bk_\bs \cdot \bA(q)}{m^*} &=& 
\frac{k_F }{m^*} \hat{\bq} \times \bA(q)~ \sin \theta(\bk_\bs, \bq) 
\nonumber \\
&=& v_F^*~ \sin(\theta)~ A_T(q) 
\end{eqnarray}
where as before $\theta(\bk_\bs, \bq)$ is the angle between $\bk_\bs$ 
and $\bq$. Now in equation \ref{hfllact} restrict the $\bq$ integral and 
let the $\bs$ sum run free; then we can write the gauge action as
\begin{equation}
S_{G}[A_\mu] = \frac{1}{2} \sum_{\bs} \sum_{\hbn_\bs \cdot \bq > 0} 
\int \frac{d\omega}{2 \pi}~ A_\mu (-q) K^{\mu \nu}(q) A_\nu(q)
\end{equation}
where the matrix ${\bf K}$ is given by
\begin{equation}
{\bf K}(q) = \left( \begin{array}{lr} \chi_0(q), &  \frac{-|\bq|}{2\pi \wp} \\
\frac{-|\bq|}{2\pi \wp}, & \chi_{T}(q) - \frac{\bq^2 V(\bq)}{(2 \pi \wp)^2}
\end{array} \right)
\end{equation}
and $\mu$ and $\nu$ are summed over $(0, T)$. Here $\chi_0$ is the longitudinal
susceptibility given by equation \ref{chi_0} and $\chi_T$, the transverse
susceptibility, is given by
\begin{eqnarray}
\chi_{T}(\bq, \omega) &=& \ves N^*(0) \int
\frac{d\theta}{2\pi} \frac{\sin^2\theta\cos\theta}{\cos\theta -x}
-\frac{n_F}{m^*}
\nonumber \\
&=& -\ves N^*(0) [x^2 - |x| \sqrt{x^2-1} \theta (x^2-1)
- i |x| \sqrt{1-x^2} \theta(1-x^2)]\ ; 
\label{chi_T}
\end{eqnarray}
as before, $x \equiv \omega / (\ve |\bq|)$.
Integrating out the gauge fields and differentiating the logarithm of the
generating functional with respect to the source fields $\xi$ and $\xi^*$ we
find the boson propagator
\begin{equation}
G_B(\bs, \bt; q) = \frac{\delta_{\bs,\bt}}{i\omega_n -v_F^* \hbn_\bs \cdot \bq} 
+ \sum_{\bs, \bt} 
\frac{\sqrt{\Omega \hbn_\bs \cdot \bq}}{i\omega -v_F^* \hbn_\bs \cdot \bq}
b_\nu^*(\bs; \bq) D^{\nu\mu}(q) b_\mu(\bt; \bq)
\frac{\sqrt{\Omega \hbn_\bt \cdot \bq}}{i\omega -v_F^* \hbn_\bt \cdot \bq}
\label{ST-prop}
\end{equation}
where the vector $b_\mu(\bs; \bq)$ has components
\begin{eqnarray} 
b_0(\bs; \bq) &=& 1
\nonumber \\
b_T(\bs; \bq) &=& iv_F^* \sin \theta(\bk_\bs, \bq) 
\end{eqnarray}
and the gauge propagator is given by
\begin{equation}
D^{\nu \mu} = \langle A^\nu(-q) A^\mu(q) \rangle = (K^{-1})^{\nu \mu}.
\end{equation}

In the $q$-limit, $\omega \ll v_F^* |\bq|$, 
the gauge propagator is dominated by its transverse (T) component and
\begin{equation}
b^*_\nu(\bs; \bq) D^{\nu\mu}(q) b_\mu(\bs; \bq) \approx 
\frac{q_\bot^2}{|\bq|^2}~ 
\frac{\ves \chi_0(q)}{\chi_0(q) \left( \chi_T(q) -
\frac{\bq^2 V(\bq)}{(2 \pi \wp)^2} \right) - \frac{\bq^2}{(2 \pi \wp)^2}}\ .
\label{transverse}
\end{equation}
In this limit the gauge propagator has an imaginary part due to quasiparticle
damping. For the Coulomb interaction, after the analytic continuation
$i \omega \rightarrow \omega$,
\begin{equation}
b^*_\nu(\bs; \bq) D^{\nu\mu}(\bq) b_\mu(\bs; \bq) \approx \ves~
\frac{q_\bot^2}{\bq^2}~ \frac{1}{i \gamma \frac{|\omega|}{|\bq|} -\chi|\bq|}
\end{equation}
where following the notation of HLR we introduce 
$\gamma \equiv n_f/\pi v_F^* N(0)$, and 
$\chi \equiv 2\pi e^2 / (4 \pi^2 \varepsilon \wp^2)$, where $\varepsilon$ 
is the dielectric constant.
The pole is located at $\omega =i(e^2/\varepsilon k_F)|\bq|^2$. If, on the
other hand, $V(\bq)$ is a short range interaction the pole is at 
$\omega \propto iv_F^* |\bq|^3 k_F^2$.  The location of the pole
in both instances agrees with the RPA result of HLR\cite{HLR}. 
In the opposite limit $\omega$-limit, $\omega \gg v_F^* |\bq|$
\begin{equation}
b^*_\nu(\bs; \bq) D^{\nu\mu}(q) b_\mu(\bs; \bq) \approx \frac{\omega^2}{\bq^2}
\frac{(n_f/m^*)(2\pi \wp)^2}{\omega^2 -(n_f/m^*)^2 2\pi\wp^2}\ . 
\label{cyclotron}
\end{equation}
In this limit, the zeros of the denominator give the characteristic frequency
of the collective modes of the system, the cyclotron modes, with frequency
$\omega = 2\pi \wp n_f/m^*$, the cyclotron frequency of a free  particle of mass
$m^*$ in an external magnetic field of magnitude $B = 2 \pi n_f \wp/e$. 
The correct result is obtained, as in Section \ref{propagator}, 
when the Fermi liquid interaction, equation \ref{FLact}, 
is included in the theory. Then the zeros shift to
\begin{equation}
|\omega|= \omega_c = 2 \pi n_f \wp \left( \frac{1}{m^*}+\frac{f_1}{4\pi} 
\right) = 2\pi n_f \wp / m 
\end{equation}
as predicted\cite{Kohn}. The details of the proof of this result are
given in Appendix \ref{Kohn_App}. 

The form of equation \ref{cyclotron} is similar to the corresponding
result found for the super long-range interaction of Bares and Wen\cite{Bares}. 
However
since a projection onto the lowest Landau level is implicit in the HLR theory,
the cutoff must be such that $v_F^* \lambda \ll \omega_c$ to disallow
inter-Landau level transitions.  This restriction is an auxiliary condition
in the definition the composite fermion Green's
function. Non-Fermi liquid behavior, if there is such behavior, arises from
the form of the gauge propagator in the opposite $q$-limit. Although there is
no obvious small parameter we can estimate the fermion quasiparticle
self-energy from equation \ref{transverse} by expanding the fermion 
Green's function in powers of $D_{\mu\nu}$. Given this caveat, 
in the case of the Coulomb
interaction, at first order in the expansion, the self-energy has the marginal
Fermi liquid form in agreement with the RPA calculation of HLR\cite{HLR}.

As the expansion in powers of the boson propagator is unreliable instead we
calculate the fermion Green's function directly from the key bosonization 
formula, equation \ref{bosonization}. The fermion Green's function 
equation \ref{key} is a product of
the free propagator and a multiplicative factor which is the exponential 
of the correction to the free $\phi$-boson propagator
\begin{equation}
G_F(\bs; \bx, \tau) = G_F^0(\bs; \bx,\tau) \exp [-\delta G_\phi(\bs; \bx, \tau)]
\end{equation}
where
\begin{equation}
\delta G_\phi(\bs; \bx,\tau) = \sum_\bq \int \frac{d\omega}{2\pi} 
(e^{i(\bq \cdot \bx - \omega\tau)} - 1) 
\frac{b_\nu^*(\bs; \bq) D^{\nu \mu}(q) b_\mu(\bs; \bq)}{(i\omega
-v_F^* \hbn_\bs \cdot \bq)^2}\ .
\end{equation}
Setting $x_\bot = 0$, the integral over $q_\parallel$ can be evaluated by 
complex integration. In the $q$-limit we find that
\begin{equation}
\delta G_\phi(\bs; \bx,\tau) = \frac{|x_\parallel|}{2\pi \chi} \int_0^\infty
\frac{d\omega}{2\pi} e^{-\frac{\omega}{v_F^*} \sgn(x_\parallel) 
(x_\parallel + iv_F^* \tau )} 
\bigg{[} \ln \bigg{(} \omega + \left( \frac{\lambda}{2}\right)^2 
\frac{\chi}{\gamma} \bigg{)}
- \ln \bigg{(} \omega +\left(\frac{\omega}{v_F^*}\right)^2 
\frac{\chi}{\gamma} \bigg{)} \bigg{]} \ .
\end{equation}
The final integration over frequency can be carried out using the integration
formula
\begin{equation}
\int_0^\infty e^{-\mu x} \ln (\beta +x) dx = \frac{1}{\mu}[ \ln
\beta-e^{\mu\beta} E_i(-\mu\beta)],
\end{equation}
where $E_i(x)$ is the exponential integral function.  The result, after
continuation to real time, $i \tau \rightarrow -t$, is
\begin{equation}
\delta G_\phi(\bs; \bx, t) 
= \frac{x_\parallel \zeta}{x_\parallel - v_F^* t} 
\bigg{[} 
\ln |(x_\parallel - v_F^* t) \delta | - e^{(x_\parallel - v_F^* t) \delta 
\sgn(x_\parallel)} E_i[-(x_\parallel - v_F^* t) \delta 
\sgn(x_\parallel)] + \gamma \bigg{]}\ . 
\end{equation}
Here
\begin{equation}
\delta \equiv \frac{e^2}{4 \varepsilon v_F^* \wp^2} \frac{\lambda^2}{k_F} 
\end{equation}
is a momentum scale,
\begin{equation}
\zeta \equiv \frac{\wp^2 \varepsilon v_F^*}{2\pi e^2} 
\end{equation}
is a dimensionless number and $\gamma \approx 0.577216$ 
is Euler's constant.  Making use of the
asymptotic expansion of the exponential integral function 
$\delta G_\phi(\bs; \bx, t)$ reduces to
\begin{equation}
\delta G_\phi(\bs; \bx, t) = \left\{ \begin{array}{ll}
\zeta|x_\parallel \delta | \ln(|x_\parallel - v_F^* t| \delta), 
& | x_\parallel - v_F^* t | \delta \ll 1 \\
\\
-\zeta \frac{x_\parallel}{x_\parallel - v_F^* t} \ln(|x_\parallel
- v_F^* t| \delta), & |x_\parallel - v_F^* t| \delta \gg 1\ .
\end{array}\right.  
\label{asymptotic}
\end{equation}

For simplicity, first we consider the equal-time propagator which at long
distances $|\hbn_\bs \cdot \bx | \gg \delta^{-1}$ is given by 
\begin{equation}
G_F(\bs; \bx) = \frac{\Lambda}{(2 \pi)^2}~
\frac{e^{i \bk_\bs \cdot \bx}}{\hbn_\bs \cdot \bx}~ 
(|\hbn_\bs \cdot \bx| \delta)^{-\zeta}
\label{equal-time}
\end{equation}
and at short distances $|\hbn_\bs \cdot \bx | \ll \delta^{-1}$ 
by the usual Fermi liquid form
\begin{equation}
G_F(\bs; \bx) = \frac{\Lambda}{(2 \pi)^2}~
\frac{e^{i \bk_\bs \cdot \bx}}{\hbn_\bs \cdot \bx}.
\end{equation}
At momentum scales less than $\delta$ the quasiparticle pole is destroyed by
transverse
gauge interactions, suggesting that the $\nu=1/2$ system is controlled by a
non-Fermi liquid fixed point. Despite the appearance of the anomalous exponent
$\zeta$, the composite quasiparticle Green's function differs in an important
way from the Green's function found in Section \ref{propagator}
for the super long-range interaction.  It can be inferred from 
equation \ref{equal-time} that the
composite quasiparticle occupancy drops abruptly from one to zero over a
momentum scale $\delta \propto \lambda^2/k_F$ which is much less than the cut
off $\lambda$, as $\lambda \rightarrow 0$. The basic consistency condition for
the bosonization scheme is well satisfied in this instance.

The carrier density of a typical Ga As / Al$_x$ Ga$_{1-x}$ As sample is of
the order of $n_f \approx 10^{11} {\rm cm}^{-2}$, 
the electron band mass $m_b \approx 0.068 m_e$, and the
dielectric constant is given by $\varepsilon \approx 12.8$. 
The effective mass
of the composite fermion quasiparticle has to be estimated. Since the electron
states in the lowest Landau level are degenerate, the electron band mass does
not enter.  Instead the energy scale is determined by the Coulomb energy at the
magnetic length scale\cite{HLR,Murthy}.  
From dimensional analysis the effective mass of the 
quasiparticle must be therefore
\begin{equation}
m^* =\frac{\varepsilon}{e^2} \frac{\sqrt{4\pi n_f}}{C} 
\end{equation}
where $C$ is a dimensionless constant to be determined either theoretically or
from experiment. Experiments at $\nu =1/2$ suggest\cite{Du,Leadley} 
that $m^* \approx 10m_b$
and hence $C \approx 10$.  This value agrees well with the theoretical
estimate of Shankar and Murthy\cite{Murthy}.
For values in this range we find 
$\zeta = C \wp^2/2\pi \simeq 0.05$, quite small and possibly difficult
to measure.  For the purpose of estimating
$\delta$, we may take the cutoff $\lambda$ to be of the order of the Fermi
momentum, $\lambda \approx k_F$, 
and find $\delta = \lambda^2 /(4 C \wp^2 k_F) \approx O (k_F)$ 
as expected since $k_F$ is the only physical momentum scale in the
problem.  Notice that the Fermi energy $\epsilon_F =k_F^2/(2m^*)$ is still much
smaller than the cyclotron energy $\omega_c = k_F^2 \wp /(2m_b)$ and hence the
requirement that $v_F^*\lambda \ll \omega_c$ is respected even for $\lambda
\approx k_F$.  

Let us consider briefly longitudinal interactions other than Coulomb.
Because longitudinal interactions control fermion density fluctuations and
hence, by the constraint, gauge fluctuations, we expect repulsive interactions
that have a longer range than Coulomb to stabilize the Fermi liquid fixed point.
Drawing from the RG approach of Nayak and Wilczek\cite{Nayak} and Chakravarty,
Norton, and Syljuasen\cite{Norton} we set $V(\bq) = g |\bq|^{y-1}$, 
as $y = 0$ (the Coulomb case) is marginal.  Repeating the 
calculation of the equal-time Green's function for $y \neq 0$, we find
\begin{equation}
\delta G_\phi(\bs; \bx) = \int\frac{dq_\bot}{2\pi} \int_0^{\lambda v_F^*}
\frac{d\omega}{2\pi} \left(\frac{x_\parallel}{v_F^*} \right) e^{i\omega
x_\parallel/v_F^*} \frac{v_F^* |q_\bot|}{i\gamma \omega - \chi |q_\bot|^{2+y}}.
\end{equation}
When $y \neq $0, the correction to the boson propagator scales 
as\cite{HKM2,Kopietz-scale} 
\begin{equation}
\delta G_\phi(\bs; \bx) \sim (|x_\parallel| \delta^\prime)^{y/(2+y)}\ ,
\end{equation}
where 
\begin{equation}
\delta^\prime \equiv (\chi /v_F^*\gamma)(v_F^*/\chi)^{(2+y)/y}\ . 
\end{equation}
The same result has been obtained in the Eikonal approach of 
Khveshchenko and Stamp\cite{Khv-Stamp}.  For $y < 0$, the gauge
interaction is irrelevant,  $\delta G_\phi \rightarrow 0$ as 
$|x_\parallel| \rightarrow \infty$ 
and the Landau fixed point is recovered. For $y > 0$,
however, there are pronounced deviations from the Landau fixed point when
$|\delta G_\phi| \gg 1$ or, equivalently, 
when $|x_\parallel| \gg (\delta^\prime)^{-1}$. The
small momentum regime is controlled by a non-Fermi-liquid fixed point about
which bosonization can tell us little as the deviations from Fermi-liquid
behavior are too large, much larger than in the case of the the longitudinal
interaction of Bares and Wen discussed in the previous Section. 
For example, the discontinuity in the quasiparticle
occupancy at the Fermi surface is completely eliminated and the formalism is
not internally self consistent. The problem of a pure gauge interaction in two
dimensions, considered by Lee and Nagaosa\cite{Nagaosa} 
and Blok and Monien\cite{Blok} as a model for the cuprate
superconductors, is the special case $y=1$. Again, this system  is outside the
scope of multidimensional bosonization.

An important feature of the bosonization procedure is the linearization
of the free fermion spectrum in the vicinity of the Fermi surface:
$\epsilon_{\bf k_S+q}-\epsilon_{\bf k_S} \approx {\bf v_S^* \cdot q}$.
Nonlinear terms in the dispersion due to curvature of the Fermi surface
are neglected and within bosonization can only be accounted for perturbatively.
This approximation may be questioned, particularly in the case of
a mediating transverse gauge interaction, when the interaction
vertex draws its most important contribution from $q_{\perp}$, the
momentum perpendicular to ${\bf k_S}$. To see this, consider the
transverse gauge propagator $D_{TT}^{-1}(q) \sim i\gamma |\omega| / |{\bf q}| -
\chi |{\bf q}|^{1+y}$; the perpendicular momentum scales as
$q_{\perp}^2 \propto |\omega|^{2\over 2+y}$ and therefore it, rather than
the pole familiar from the linearized approximation, 
$q_\parallel = \omega / \ve$, dominates the physics of quasiparticle
propagation for $y > 0$.
The question of the reliability of multidimensional bosonization 
has been addressed\cite{HKM2} in the fermion basis by
the use of the Ward identity approach derived by
Castellani, Di Castro and Metzner\cite{Castellani}.  As mentioned in Section
\ref{essentials} the Green's function derived using the Ward identities 
is identical to that found by bosonization provided that the relationship
between density and current vertices, 
${\bf \Lambda}(\bs) = {\bf v^*_\bs} \Lambda^0(\bs)$,
which becomes exact in one dimension, remains accurate in
dimensions greater than one.  In this context, we can 
ask whether deviations from this relationship
are so strong that they destroy the form of the
propagator found within the multidimensional bosonization approach.
For the Chern-Simons gauge theory with $y < 0$, multidimensional
bosonization is internally self-consistent as the quadratic parts of
the dispersion cancel out and the vertex corrections are subleading\cite{HKM2}.
The marginal case of the Coulomb interaction, $y = 0$, is delicate and must
be treated with special care. Whether or not multidimensional 
bosonization is accurate for $y = 0$ is a question we address below. 

We return now to an examination of the single particle properties 
of composite fermions, in particular the physically interesting case of
the marginal Coulomb interaction, $y = 0$.  
The starting point is the single particle propagator in
real time which is obtained from equation \ref{asymptotic}. 
In particular we show that the Green's function obtained
from multidimensional bosonization is similar, but not identical, to 
the marginal Fermi liquid (MFL) form\cite{MFL}.  
For simplicity we set $v_F^* = 1$ and denote $x_\parallel$ by $x$; we 
also drop the prefactor of $\Lambda e^{i \bk_\bs \cdot x} / (2 \pi)^2$. 
Then the Green's function in the two limits $|x-t| \delta \gg 1$ and
$|x-t| \delta \ll 1$ can be written as
\begin{equation}
G_F(\bs; \bx, t) =  \left\{ \begin{array}{ll}
\frac{1}{x - t + i a~ \sgn(x)} \exp \left\{ \zeta |x| \delta
\ln |x-t|\delta \right\}; & |x-t|\delta \ll 1 \\ 
\\
\frac{1}{x-t + i a~ \sgn(x)}\exp \left\{-\zeta \frac{x \ln
|x-t|\delta}{x-t}\right\}; & |x-t| \delta \gg 1
\end{array}\right.
\label{two-limits}
\end{equation}
Ideally we would Fourier transform the Green's function directly into momentum
and frequency space to extract the self-energy.  Unfortunately, this task is
quite difficult, so we begin with the observation that the pole of the free
fermion Green's function at $x=t$ has been eliminated by the transverse
interactions, because as $x \rightarrow t$, 
$G_F(\bs; \bx, t) \rightarrow \ |x-t|^{\zeta \delta |x|}/(x-t) \rightarrow 0$. 
To find where the spectral weight is concentrated, we
expand the Green's function in the limit of long time: 
$|t| \gg |x + \zeta x \ln(|x - t|\delta)$.  Since in the long-time
limit $|x / (x-t)| \ll 1$ we may expand the exponential appearing in the second
line of equation \ref{two-limits} in Taylor series; 
keeping the leading terms we obtain
\begin{eqnarray}
G_F(\bs; \bx, t) &=& \frac{1}{x - t + i a~ \sgn(x)} \times 
\frac{1}{1+ \frac{\zeta x}{x-t} \ln (|x-t|)\delta) + \ldots}
\nonumber \\
&=& \frac{1}{(x-t) + \zeta x \ln (|x-t| \delta) + i a~ \sgn(x)}
\label{Taylor}
\end{eqnarray}
where the ellipses denote higher order terms that are negligible in the
$|t| \gg x$ limit. We may now Fourier transform the leading term to obtain the
self-energy.  If we make the approximation that near the pole in 
equation \ref{Taylor},
$\ln (|x-t|) \delta) \simeq \ln (|x|\delta)$ the integral over time can be
done and
\begin{eqnarray}
G_F(\bs; \bk, \omega) &=& \int dxdt \frac{e^{i(\omega t - \bk \cdot
\bx)}}{(x-t)+ \zeta x \ln(|x-t|\delta) + i a~ \sgn(x)} 
\nonumber \\
&\approx& 2\pi i  \int dx [\theta (-x) \theta(\omega) -\theta (x)
\theta(-\omega)]e^{i(\omega - k)x}  \exp [i\zeta \omega \ln |x| \delta]
\nonumber \\
&=& 2\pi i \int dx [\theta (-x) \theta (\omega )-\theta (x) \theta(-\omega) ]
e^{i(\omega -k)x} \sum_{n=0}^\infty \frac{[i\zeta \omega x \ln (|x|\delta
)]^n}{n!}\ .
\end{eqnarray}
Now each term coming from the Taylor-series expansion of the exponential can be
integrated separately and the series resummed. It is readily shown that
\begin{equation}
\int_0^\infty dx~ e^{(i k - \eta) x} [x \ln (|x|\delta)]^n =
\frac{-1}{(i k - \eta)^{n+1}} \left\{ n! \ln^n [(\eta - i k)/\delta] 
+ O (\ln^{n-1} [(\eta - i k)|\delta]) \right\}
\end{equation}
where $\eta >0$ is a small term which ensures convergence. Up to subleading
corrections, the Green's function is
\begin{eqnarray}
G_F(\bs; \bk, \omega) &=& \frac{2\pi}{(\omega -k)} \sum_{n=0}^\infty 
\left[ \frac{\zeta\omega}{\omega-k}
\ln [ i(\omega -k) \sgn(\omega) / \delta ]^n \right. 
\nonumber \\
&=& \left. \frac{2\pi}{(\omega -k) -\zeta \omega \ln [ i(\omega -k) 
\sgn(\omega) /\delta]} \right]
\end{eqnarray}
and hence the leading contribution to the quasiparticle self-energy is
\begin{equation}
\Sigma(\bs; \bk, \omega) = \zeta \omega \ln (|\omega - k| / \delta)+
\frac{i\pi}{2} \zeta \omega~ \sgn[(\omega -k)\omega ]\ .
\end{equation}
At the pole $k \equiv \omega - \zeta \omega \ln (|\omega | / \delta), w \ll k$
and to logarithmic accuracy the self-energy is given by
\begin{equation}
\Sigma(\bs; \bk, \omega ) = \zeta \omega \ln (|\omega | /v_F^*
\delta ) - i\frac{\pi}{2} \zeta \omega
\label{marginal}
\end{equation}
the marginal Fermi liquid form\cite{MFL}. In equation \ref{marginal} 
we have restored $v_F^*$. It
should be emphasized that $v_F^*$ is the Fermi velocity obtained by
integrating out the high energy degrees of freedom; but in a marginal Fermi
liquid such as that 
found here there is no definite velocity; rather the velocity is
a logarithmic function of the energy scale and is of the order of, but not
equal to, $v_F^*$. At half filling $\wp = 2$ and $\zeta = 2v_F^*/\pi e^2$. Upon
setting $\lambda \equiv k_F$ for the purpose of  estimating the size of the
self-energy we find $v_F^* \delta = O(\epsilon_F)$, the Fermi energy. 
It should be emphasized that this result is incomplete, as the equal-time
composite fermion Green's function, equation \ref{equal-time}, 
exhibits an anomalous exponent.  This equal-time behavior is subleading,
however, as the Green's function decays more rapidly as a function of 
increasing spatial separation than at large temporal separation.
The approximate self-energy calculated above, equation \ref{marginal},
captures only the long-time behavior of the Green's function. 

\begin{figure}
\center
\epsfxsize = 10.0cm \epsfbox{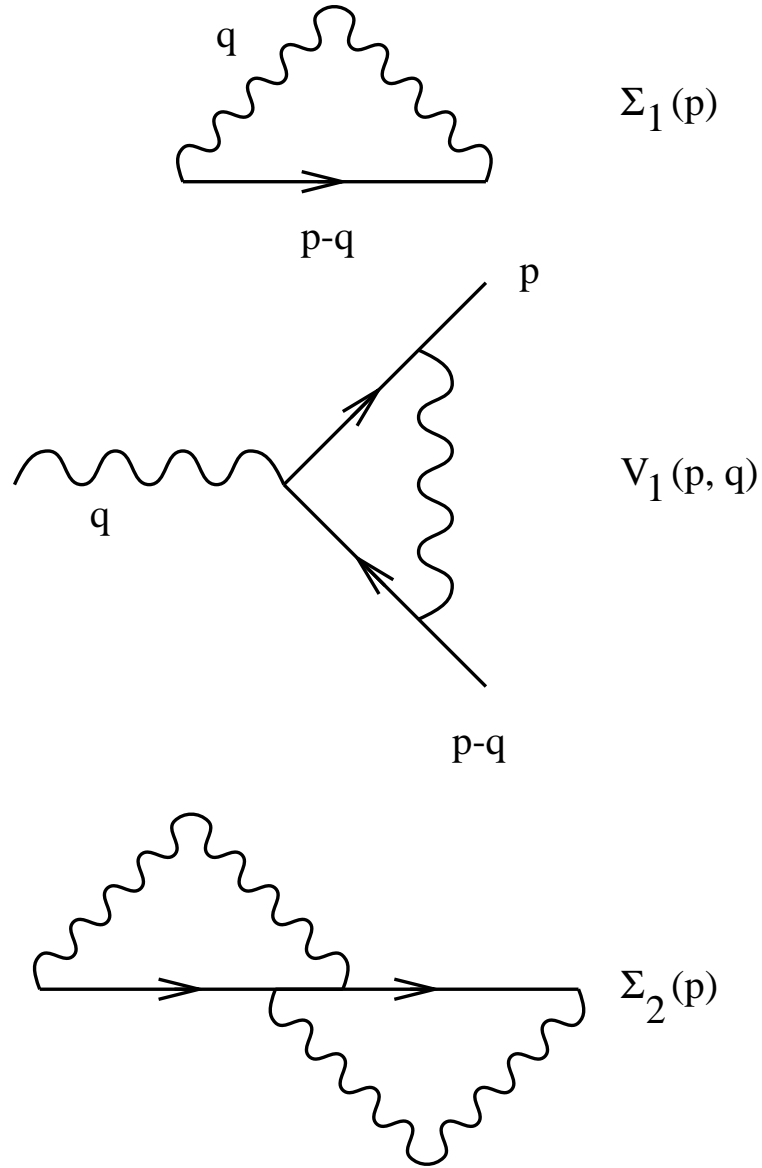}
\caption{
\label{vertex}
The one-loop contribution to the self-energy $\Sigma_1(p)$, 
the one-loop vertex correction $V_1(p; q)$, and the two-loop contribution
to the self-energy, $\Sigma_2(p)$.  The calculation of $\Sigma_2(p)$ can be
broken down into two steps: first calculate $V_1(p; q)$ and then replace
one of the two bare vertices $V_0$ which appears in $\Sigma_1(p)$ 
by $V_1$.}
\end{figure}

Equation \ref{marginal} is identical to the self-energy found by 
Altshuler, Ioffe, and Millis via 
a perturbative loop expansion\cite{Altshuler}.  
These authors argue that the multi-loop contributions to the self-energy only
contribute sub-leading corrections to the one-loop result.  This claim 
is clearly at variance with the anomalous power law decay of the equal-time
propagator found within multidimensional bosonization, 
equation \ref{equal-time}.
However, a recent calculation by Kopietz\cite{Kopietz-Vertex} 
shows that the two-loop contribution to the self-energy, $\Sigma_2(\epsilon)$, 
is actually more singular than the one-loop self-energy 
$\Sigma_1(\epsilon) \sim \epsilon^{2/(2+y)}$, 
casting doubt on the reliability of perturbative loop expansions.
Specifically, Kopietz evaluates the two-loop Feynman diagram (see
figure \ref{vertex}), employing bare
fermion propagators with the full non-linear quasiparticle dispersion:
\begin{equation}
\Sigma_2(\bp, \epsilon) = i \int \frac{d\omega d^{2}q}{(2\pi)^3}~
V_{0}(\bp-\bq; -\bq)~ 
G_0(\bp-\bq, \epsilon-\omega)~ D_{TT}(\bq,\omega)~ 
V_{1}(\bp, \epsilon; \bq, \omega)\ .
\label{two-loop}
\end{equation}
here $G_0$ is the bare fermion Green's function given, 
setting $\ve = k_F = \gamma = \chi = 1$ for simplicity, by
\begin{equation}
G_0(\bp, \nu) = \frac{1}{i \nu - \frac{1}{2}~ (\bp^2 - 1) 
+ i \eta~ \sgn(\nu)}\ .
\end{equation}
$D_{TT}$ is the transverse gauge propagator which does not
renormalize significantly\cite{Gan,Furusaki} and is therefore given by the
formula obtained above using multidimensional bosonization:  
\begin{equation}
D_{TT}(\bq, \omega) = \frac{1}{i \frac{|\omega|}{|\bq|} - |\bq|^{1+y}}\ .
\end{equation}
$V_1$ is the one-loop correction to the bare current vertex $V_0$. 
Unlike $V_1$, $V_0$ does not depend on the external frequencies, and is given
simply by:
\begin{equation}
V_0(\bp; \bq) = \hat{\bq} \times \bp\ .
\end{equation}
Kopietz finds that $\Sigma_2(\epsilon) \sim \epsilon^{2/(2+y)}
\ln(\epsilon)$ which is bigger than $\Sigma_1(\epsilon)$ by the logarithmic
factor $\ln(\epsilon)$.  
This result contradicts equation 8 of Reference \cite{Altshuler}.   

To check the result, the one-loop correction to the current vertex
has been computed numerically\cite{Arnaldo} on lattices up to size 
$1024 \times 1024 \times 1024$:
\begin{eqnarray}
V_{1}(\bp, \epsilon; \bq, \omega)
&=& i \int \frac{d\omega^\prime d^{2}q^\prime}{(2\pi)^3}
G_{0}(\bp-\bq-\bq^\prime, \epsilon-\omega-\omega^\prime)~
V_0(\bp-\bq; \bq^\prime)~
\nonumber \\
&\times& D_{TT}(\bq^\prime, \omega^\prime)~
V_0(\bp-\bq-\bq^\prime; -\bq^\prime)~
G_{0}(\bp-\bq^\prime, \epsilon-\omega^\prime).
\end{eqnarray}
An important test of the numerical calculation can be performed by 
invoking the Ward identity.
For external momentum $\bq = 0$, the Ward identity (which reflects the
conservation of the total number of particles\cite{Schrieffer,Castellani}) 
relates the one-loop vertex correction $V_1$ 
to the quasiparticle Green's function $G_1$ dressed with the
one-loop self-energy $\Sigma_1(\epsilon) \sim \epsilon^{2/(2+y)}$:
\begin{equation}
V_{1}(\bp, \epsilon; {\bf 0}, \omega) = \frac{G_{1}^{-1}(\bp, \epsilon)-
G_{1}^{-1}(\bp, \epsilon - \omega)}{\omega}.
\end{equation}
The numerical results agree with this exact relation for $0 \leq y \leq 1$,
demonstrating their reliability.
The vertex can also be evaluated in the kinematic region relevant for the
calculation of the two-loop contribution of the self-energy when the fermion 
propagators are dressed at one-loop order, $\Sigma_2^{(1)}$: 
\begin{equation}
\Sigma_2^{(1)}(\bp, \epsilon) = i \int \frac{d\omega d^{2}q}{(2\pi)^3}
V_0(\bp-\bq; -\bq)~ G_1(\bp-\bq, \epsilon-\omega)~ D_{TT}(\bq,\omega)~ 
V_{1}(\bp, \epsilon; \bq, \omega).
\end{equation}
For the case of very strong gauge fluctuations, $y = 1$, 
the important region is given by 
$q_\parallel \sim \omega^{2/3}$ and $q_\perp \sim \omega^{1/3}$, 
the vertex correction is finite and nearly constant as a function of
the gauge frequency $\omega$:
\begin{equation}
V_{1}(\bp, \epsilon;~ q_{\parallel} \sim \omega^{2/3}, 
q_{\perp} \sim \omega^{1/3}, \omega)~ \approx~ {\rm const}.
\end{equation}
If this were the whole story, it would be a 
self-consistent demonstration that the leading behavior of the full self-energy
is given simply by the one-loop form, as stated by Altshuler, Ioffe and
Millis\cite{Altshuler} and by Stern and Halperin\cite{AStern}.
The Fermi liquid fixed point would be destroyed, the quasiparticle weight
would vanish, and a type of Migdal theorem\cite{Migdal,AGD} would hold
as vertex corrections would make only subleading contributions. However,   
unlike the usual
Migdal theorem where the electron mass divided by the ion mass is the small
parameter which limits the size of the vertex corrections, 
in the gauge theory there is no obvious small parameter. 

Kopietz\cite{Kopietz-Vertex} points out, however, that in the kinematic
region relevant for {\it bare} fermion propagators,  
$q_\parallel \sim \omega$ and $q_\perp \sim \omega^{1/2}$, 
there is a contribution to the one-loop vertex correction $V_1$ 
which, in contrast to the bare vertex $V_0$, changes sign under: 
\begin{equation}
\epsilon \rightarrow -\epsilon
\end{equation}
or
\begin{equation}
\omega \rightarrow -\omega \ .
\end{equation}
This odd part of the vertex correction has the form:
\begin{equation}
V_{1}(\bp,\epsilon; \bq,\omega) =
-c~ \sgn(\epsilon)~ \frac{|\epsilon|^{\frac{2}{2+y}}}{\omega}
\label{odd-vertex}
\end{equation}
where $c$ is a constant; this behavior is found in the limit
$|\omega| > |\epsilon|^{\frac{1}{2+y}}$.  Numerical calculations\cite{Arnaldo} 
confirm the existence of this feature of the vertex,
tracking the fermion energy $|\epsilon|$ as predicted.
Linearization of the quasiparticle spectrum does not change the
feature qualitatively. 
Naively equation \ref{odd-vertex} would appear to be a subleading correction, 
but, because it has a different symmetry than the bare vertex (which has
no frequency dependence), Kopietz\cite{Kopietz-Vertex} finds that 
it makes the leading contribution to the 2-loop self-energy, at least
when bare fermion propagators $G_0$ are used, equation \ref{two-loop}.  
If this is correct, 
there is no obvious sense in which a Migdal-type theorem holds.
On the other hand, it would be surprising if Fermi liquid behavior
were restored, as claimed by Kopietz and Castilla\cite{Kopietz-Castilla}.  
In particular, it seems unlikely that the $\ln(\epsilon)$ term at two-loop
order resums into a compensating power law as suggested in Reference
\cite{Kopietz-Vertex}.  We know that Fermi liquid behavior is {\bf not} restored
in the case of a linearized fermion spectrum, 
where multidimensional bosonization can be applied with confidence\cite{Millis}.

The form of the composite quasiparticle propagator therefore 
remains controversial.  In any case, as the propagator is not gauge-invariant, 
it is unclear whether or not physical meaning can be ascribed to it.   
We turn now to an examination of some gauge-invariant properties. 

\section{Two-Particle Response of Gauge-Fermion Systems}
\label{response}
\setcounter{equation}{0}
In the previous section we focused attention exclusively on the single
particle properties of the composite fermions of the Chern-Simons gauge
theory. Using multidimensional bosonization we derived the composite
fermion propagator which was found to have non-Fermi liquid form. However 
the composite fermion Green's function is not gauge invariant, 
and does not have a clear physical meaning. In fact it has been shown by Kim 
{\it et al.}\cite{Furusaki} that, depending on the choice of gauge,
different types of singularity appear.
These same authors also used the perturbative loop expansion to study such gauge
invariant quantities as the density and current response at small momentum
transfer. These two particle response functions are gauge-invariant because
the physical electron density, which is of course gauge-invariant, equals
the composite fermion density.  Kim {\it et al.} 
were able to show that the two-particle properties of the interacting gauge
system are in fact Fermi liquid like: the singular corrections to the
quasiparticle self-energy were canceled by singular vertex corrections, at 
least up to two-loop order.
However the possibility of singular behavior at large momentum transfer could
not be ruled out.  In fact, Altshuler, Ioffe and Millis\cite{Altshuler} have
argued that the vertex correction at momentum transfer $2k_F$ should be singular
since the gauge fields mediate an attractive interaction between currents
moving in the same direction. In this section we use the multidimensional
bosonization
formalism to determine the density and current response for both small momentum
transfer $|\bq| \approx 0$ and $|\bq| \approx 2k_F$.

The density-density correlation function is given in all generality by equation
\ref{rhorho}, in particular at long wavelengths the density response is given by
\begin{equation}
\Pi_{00} (\bq, \omega) = -\sum_{\bs, \bt} 
\langle J(\bs; \bq, \omega ) J(\bt; -\bq, -\omega) \rangle 
\end{equation}
where in complete analogy to equation \ref{JJ} the correlation function for the
gauge theory can be written as
\begin{eqnarray}
&& \langle J(\bs; \bq, \omega ) J(\bt; -\bq, -\omega) \rangle  
\nonumber \\
&=& -\Omega \left[ \frac{\hbn_\bs \cdot \bq}
{i \omega - v_F^* \hbn_\bs \cdot \bq}~ \delta^{D-1}_{\bs, \bt} \right.
\nonumber \\
&+& \left. \frac{\Lambda}{(2\pi)^2} \frac{(\hbn_\bs \cdot \bq) (\hbn_\bt \cdot
\bq)}{(i\omega -\ve \hbn_\bs \cdot \bq)(i\omega -\ve \hbn_\bt \cdot \bq)}
b_\nu^*(\bs; \bq) D^{\nu \mu}(q) b_\mu(\bt; \bq) \right]\ .
\end{eqnarray}
Recall that ${\bf D}$ is the gauge field propagator, the vector
$b_\nu(\bs; \bq) = (1, i \ve \frac{\bq \times \hbn_\bs}{|\bq|})$,
and the index $\nu$ runs over $(0,T)$.  Now as
\begin{eqnarray}
b_\nu^*(\bs; \bq) D^{\nu\mu}(q) b_\mu(\bt; \bq) &=& D_{00}(q) 
- i \ve \frac{\bq \times \hbn_\bs}{|\bq|} D_{T0}(q) 
+ i \ve \frac{\bq\times \hbn_\bt}{|\bq|} D_{0T}(q)
\nonumber \\
&+& \frac{(\bq \times \hbn_\bs)}{|\bq|}\frac{(\bq \times \hbn_\bt)}{|\bq|} 
D_{TT}(q) 
\end{eqnarray}
only the term proportional to $D_{00}$ survives the summation over 
$\bs$ and $\bt$ and therefore
\begin{equation}
\Pi_{00}(\bq, \omega) = -\chi_0(q) \left[ 1 - D_{00}(q) \chi_0(q) \right] 
\end{equation}
where
\begin{equation}
D_{00}(q) = \left[\chi_T(q) - \frac{\bq^2 V(\bq)}{(2\pi \wp)^{2}}\right] 
({\rm det} {\bf K})^{-1} 
\end{equation}
and
\begin{equation}
{\rm det} {\bf K} = \chi_0(q) \left[ 
\chi_T(q) - \frac{\bq^2 V(\bq)}{(2\pi \wp)^{2}} \right] 
- \frac{\bq^2}{(2\pi \wp)^{2}} \ .
\end{equation}
Here
$\chi_0$ and $\chi_T$ are the longitudinal and transverse susceptibilities of
the free electron gas defined in equations \ref{chi_0} and \ref{chi_T}. 
Combining these results, the long wavelength density response can be written as
\begin{equation}
\Pi_{00}(\bq, \omega) = \frac{\Pi_{00}^0(\bq, \omega) 
\bq^2/(2\pi \wp)^2}{\bq^2/(2\pi \wp)^2 - \Pi_{00}^0(\bq, \omega)
[n_f/m^* + q^2 V(\bq) / (2\pi\wp)^2 - \Pi_{11}^0(\bq, \omega)]}
\label{longitudinal-response}
\end{equation}
where we have introduced the paramagnetic component of the current response
\begin{equation}
\chi_T(q) =\Pi_{11}^0(q) - n_f/m^* 
\end{equation}
and $\Pi_{00}^0$ is the density response of the free electron gas. The
transverse current response can be found in a similar way
\begin{equation}
\Pi_{11}^0(\bq, \omega) = \Pi_{11}^0(\bq, \omega) \left [1 + D_{TT}(\bq, \omega)
\Pi_{11}^0(\bq, \omega) \right]
\label{transverse-response}
\end{equation}
where
\begin{equation}
D_{TT}(\bq, \omega)= \frac{\Pi_{00}^0(\bq, \omega)}
{\bq^2 / (2\pi\wp)^2 - \Pi_{00}^0(\bq, \omega) 
[n_f/m^* + \bq^2 V(\bq) /(2\pi \wp)^2 -\Pi_{11}^0(\bq, \omega) ]}. 
\end{equation}
The results found here for the longitudinal and transverse response, equations
\ref{longitudinal-response} and \ref{transverse-response},
are identical to the results found within RPA\cite{HLR}.  They also agree 
with the leading two-loop result of Kim {\it et al.}\cite{Furusaki} 
from which we can conclude that the cancellation
between divergent vertex corrections and divergent self-energy corrections is
inherent in bosonization.  The zero-frequency density response, the 
compressibility, vanishes as
\begin{equation} 
\Pi_{00}(\bq, 0) \propto |\bq| \ ,
\end{equation}
instead of $|\bq|^2$ or faster,
so the electron system at $\nu = 1/2$ is in fact compressible, a result
reproduced recently within the formalism of Shankar and Murthy\cite{Murthy}
by Halperin and Stern\cite{Halperin-Stern}.

We now turn to consider the real space density response function in two
dimensions at momentum transfer $2k_F$.  With the gauge interaction, currents
moving in the same direction are attracted to each other so we might anticipate
nonanalytic behavior in $\Pi_{2k_F}$. This nonanalyticity has a quite
different physical origin than the Kohn anomaly. The Kohn anomaly is due to the
reduced availability of phase space at low energies. By contrast the behavior
we expect in the present context should be apparent even if we only consider
scattering between two patches at exactly opposite points on the Fermi surface.
Therefore we need only consider terms with
$\bk_\bs = -\bk_\bt$ in the $2k_F$ response function, equation \ref{2Pi}.
\begin{eqnarray}
\Pi_{2k_{F}} (\bx, \tau)
&=& -\left(\frac{\Omega}{Va}\right)^2 \sum_\bs \bigg{\langle} \exp \left[
i\frac{\sqrt{4\pi}}{\Omega} (-\phi(\bs; x) +\phi(-\bs; x))\right] 
\nonumber \\
&\times& \exp \left[i\frac{\sqrt{4\pi}}{\Omega}(-\phi(\bs; 0) 
+ \phi(-\bs; 0)) \right] \bigg{\rangle} 
\end{eqnarray}
which can be transformed using the generalized Baker-Hausdorff formula,
equation \ref{Baker-Hausdorff}, into
\begin{eqnarray}
\Pi_{2k_{F}} (\bx, \tau) &=& -\left(\frac{\Omega}{Va}\right)^2 \exp
-\frac{4\pi}{\Omega^2} \bigg{(} [\langle \phi(\bs; \bx) \phi(-\bs; 0) \rangle 
- \langle \phi(\bs; 0) \phi(-\bs; 0) \rangle ]
\nonumber \\
&-& [\langle \phi(\bs; \bx) \phi(\bs; {\bf 0}) \rangle 
- \langle \phi^2(\bs; {\bf 0})\rangle] + [\bs \rightarrow -\bs] \bigg{)} 
\label{2kF-gauge}
\end{eqnarray}

The correlation function connecting patches at opposite points on the Fermi
surface contains the new physics; it can be expressed in terms of the canonical
boson correlation functions as:
\begin{eqnarray}
&&\left[ \langle \phi(\bs; x) \phi(-\bs; 0) \rangle 
- \langle \phi(\bs; 0) \phi(-\bs; 0) \rangle \right]
= -\frac{\Omega}{4\pi} \sum_{\bp, \hbn_\bs \cdot \bp > 0} \int
\frac{d\omega}{2\pi} \frac{1}{|\hbn_\bs \cdot \bp|} 
\nonumber \\
&& \times \bigg{[} (e^{i(\bp \cdot \bx - \omega \tau)}-1) 
\langle a(\bs; p) a(-\bs; -p) \rangle 
+ (e^{-i(\bp \cdot \bx -\omega \tau)} - 1) 
\langle a^*(\bs; p) a^*(-\bs; -p) \rangle \bigg{]}  
\label{correlated-bosons}
\end{eqnarray}
The boson correlation functions are then determined from the generating
functional equation \ref{hfllgenfunc}:
\begin{eqnarray}
\langle a^*(\bs; p) a^*(-\bs; -p) \rangle 
&=& \frac{\Omega |\hbn_\bs \cdot \bp|}{(\ve \hbn_\bs \cdot \bp)^2 -
(i\omega)^2} \bigg{[} - D_{00}(p) + i \ve \frac{{\bf p} \times \hbn_\bs}{|\bp|}
(D_{0T}(p) + D_{T0}(p)) 
\nonumber \\
&+& \ves \left( \frac{{\bf p} \times \hbn_\bs}{|\bp|} \right)^2 D_{TT}(p) 
\bigg{]}
\end{eqnarray}
The other correlation function, $\langle a(\bs; p) a(-\bs; -p) \rangle$, 
is given by this same expression upon replacing $i \ve \rightarrow -i \ve$.
Combining these results in equation \ref{correlated-bosons} 
and adding the corresponding terms
with $\bs \rightarrow -\bs$ we find that the contribution to the exponent of
equation \ref{2kF-gauge} from the interpatch correlations, $(\bs, -\bs)$, 
is given by
\begin{eqnarray}
&&-\frac{4\pi}{\Omega^2} \left\{ [\langle \phi(\bs; x) \phi(-\bs; 0) \rangle -
\langle \phi(\bs, 0) \phi(-\bs, 0) \rangle] + [\bs \rightarrow -\bs] \right\}
\nonumber \\
&&= 2 \sum_\bp \int \frac{d\omega}{2\pi} 
\frac{[e^{i(\bp \cdot \bx - \omega\tau)} - 1]}{\omega^2 
+ (\ve \hbn_\bs \cdot \bp)^2} 
\left[-D_{00}(\bp, \omega) + \left(\frac{\ve p_\bot}{|\bp|}\right)^2 
D_{TT} (\bp, \omega)\right] 
\nonumber \\
\end{eqnarray}
The contribution of the intrapatch correlations can be obtained directly 
from the boson propagator, equation \ref{ST-prop}:
\begin{eqnarray}
&&\frac{4\pi}{\Omega^2} [\langle \phi(\bs; \bx) \phi(\bs; 0) \rangle 
- \langle \phi^2(\bs; 0) \rangle] + [\bs \rightarrow -\bs] 
= \ln \frac{a^2}{(\hbn_\bs \cdot \bx + i\ve \tau)
(\hbn_\bs \cdot \bx - i\ve \tau)}  
\nonumber \\
&-& \sum_\bp \int \frac{d\omega}{2\pi} \left[ e^{i(\bp \cdot \bx - \omega
\tau)}-1\right] \left[\frac{1}{(i\omega -\ve \hbn_\bs \cdot \bp)^2 } +
\frac{1}{(i\omega +\ve \hbn_\bs \cdot \bp)^2}\right] 
\nonumber \\
&\times& \left[D_{00}(\bp, \omega) +
\left(\frac{\ve p_\bot}{|\bp|}\right)^2 D_{TT}(\bp, \omega) \right]
\end{eqnarray}
and therefore
\begin{eqnarray}
\Pi_{2k_{F}}(\bx, \tau) &=& -\left[ \frac{\Lambda}{(2\pi)^2}\right]^2
{\sum_{\bs}}^\prime
\frac{1}{(\hbn_\bs \cdot \bx + i \ve \tau)(\hbn_\bs \cdot \bx - i \ve \tau)}
\nonumber \\
&\times& \exp - 4 \sum_\bp \int \frac{d\omega}{2\pi}
[ e^{i(\bp \cdot \bx -\omega\tau)} - 1] \left[ \frac{v_F^2 p_\parallel^2 
D_{00}(p) + (i\omega)^2 (\frac{v_F \bp_\bot}{|\bp|})^2 
D_{TT}(p)}{[(i\omega)^2 -(\ve \hbn_\bs \cdot \bp)^2]^2}\right]. 
\end{eqnarray}
where the prime on the sum over patches indicates that the sum over $\bs$ is
only over patches for which $|\bx \times \hbn_\bs| \lambda < 1$. 
As in the case of the two-point function the most singular contribution comes 
from the $p$-limit, the exponent can be evaluated and we find
\begin{eqnarray}
\Pi_{2k_{F}}(\bx, \tau) &=& -\left[\frac{\Lambda}{(2\pi)^2}\right]
{\sum_\bs}^\prime \frac{\exp {\cal E}(\hbn_\bs \cdot \bx, \tau)}
{(\hbn_\bs \cdot \bx + i \ve \tau)(\hbn_\bs \cdot \bx - i\ve\tau)}
\nonumber \\
&\propto& -\frac{\exp {\cal E}(|\bx|,\tau)}{(|\bx|+i\ve\tau)(|\bx|-i\ve\tau)}
\end{eqnarray}
where
\begin{eqnarray}
{\cal E}(x, \tau) &\equiv& 
\zeta \left[ \frac{x}{(x-i\ve \tau)} \ln |(x-i\ve \tau) \delta|
+ \frac{x}{(x+i\ve \tau)} \ln |(x+i\ve \tau) \delta| \right.
\nonumber \\
&+& \left. \frac{1}{2} \ln^2  |(x-i\ve \tau)\delta|+\frac{1}{2} 
\ln^2 |(x + i \ve \tau) \delta |\right] 
\end{eqnarray}
for $| x - i\ve \tau | \delta \gg 1$ and $| x + i \ve \tau | \delta \gg 1$. 
The log-squared terms appearing in the exponent 
${\cal E}(|x|, \tau)$ have their origin
in the correlations induced by the gauge interaction between bosons in opposite
hemispheres.  Although it is technically difficult to Fourier transform the
response function into frequency space, the leading singular behavior can be
estimated by a rough scaling argument, and it is found to be
\begin{equation}
\Pi_{2k_{F}}(\omega) \propto \exp [\zeta \ln^2|\omega/ \ve \delta |]
\end{equation}
where again $\ve \delta = O(\epsilon_F)$. $\Pi_{2k_F}$ is strongly singular as
$\omega \rightarrow 0$, in qualitative agreement with results from an exact 
diagonalization study\cite{Rezayi}.

\section{Nested Fermi Surfaces}
\label{nest}
\setcounter{equation}{0}
Understanding the cuprate superconductors, both in the normal state and in
the superconducting state, is one of the most challenging problems in 
condensed matter physics.  Many interesting phenomena emerge, 
especially when the hole doping is less than optimal.
As the doping is decreased superconductivity is quenched and eventually the 
material becomes an antiferromagnetic insulator; hence in the underdoped regime
the interplay between the antiferromagnetic
exchange channel and the superconducting channel is likely to be important.  
This is the common operating assumption of spin fluctuation\cite{spin-fluct}, 
gauge\cite{Baskaran,AM,su2}, SO(5) symmetry\cite{so5}, and 
stripe\cite{stripe} models of high-temperature superconductivity, 
to mention just some of the theories.
Indeed since angle-resolved photoemission experiments\cite{Dessau} indicate that
the Fermi surface of the copper-oxygen plane is nested or nearly so; 
it seems clear that the tendency towards antiferromagnetism must
be taken into account. 

The interplay between nesting, umklapp processes, and the formation of 
charge-density-wave (CDW), spin-density-wave (SDW), 
and BCS instabilities has been illustrated in a simple model of the Fermi
surface which reduces the nested planes to just four Fermi points\cite{Tony}.
In a recent paper, Furukawa and Rice use a generalized version of this
model to argue that the spin gap in the cuprate superconductors 
can be understood as a consequence of umklapp
processes\cite{Rice}.  In this section, however, we retain the whole Fermi
surface.  Wilson's RG algorithm, which we introduced in 
Section \ref{stability}, 
integrates out high-energy degrees of freedom to derive a low-energy effective
theory.  In degenerate fermion systems this means 
that we shrink successively the 
inner and outer sides of the momentum shell which encloses the 
Fermi surface\cite{Shankar,Chitov}.  As in Section \ref{stability}
we investigate the RG flow of the effective interactions by
integrating out the high-energy modes of the bosonic field $\phi(\bs; x)$.  
The advantage of the bosonized RG calculation is that the existence of 
well-defined quasiparticles is not assumed {\it a priori}.  Furthermore,
interaction channels which possess $U(1)^\infty_k$ symmetry 
can be diagonalized exactly at the outset\cite{HKM}.

First we introduce the prototypical low-energy model of
a two-dimensional nested fermion liquid.  Then multidimensional bosonization
is used to incorporate the direct and exchange channels and to show that 
the resulting low-energy fixed point is a marginal Fermi liquid, which has
obvious implications for the normal state of the cuprate 
superconductors\cite{MFL}.  Next we derive the RG equations for the 
BCS channel and discuss
the BCS instability.  An analog of the Landau theorem, here made applicable to 
anisotropic Fermi surfaces, is invoked.
We include the marginally relevant antiferromagnetic interactions and evaluate 
their RG flow and their contributions to the different BCS channels. 
The flow in the $d_{x^2-y^2}$-wave channel dominates the other BCS channels.
This phenomena is similar to the 
Kohn-Luttinger effect\cite{KL,Baranov}. The difference here being that 
nesting of the Fermi surface can induce BCS instabilities at leading order,
rather than at subleading order as in the Kohn-Luttinger effect, and 
therefore can enhance the tendency toward superconductivity.

The Hubbard model is a deceptively simple, but convenient, 
description of many
strongly correlated fermion systems\cite{PWAScience}.  It serves
as the prototype for our analysis\cite{Kwon}.
A RG investigation of this model, of course, must take into account
a realistic band structure near the Fermi surface.  On a square lattice, the 
two-dimensional Hubbard model with nearest-neighbor hopping 
has a low-energy description which 
is complicated by several features of its anisotropic nested Fermi surface.
Near half-filling, low-energy umklapp processes are permitted.  
The Fermi speed varies along the
Fermi surface, and there are van Hove singularities in the density of states.  
Due to nesting, the system certainly cannot be a standard Landau Fermi liquid. 
The quasiparticles are expected to have a short lifetime, $\tau \sim 1/T$,
and even the existence of a well-defined Fermi surface 
may be questioned\cite{Rice,Virosztek}.  

Consider the Hamiltonian of the two-dimensional Hubbard model: 
\begin{equation}
H = -t \sum_{<\bx, \by>} c^{\dag \alpha}_\bx c_{\by \alpha} 
+ U \sum_\bx n_{\bx \uparrow} n_{\bx \downarrow}~,
\label{hubbard}
\end{equation}
where $<\bx, \by>$ are nearest neighbors on a square lattice
and $U$ is the on-site interaction strength.
We assume $|U / t| \ll 1$ so that a weak-coupling 
perturbative expansion is justified.  Of course in real systems typically
$|U / t| \gg 1$, but we assume as an operating hypothesis that the behavior
at strong coupling
is qualitatively the same as at weak coupling.  This is equivalent to 
assuming that, at intermediate values of the coupling, no other fixed points 
intervene between the weak and strong coupling regimes.  
To begin, as usual, first we turn off the on-site 
interaction term and discuss the nature of the Fermi surface resulting
from the hopping term alone.  The spectrum of the fermions is
\begin{equation}
\epsilon({\bf k}) = -2t~ (\cos k_x a + \cos k_y a),
\label{spec}
\end{equation}
where $a$ is the lattice spacing, not to be confused with $1/\lambda$, 
the ultraviolet cutoff of multidimensional bosonization.
At half-filling, due to particle-hole symmetry, 
the Fermi surface is the set of points in $\bk$ space with zero energy, and
this set satisfies:
\begin{equation}
\cos k_x a + \cos k_y a = 0 ~.
\label{trace}
\end{equation}
The Fermi velocity depends on $\bf k$ and is given, on the Fermi surface, by
\begin{eqnarray}
{\bf v}_F(\bk) &=& 2 t a~ (\sin k_x a, \sin k_y a)
\nonumber \\
&=& v~ (\sin k_x a, \sin k_y a)
\label{vel}
\end{eqnarray}
where $v \equiv 2 t a$.
Van Hove singularities in the density of states occur at the four corners of
the Fermi surface, ${\bf k} = (\pm \pi / a, 0)$ and 
$\bk = (0, \pm \pi / a)$, as the Fermi velocity vanishes at those points.

\begin{figure}
\epsfxsize=10.0cm \epsfbox{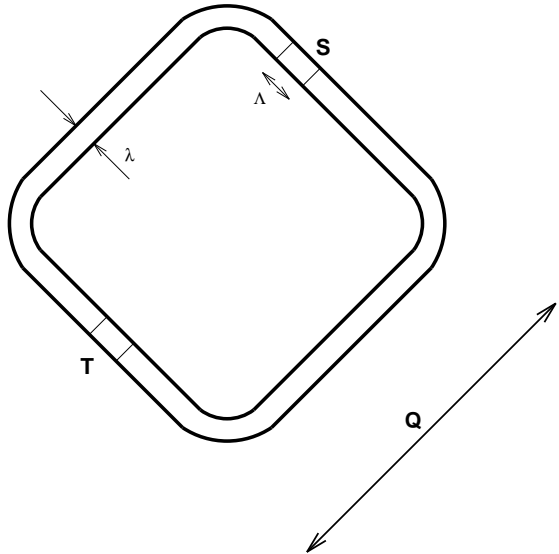}
\label{fig:nest1}
\caption{The low-energy effective theory in the vicinity of the Fermi
surface.  The faces are flat, but the Fermi speed varies along each face.
The corners have been rounded out into quarter-circles of radius
$ \kappa$ to smooth out the Van Hove singularities.
The thin momentum shell is coarse-grained
into squat boxes of height $\lambda(\bs)$ and width $\Lambda$.
Again we require $|\bq| < \lambda(\bs) \ll \Lambda \ll 1/a$ 
for bosonization to work.
Momentum $\bf Q$ is one of four commensurate nesting vectors.}
\end{figure}

Now turn the interaction $U$ back on and perform RG transformations 
until the low energy scale $\epsilon_c$ is reached.   
Assuming that the weak interaction does not alter the shape of the Fermi surface
or change the spectrum qualitatively, we may linearize the
quasiparticle spectrum:
\begin{equation}
\epsilon(\bs; \bq) = v^*~ |\sin (\bk_\bsx a) |~ q_{\parallel}(\bs)~,
\label{renspec}
\end{equation}
which agrees with equation \ref{spec} close to the Fermi surface.
Here $q_{\parallel}(\bs) \equiv \hbn_\bs \cdot (\bk - \bk_\bs)$, where 
as usual $\bk_\bs$ is the Fermi wavevector for patch $\bs$ and $\bk_\bsx$
is its x-component. This Fermi surface is difficult to bosonize since at the
four corners, the van Hove points, the Fermi velocity is zero.  However, 
away from half-filling the van Hove singularities are no longer in the 
low-energy zone.  Therefore we introduce a simple spectrum in which the 
van Hove singularities are smoothed out. 
If we distort the four corners of the Fermi surface into four
quarter-circles with a radius $\kappa$ as shown in 
figure \ref{fig:nest1}), the modified spectrum is given by: 
\begin{equation}
\epsilon(\bs; \bq) = \left\{ \begin{array}{ll} 
v^*~ | \sin (\bk_\bsx a) |~ q_{\parallel}(\bs),\ & 
\kappa < |\bk_\bsx| < \pi - \kappa, \\
\\
v^*~ \sin (\kappa a)~ q_{\parallel}(\bs),\ & {\rm otherwise}
\end{array}\right.
\label{modspec}
\end{equation}
This is a reasonable choice because the number of states near
the van Hove points is of order $\kappa a / \sin(\kappa a) \sim O(1)$, 
which is finite even upon taking the limit $\kappa \rightarrow 0$.
The smoothed quarter-circles contain a fixed fraction of the
total number of states, as do the original van Hove points. 
From now on we set $a = 1$ and $v^* = 1$ for convenience.
Factors such as $1/a$ may be recovered whenever a comparison with
other scales is needed.  For example, with this convention 
the nesting vectors are 
${\bf Q} = \pm (\pi ,\pi )$ and ${\bf Q} = \pm (\pi, -\pi)$ and the
density of states at the Fermi energy, $N^*(0)$,
which is of order $(v^* a)^{-1}$ becomes a factor of order unity.

Now that the Fermi velocity is well-defined over the
entire Fermi surface, the model can be bosonized. 
As before, we coarse-grain the Fermi surface with small squat boxes
of width $\Lambda$ along the Fermi surface
and height 
\begin{equation}
\lambda(\bs) \equiv \frac{\epsilon_c}{v_F(\bs)}
\end{equation}
normal to the Fermi surface, as shown in
figure \ref{fig:nest1}.  The Fermi speed in patch $\bs$ is 
\begin{equation}
v_F(\bs) = \sin(\bk_\bsx)\ .
\end{equation} 
The two momentum scales $\Lambda$ and $\lambda(\bs)$ are taken to be small
so that the inequality $1 \gg \Lambda \gg \lambda(\bs)$ is obeyed in each
patch.  As we stressed in Section \ref{essentials}, these limits are imposed 
to minimize the scattering of fermions outside
of the squat box to conserve the number of fermions in each patch;
the $U(1)$ current algebra equation \ref{current_algebra} holds in this limit. 
The effective action of the bosonized low-energy theory is
\begin{equation}
S[a^*, a] = \sum_{\bq} \sum_{\lambda(\bs)>{\hbn_\bs \cdot \bq}>0}
\int {d\omega \over 2\pi } [i \omega - v_F(\bs){\hbn_\bs \cdot \bq}]~
a^{* \alpha}(\bs; q)~ a_{\alpha}(\bs; q) + S_I[a^*, a] ~.
\end{equation}
Here $a_\alpha(\bs; q)$ and  $a^{* \alpha}(\bs; q)$ are the familiar 
canonical bosonic fields and $S_I$ is the interaction term.
Now consider the renormalization of the fermion interactions.
As in the case of a circular Fermi surface, the marginal interactions
are four-fermion (two-body) interactions, which can be
verified by checking the engineering dimension.
However, unlike the case of a circular Fermi surface, nesting permits additional
marginal scattering channels at large momentum. 
In addition to the BCS instability, these channels 
could induce spin-density-wave (SDW) or charge-density-wave (CDW) instabilities;
but first, before we discuss these logarithmically renormalized channels
we must consider the forward scattering channel.
We recall that, in the case of a circular Fermi surface, the interacting Fermi
system with non-singular forward scattering is a 
Landau Fermi liquid in $D > 1$.  In one dimension, however, forward scattering
has a dramatic effect leading to a Luttinger liquid.
Clearly the nested Fermi surface has some similarities to the
one-dimensional system; it is important, therefore,
to understand the effect of the forward scattering channel
before the large-momentum scattering channels are included perturbatively.

For simplicity, we suppress the momentum dependence of the coupling function
and drop the spin indices. Inclusion of the spin exchange channel is not 
expected to change the result qualitatively.  
The action for the forward scattering channel in the fermion basis is 
\begin{equation}
S_{\rm fs}[\psi^*, \psi]
= -f_0~ \sum_{\bs, \bt}~ \int d^2x~ d\tau~ \psi^*(\bs; \bx, \tau)~
\psi(\bs; \bx, \tau)~ \psi^*(\bt; \bx, \tau)~ \psi(\bt; \bx, \tau)~.
\end{equation}
In bosonized form it is written compactly as
\begin{equation}
S_{\rm fs} = -{f_0\over V}\sum_{\bs ,\bt } \sum_\bq \int {d\omega \over 2 \pi}
J(\bs; q)~ J(\bt; -q)~,
\end{equation}
which is clearly marginal since it is bilinear in the boson fields.
Here as usual $q$ denotes the 2+1 dimensional vector $({\bf q}, \omega)$.
Focusing our attention on the fermions which lie on the 
nested part of the Fermi surface, since
that is where the most singular behavior arises, we obtain the
in-patch boson correlation function
\begin{equation}
\langle a(\bs; q) a^{\dag}(\bs; q) \rangle
= \frac{1}{i \omega - v_F(\bs) \hbn_\bs \cdot \bq} +
{\Lambda \over (2 \pi)^2}~ {{\hbn_\bs \cdot \bq} \over
{(i \omega - v_F(\bs) \hbn_\bs \cdot \bq)^2}}~ \big{[} K_0(q)\big{]}^{-1}~ .
\label{aac}
\end{equation}
After analytic continuation to real frequencies,
\begin{eqnarray}
K_0(q) &=& {1\over f_0}-{2\Lambda \over (2\pi )^2} \sum_\bs 
{\theta(\hbn_\bs \cdot \bq) ~\hbn_\bs \cdot \bq \over
\omega - v_F(\bs) \hbn_\bs \cdot \bq} \nonumber \\
&=& {1\over f_0}+ {\sqrt{2} \over (2\pi )^2 }\Big{[}
(2\pi )^2~ \chi_0(\sin \kappa ~\bf q, \omega)
+ \chi_N(\bq, \omega) \Big{]} ~,
\label{K0}
\end{eqnarray}
where $\chi_N$ is the contribution from the flat part of the Fermi
surface:
\begin{eqnarray}
\chi_N(q) &=& {4\over \sqrt{1-\omega ^2/
q_{\parallel} ^2}} \ln \bigg{[}{\sqrt{1-\omega ^2/q_{\parallel} ^2}
+ \cos \kappa \over \sqrt{1-\omega ^2/q_{\parallel} ^2}
- \cos \kappa} \bigg{]} \nonumber \\
&&+{4\over \sqrt{1 - \omega ^2 /
q_{\perp} ^2}} \ln \bigg{[}{\sqrt{1-\omega ^2/q_{\perp} ^2}
+\cos \kappa \over\sqrt{1-\omega ^2/q_{\perp} ^2}-\cos \kappa }
\bigg{]} ~.
\label{chiN}
\end{eqnarray}
As before $q_{\parallel}$ is the component of $\bf q$ parallel to the
Fermi surface normal $\hbn_\bs$ and $q_{\perp}$ is the perpendicular component.
The contribution to the boson correlation function from the curved part of 
the Fermi surface, the term proportional to $\chi_0$ in
equation \ref{K0}, as shown in Section \ref{propagator}, 
contributes a term to the fermion self-energy, 
${\rm Im} ~\Sigma \sim \omega ^2 \ln \omega$, and is therefore uninteresting.

We can estimate perturbatively the imaginary part of the quasiparticle
self-energy arising from the susceptibility $\chi_N(q)$.
In the limits $|q_\parallel| > |\omega| > |q_\parallel| ~\sin \kappa$
or $|q_{\perp}| > |\omega| > |q_{\perp}| ~\sin \kappa$, the susceptibility
has an imaginary part.  The boson correlation function then can be expanded 
to second order in $f_0$:
\begin{eqnarray}
\langle a(\bs; q) a^{\dag}(\bs; q) \rangle
&\approx & {i\over \omega - v_F(\bs) \hbn_\bs \cdot \bq} 
\nonumber \\
&+& i{\Lambda \over (2\pi)^2}~ {\hbn_\bs \cdot \bq \over
(\omega - v_F(\bs) \hbn_\bs \cdot \bq)^2}~\Big{(}
f_0 - f_0^2 {\sqrt{2} \over (2\pi )^2}~ \chi_N(q)
\Big{)}~.
\label{sopt}
\end{eqnarray}
Just as for the isotropic Fermi liquid of Section \ref{propagator},
the first-order term in $f_0$ renormalizes the Fermi velocity,
and the velocity correction, $\delta v = f_0 \Lambda$, 
is infinitesimal in the $\Lambda \rightarrow 0$ limit.  
Spin-charge velocity separation therefore does not
occur, as can be verified easily by inclusion of the spin indices. 
Inserting equation \ref{sopt} into equation \ref{back-transformed}, 
exponentiating, expanding the
exponential to second order in $f_0$, and finally performing a
Fourier transform into $({\bf k}, \omega)$ space gives the
imaginary part of the quasiparticle self-energy\cite{HKM},
${\rm Im} ~\Sigma ({\bf k}, \omega) \approx
\omega~ f_0 ^2 \Lambda$, which has the marginal Fermi liquid
form\cite{MFL}.  Like the $D=3$ case of a spherical Fermi surface, the 
prefactor $\Lambda$ which appears in the self-energy obtained by bosonization
is replaced by a number 
of order unity in an RPA calculation of the self-energy.  
Corrections at higher orders in $f_0$ can be
estimated and are also consistent with the marginal Fermi
liquid form, in agreement with expectations of others who found that 
the inverse quasiparticle lifetime, $1/\tau$, is proportional to 
the temperature\cite{Virosztek}.
Since the quasiparticle weight in a marginal Fermi liquid
vanishes logarithmically at the Fermi surface,
the Fermi liquid fixed point breaks down but the occupancy still changes
rapidly enough near the Fermi surface for 
multidimensional bosonization to be applied.

Now we include the large-momentum scattering channels
by the perturbative RG method.  We can 
discard the forward scattering channel since we have shown that it
contributes only sub-leading corrections to the RG equations.
We study the RG flow of the BCS channel 
and derive an analog of the Landau theorem\cite{Lifshitz}.
Like the circular Fermi surface, the nested Fermi surface has reflection
symmetry about the origin and therefore the BCS channel flows logarithmically
with decreasing energy scale.
For a circular Fermi surface, the Landau theorem states that there is a
BCS instability if any one of the $V_\ell$ is negative, where
$V_\ell$ is the $\ell$-th orbital angular momentum component of the BCS vertex
function defined in equation \ref{Nangl}.  However, for nested Fermi surfaces, 
the BCS vertex is not rotationally invariant, and
$V_{BCS}(\bs, \bt)$ must be expanded in a double Fourier series\cite{Baranov}
and the Fourier coefficients are now coupled.  
The RG equation for the BCS flows,   
equation \ref{BCS-beta}, is also modified as the Fermi speed $v_F(\bu)$ 
is not uniform.
\begin{equation}
{dV_{BCS}(\bs, \bt)\over d\ln(s)}= -{\Lambda \over (2\pi )^2}~ \sum_\bu~
\frac{V_{BCS}(\bs, \bu)~ V_{BCS}(\bu, \bt)}{v_F(\bu)}~.
\label{RGBCS2}
\end{equation}
Note that the weight function $v_F(\bu)^{-1}$ now appears in the sum. 

In order to analyze the RG equation, we must Fourier-decompose the
BCS vertex function.  The vertex function has $D_4$ symmetry, and there are five
sets of basis functions which generate the two-dimensional
function space to which it belongs.  The five sets are: 
$A_1$ and $A_2$ ($s$-wave), $B_1$ and $B_2$ ($d$-wave),
and $E$ ($p$-wave)\cite{Baranov}.
The basis functions for each set are as follows:
\begin{eqnarray}
A_1 &:&~ a_1(n, {\bf S}) \equiv  \bigg{[}{v_F({\bf S})\over 2\pi \sqrt{2}}
\bigg{]}^{1/2}~\cos (2n \bk_\bsx)~,\ \ a_1(0, {\bf S})
\equiv \bigg{[}{v_F({\bf S})\over 4\pi \sqrt{2}}\bigg{]}^{1/2}~,
\nonumber \\
A_2 &:&~a_2(n, {\bf S}) \equiv {\rm sgn} (\bk_\bsy)~ \bigg{[}{v_F({\bf S})
\over 2 \pi \sqrt{2}}\bigg{]}^{1/2}~\sin (2n \bk_\bsx) ~,\nonumber \\
B_1 &:&~ b_1(n, {\bf S}) \equiv  \bigg{[}{v_F({\bf S})
\over 2 \pi \sqrt{2}}\bigg{]}^{1/2}~ \cos [(2n+1) \bk_\bsx]~ , \nonumber \\
B_2 &:&~ b_2(n, {\bf S}) \equiv {\rm sgn} (\bk_\bsy)~ \bigg{[}{v_F({\bf S})
\over 2\pi \sqrt{2}}\bigg{]}^{1/2} ~ \sin [(2n+1) \bk_\bsx]~ ,
\nonumber \\
E &:&~e(n, {\bf S}) \equiv \bigg{[}{v_F({\bf S})
\over 2 \pi \sqrt{2}}\bigg{]}^{1/2} {}~\bigg{(} c_1~\sin [(n+1/2) \bk_\bsx] 
\nonumber \\
&+& c_2~{\rm sgn} (\bk_\bsy)~ \cos [(n+1/2) \bk_\bsx] \bigg{)}\ .
\label{modbas}
\end{eqnarray}
A factor of $\sqrt{v_F({\bf S})}$ has been inserted to render
the functions orthonormal under summation over the Fermi surface
with the weight function $v_F(\bs)^{-1}$.
In general the BCS vertex function can be expressed as a double Fourier 
expansion in terms of these orthonormal basis functions; there are 
no cross terms between basis functions of different symmetry:
\begin{eqnarray}
V_{BCS}({\bf S, T}) &=& \sum_{n,m} \Big{\{}
C_1^{nm}~ a_1(n, {\bf S})~ a_1(m, {\bf T})
+C_2^{nm}~ a_2(n, {\bf S})~ a_2(m, {\bf T}) \nonumber \\
&&+C_3^{nm}~ b_1(n, {\bf S})~ b_1(m, {\bf T})
+C_4^{nm}~ b_2(n, {\bf S})~ b_2(m, {\bf T}) \nonumber \\
&&+C_5^{nm}~ e(n, {\bf S})~ e(m, {\bf T}) \Big{\}}~.
\label{GFE}
\end{eqnarray}
This expansion is valid as long as $v_F({\bf S}) \neq 0$
which is the case for the model considered here as the Van Hove singularities
have been smoothed out.

Inserting equation \ref{GFE} into equation \ref{RGBCS2}
the RG equations for the Fourier coefficients are found:
\begin{equation}
{d C_i^{nm}\over d \ln(s)} = -~{1\over (2\pi )^2}
\sum_{l=0}^{\infty} C_i^{nl}~C_i^{lm}
\end{equation}
where $i=1,...,5$ labels the 5 sets of coefficients.  The analog of the Landau 
theorem for the nested fermion liquid is obtained by considering the diagonal
Fourier coefficients which obey the equations
\begin{equation}
{d C_i^{nn}\over d \ln(s)} = -~{1\over (2\pi )^2}
\sum_{l=0}^{\infty} |C_i^{nl}|^2~.
\end{equation}
The diagonal coefficients decrease as the length scale $s$ increases, and
therefore, if any one value of $C_i^{nn}$ is negative initially,
a BCS instability occurs.  This sufficient condition 
for an instability allows us to identify the possible symmetry
of the resulting order parameter, since the gap function $\Delta(\bs)$
is related to the BCS vertex function as usual by:
\begin{equation}
\Delta(\bs)= -\sum_{\bt} V_{BCS}(\bs, \bt) {{\Delta(\bt)}\over{2E(\bk_\bt)}}~,
\end{equation}
where $E({\bf k})$ is the dispersion of the low-lying quasiparticle
excitations in the superconducting state.

With these criteria we can study the possibility of a
BCS instability in the presence of a repulsive Hubbard-$U$ interaction.
First, we consider the RG flow of the channels generated directly by the
repulsive-$U$ interaction.
We consider only those channels which flow logarithmically in length scale $s$,
as all other channels are irrelevant.  In addition to the BCS channel,
there are three other classes of interaction channel which renormalize
logarithmically in the presence of nesting.  The first class is the
CDW or SDW channel, which we denote simply by DW (density wave). 
In the physically relevant case of a repulsive bare Hubbard
interaction, the DW channel leads to a SDW instability.
However, this channel is decoupled from the BCS channel. 
The second and third classes are the $\bQ = {\bf 0}$ and the 
umklapp $\bQ \neq {\bf 0}$ repulsive exchange 
channels which do renormalize the BCS channel, and which therefore can
induce superconducting instabilities.  

\begin{figure}
\center
\epsfxsize = 6.0cm \epsfbox{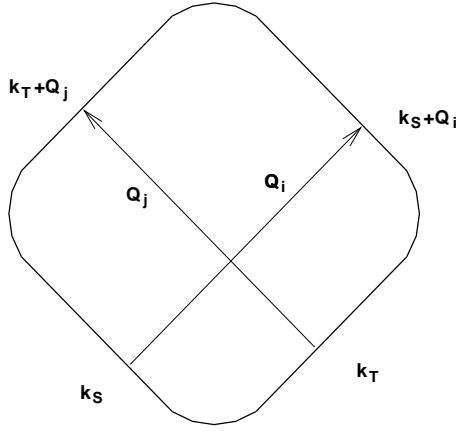}
\caption{A typical density-wave scattering channel which is permitted
because $\bQ_i + \bQ_j$ is a reciprocal lattice wavevector.}
\label{fig:nest2}
\end{figure}

Expressed in terms of the fermion fields, the action for the DW channel is: 
\begin{eqnarray}
S_{DW}[\psi^*, \psi] &=&
\frac{f_{DW}}{2}~ \int d\tau~ d^2x~ \sum_{\bs, \bt} \sum_{i=1,j=1}^4
\psi^{* \alpha}(\bs; x)~ \psi_{\beta}(\bs+\bQ_i; x)~
\nonumber \\
&\times& \psi^{* \gamma}(\bt; x)~ \psi_{\delta}(\bt+\bQ_j; x)~
[\delta ^{\beta}_{\alpha}\delta ^{\delta}_{\gamma}-\delta ^{\beta}
_{\gamma}\delta ^{\delta}_{\alpha} ]~.
\label{DW}
\end{eqnarray}
Here $\bQ_i$ and $\bQ_j$ are two of the four nesting vectors shown in  
figure \ref{fig:nest2}, $i, j = 1, 2, 3, 4$, 
and the notation $\bs + \bQ_i$ is shorthand for momentum $\bk_\bs + \bQ_i$.
Umklapp scattering, $\bQ_i + \bQ_j \neq {\bf 0}$, is permitted because
$\bQ_i + \bQ_j$ is a reciprocal lattice wavevector.
The spin indices in equation \ref{DW} have been 
symmetrized consistent with Fermi statistics\cite{Chitov}.
Due to nesting, this channel flows logarithmically with length scale
and leads to either a CDW or a SDW instability, depending on the sign of
$f_{DW}$.  The action can be decoupled into CDW and SDW channels,
\begin{equation}
S_{DW} = {1\over 2}\int d\tau d^2x~
{\sum_{\bs, \bt}}^\prime \sum_{i=1,j=1}^4
\big{\{} f_{SDW}~ \vec{S}(\bs, \bQ_i; x) \cdot \vec{S}(\bt, \bQ_j; x)
+ f_{CDW}~ n(\bs, \bQ_i; x)~ n(\bt, \bQ_j; x) \big{\}}~,
\label{Hnest}
\end{equation}
where
\begin{equation}
\vec{S}(\bs, \bQ_i; x) = {1\over 2}~\psi^{* \alpha}(\bs; x)~
\vec{\sigma}^{\beta}_{\alpha} ~\psi_{\beta}(\bs + \bQ_i; x)
\end{equation}
and
\begin{equation}
n(\bs, \bQ_i; x) = \psi^{* \alpha}(\bs; x)~ 
\psi_{\alpha}(\bs + \bQ_i; x)
\end{equation}
are the spin density and the charge density, respectively.
The primed summation over $\bs, \bt$ indicates
that the sum is only over the flat parts of the Fermi surface.

Consider the RG flow of the CDW term; for simplicity again we
drop the spin indices.
From the bosonization formula, equation \ref{bosonization}, 
the bosonized form of the CDW action is 
\begin{eqnarray}
S_{CDW}[\phi] &=& {f_{CDW} \over 2}~ \frac{\Lambda^4}{(2\pi)^4}~
{\sum_{\bs, \bt}}^\prime \sum_{i=1,j=1}^4  \int d\tau~ d^2x~ 
{\lambda(\bs)~ \lambda(\bt) \over 4}
\nonumber \\&&\times \exp \Big{\{}i {\sqrt{4\pi} \over \Omega}
[ \phi ({\bf S + Q}_i; x)+\phi ({\bf T + Q}_j; x)
-\phi ({\bf S}; x) - \phi ({\bf T}; x) ]
\Big{\}}~ .
\label{SU}
\end{eqnarray}
Under rescaling, $t \rightarrow t^\prime / s$, $x_{\parallel} \rightarrow
x^{\prime}_{\parallel} /s$, and $\epsilon_c \rightarrow s \epsilon_c^\prime$;
as expected $S_{CDW}$ is scale invariant and therefore marginal.
Next the fields are split into high and low-energy modes
$\phi({\bf S}; x) \rightarrow \phi^\prime({\bf S}; x) + h({\bf S}; x)$ 
where, as in Section \ref{stability}, $h(\bs; x)$ 
is the high-energy field with momentum in the range 
$\lambda^\prime(\bs) / 2 < |{\bf p}| < \lambda(\bs) / 2$ 
with $\lambda^\prime(\bs) \equiv \epsilon_c^\prime / v_F(\bs)$.
The correlation function of the $h(\bs; x)$ fields is given by
\begin{equation}
\langle h({\bf S; x}, \tau)~ h({\bf T; 0}, 0)\rangle = \left\{ \begin{array}{ll}
\delta_{\bs,\bt}^{D-1}~{\Omega ^2\over 4\pi}~\ln \bigg{(}
{{\hbn_\bs \cdot \bx + i v_F(\bs) \tau + 2is/\lambda(\bs)}\over
{\hbn_\bs \cdot \bx + i v_F(\bs) \tau + 2i/\lambda(\bs)}}\bigg{)}~, & \quad~
|{\bf \hat{n}_S \times x}| \Lambda \ll 1  \\
\\
0~, & \quad |{\bf \hat{n}_S \times x}| \Lambda \gg 1~,
\end{array}\right.
\label{hcorr}
\end{equation}
similar to equation \ref{Nhh}, but with explicit patch-dependence
in the speed $v_F(\bs)$ and in the cutoff $\lambda(\bs)$.
We then determine how the partition function renormalizes as the fast
modes are integrated out:
\begin{eqnarray}
\langle \exp \big{[} -S_{CDW} \big{]} \rangle &=& 
\langle 1 - S_{CDW} + {1\over 2} S_{CDW}^2 + \ldots \rangle 
\nonumber \\
&=& \exp \big{[} -S_{CDW}^\prime \big{]}~.
\label{expsu}
\end{eqnarray}
Here the average in the brackets is taken over only the high-energy fields 
$h(\bs; x)$ and $S_{CDW}^\prime$ is the renormalized CDW action.
We evaluate $\langle S_{CDW} \rangle$ using 
equation \ref{Nhh0} and the formula
$\langle e^{ih} \rangle = e^ {-\langle h^2/2 \rangle}$ to obtain
\begin{eqnarray}
\langle S_{CDW} \rangle &=&{f_{CDW}\over 2}\big{(}{ \Lambda \over 
(2\pi)^2}\big{)}^2 {\sum_{\bs, \bt}}^\prime \sum_{i=1,j=1}^4 
\int d\tau~ d^2x~ {\lambda(\bs)~\lambda(\bt)\over 4s^2}
\nonumber \\&&\times \exp \Big{\{}i {\sqrt{4\pi} \over \Omega}
[ \phi^\prime({\bf S+Q}_i; x) + \phi^\prime({\bf T+Q}_j; x) -
\phi^\prime({\bf S}; x) - \phi^\prime({\bf T} ;x) ]
\Big{\}}~ .
\end{eqnarray}
Replacing $\lambda(\bs) $ with $s\lambda ^\prime(\bs)$ 
[$\lambda^\prime(\bs)$ is the new momentum cutoff] and $\phi^\prime 
\rightarrow \phi$ we recover equation \ref{SU} with the new cutoff.
The third term on the right hand side of equation \ref{expsu}
gives the second order correction to $S_{CDW}$.
Using the relation $\langle e^A~e^B\rangle = e^{\langle
AB+(A^2+B^2)/2\rangle }$ and equation \ref{hcorr}, we find
\begin{eqnarray}
&&{1\over 2}\Big{[}\langle S_{CDW} ^2\rangle -\langle S_{CDW}\rangle ^2 \Big{]}
= {f^2_{CDW} \over 2}\bigg{[}{ \Lambda  \over (2\pi )^2}\bigg{]}^4
{\sum_{\bs, \bt, \bu}}^\prime \sum_{i,j,l} \int d\tau~ d^2x~
{\lambda(\bs)~\lambda(\bt)~\lambda^2(\bu) \over 16s^4} 
\nonumber \\
&& \times
\exp \Big{\{} i{\sqrt{4\pi} \over \Omega}
[\phi^\prime({\bf S + Q}_i;x) + \phi^\prime({\bf T + Q}_j;x)
-\phi^\prime({\bf k}; x) - \phi^\prime({\bf T}; x) ]
\Big{\}} \nonumber \\
&&\times\int d\sigma~ d^2y~ \exp \Big{\{} {{4\pi}\over \Omega ^2}
\big{[}\langle h({\bf U}; {\bf y}, \sigma)~ h({\bf U}; {\bf 0}, 0)\rangle
+\langle h({\bf U + Q}_l; {\bf y}, \sigma)~ h({\bf U + Q}_l; {\bf 0}, 0)
\rangle \big{]}\Big{\}} \nonumber \\
&&+~{\rm subleading ~terms} \nonumber \\
&\approx & {f^2_{CDW}\over 2}\bigg{[}{ \Lambda \over (2\pi)^2}\bigg{]}^4
{\sum_{\bs, \bt}}^\prime \sum_{i=1,j=1}^4 \int d\tau~ d^2x~
{\sum_\bu}^\prime {\lambda^2(\bu)\over 4s^2}
{2\pi \over \Lambda}\int d\sigma~ d({\bf \hat{n}_U \cdot y})~ 
{4s^2/\lambda^2(\bu)
\over [v_F(\bu) \sigma]^2 + ({\bf \hat{n}_U \cdot y})^2} \nonumber \\
&&\times {\lambda(\bs)~\lambda(\bt)\over 4s^2}
{}~\exp\Big{\{} i{\sqrt{4\pi} \over \Omega}
[\phi^\prime({\bf S + Q}_i; x) + \phi^\prime({\bf T + Q}_j; x) 
- \phi^\prime({\bf S}; x) - \phi^\prime({\bf T}; x) ]
\Big{\}} \nonumber \\
&&+~{\rm subleading~terms}~.
\end{eqnarray}
Performing the integrals over $\bf \hat{n}_U \cdot y$ and $\sigma$, 
we obtain the RG equation for the CDW coupling function
\begin{equation}
{d f_{CDW}(s)\over d \ln(s)}= - ~{\Lambda \over 2(2\pi )^2 }
{\sum_\bu}^\prime {1\over v_F(\bu)}~ f_{CDW}^2(s) ~.
\end{equation}
This channel is unstable if $f_{CDW}$
is negative, which is the case for a bare attractive interaction, 
$U < 0$, leading to 
charge-density ordering at wavevector $(\pi ,\pi)$ or $(\pi, -\pi)$.

Similarly, a RG transformation can be performed to find the
flow equation in the SDW channel yielding
\begin{equation}
{d f_{SDW}(s)\over d \ln(s)}= - ~{\Lambda \over (2\pi )^2 }
{\sum_\bu}^\prime {1\over v_F(\bu)}~ f_{SDW}^2(s)~ .
\label{RGFS}
\end{equation}
There is an instability if $f_{SDW} < 0$, which
corresponds to a bare repulsive interaction $U > 0$.
The SDW instability leads to antiferromagnetic spin-density ordering, also with
wavevectors $(\pi, \pi)$ and $(\pi,-\pi)$ as the repulsive
interaction induces an antiferromagnetic exchange interaction.
We can determine easily the energy scale of the SDW instability.
Integrating equation \ref{RGFS}, we obtain the critical energy at which the 
SDW instability occurs by equating it to the energy scale at which the coupling
constant $f_{SDW}$ grows to be of order one:
\begin{equation}
E_{SDW} \sim E_F ~\exp \bigg{[}- ~{1\over {8\sqrt{2}\over (2\pi)^2}
\ln {2\over \epsilon_c}~|f_{SDW}|} \bigg{]}~.
\label{ESDW}
\end{equation}

\begin{figure}
\center
\epsfxsize = 6.0cm \epsfbox{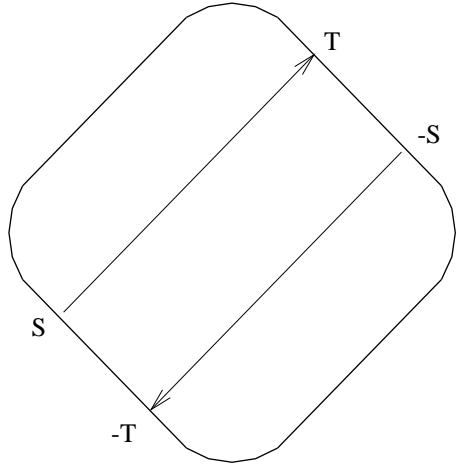}
\caption{A typical overlap between density-wave and BCS channels.}
\label{fig:nest3}
\end{figure}

Now we can examine the overlap between the BCS and the DW channels.
In figure \ref{fig:nest3} we show a typical scattering process 
belonging to both channels.
The contribution of these processes to the action can be written in terms of 
the BCS vertex function and the fermion fields as:
\begin{equation}
S_{\rm overlap} = \sum_{\bs, \bt} \int d\tau~ d^2x~ V_{BCS}({\bf S, T})~
\psi^{* \alpha}({\bf S}; x)~ \psi^{* \beta}({\bf -S}; x)~
\psi_{\beta}({\bf -T}; x)~\psi_{\alpha}({\bf T}; x)~
\delta^2_{\bf T, S+Q}~.
\end{equation}
Clearly, due to the factor of $\delta^2_{\bf T, S+Q}$, there is a severe
phase space restriction in the overlapping channel, which
gives rise to an overall factor of $\Lambda a \ll 1$. Therefore, 
at leading order the DW channel is decoupled from the BCS channel.

\begin{figure}
\begin{multicols}{2}
\center
\epsfxsize = 7.0cm \epsfbox{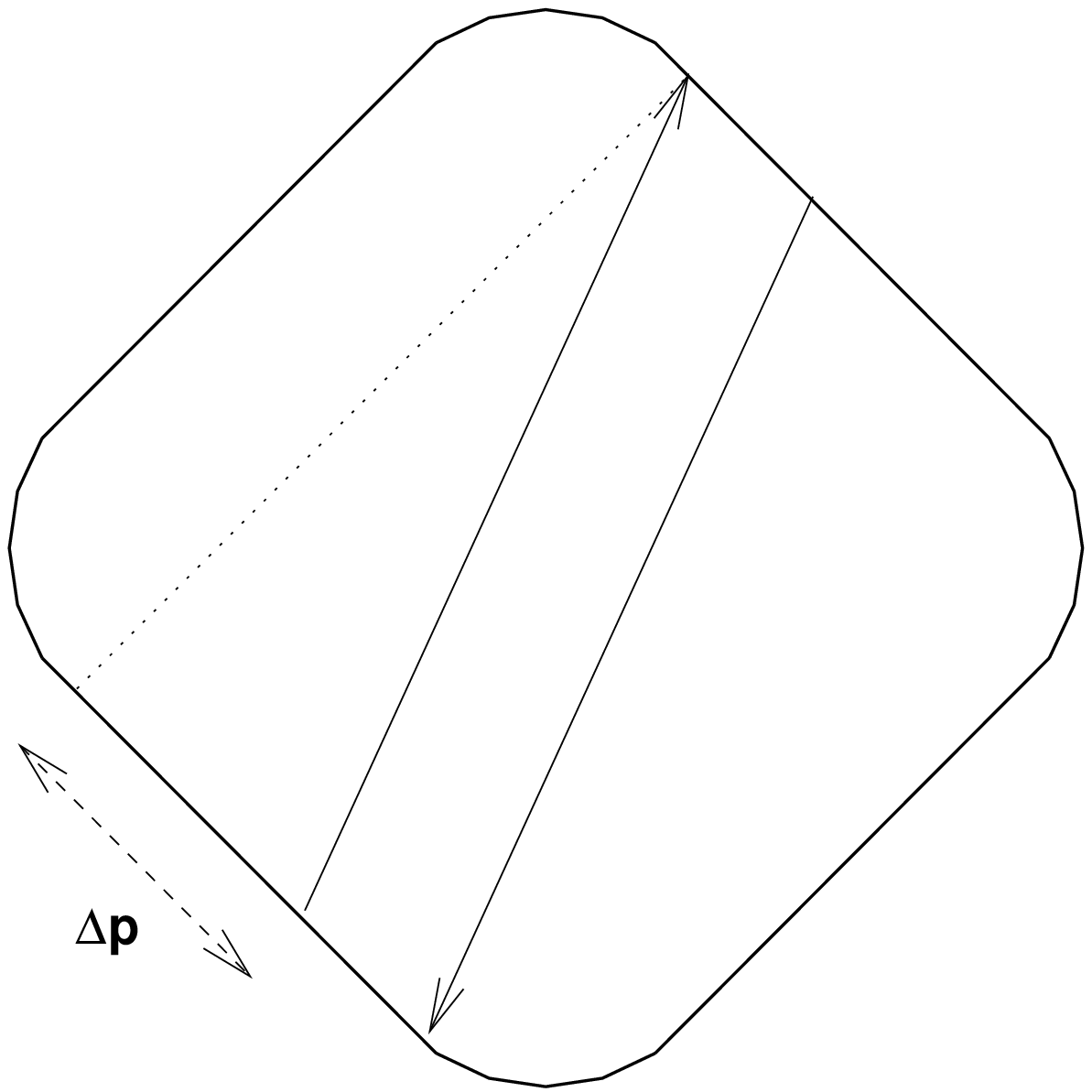}
\epsfxsize = 7.0cm \epsfbox{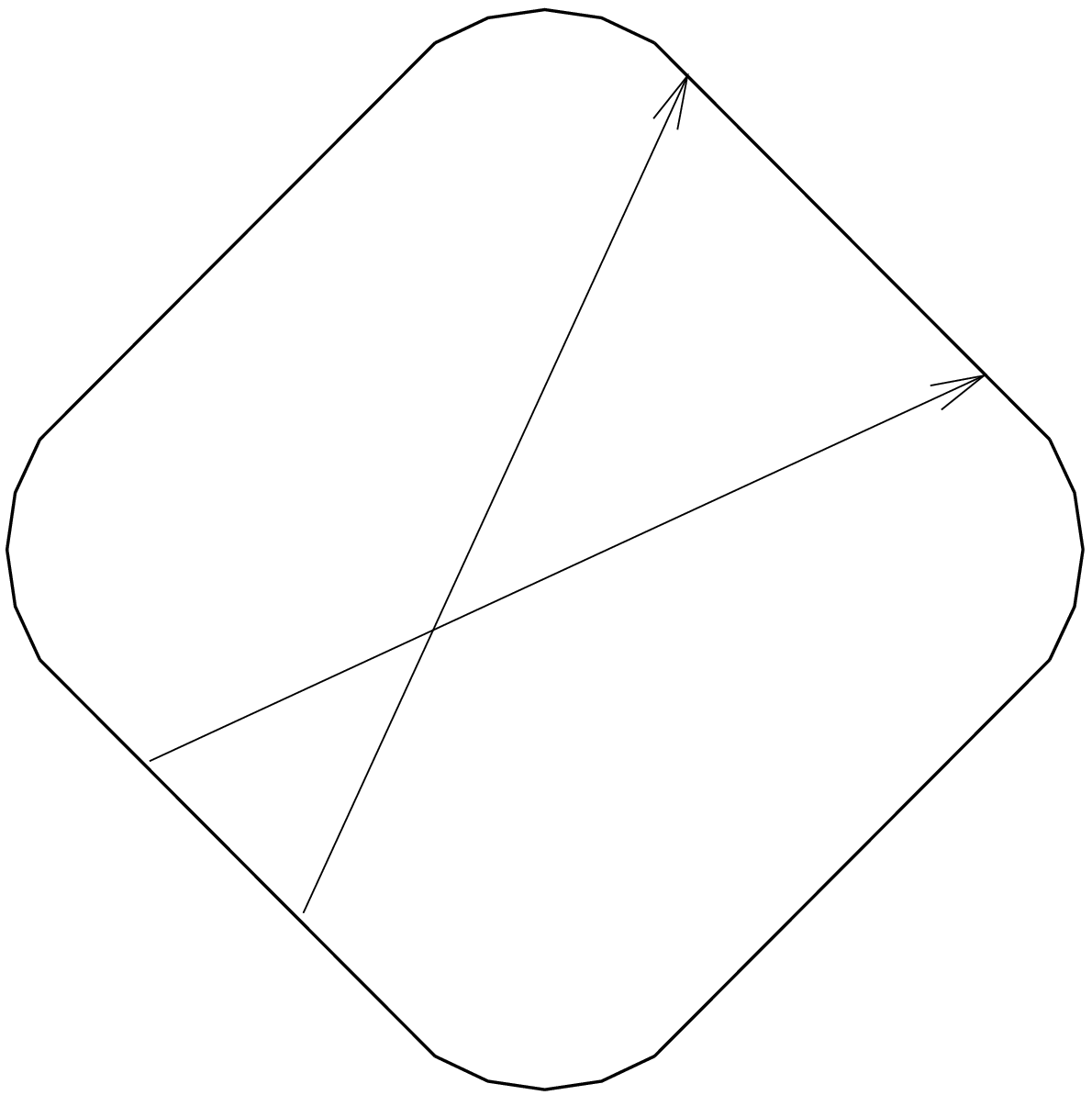}
\end{multicols}
\caption{(a) An exchange channel with zero momentum
transfer which logarithmically renormalizes the BCS channel. (b) Same,
but with momentum transfer $2 \bQ$.  This is an umklapp process.}
\label{fig:nest4}
\end{figure}

The two classes of repulsive channels which do renormalize the BCS channel are
shown in figure \ref{fig:nest4}. 
Their contributions to the action, written in terms of 
the fermion variables, are respectively:
\begin{eqnarray}
S_R[\psi^{* \alpha}, \psi_\beta] &=& 
{f_R \over 2}~ \sum_{\bs, \bt} \sum_{i=1}^4 
\sum_{\bf \Delta p} \int d\tau~ d^2x~
\psi^{* \alpha}(\bs; x)~ \psi_{\beta}({\bf S + \Delta p + Q}_i ;x) 
\nonumber \\
&&\times
\psi^{* \gamma}({\bf T}; x)~ \psi_{\delta}({\bf T - \Delta p - Q}_i; x)~
[\delta ^{\beta}_{\alpha} \delta ^{\delta}_{\gamma} -\delta ^{\beta}_{\gamma}
\delta ^{\delta}_{\alpha}] ~ ,
\end{eqnarray}
which is a zero momentum process and
\begin{eqnarray}
S_U[\psi^{* \alpha}, \psi_\beta] &=& 
{f_U \over 2}~ \sum_{\bs, \bt} \sum_{i=1}^4 \sum_{\bf \Delta p}
\int d\tau~ d^2x~ \psi^{* \alpha}({\bf S}; x)~ 
\psi _{\beta}({\bf S + \Delta p + Q}_i; x) \nonumber \\
&&\times
\psi^{* \gamma}(\bt; x)~ \psi_{\delta}({\bf T - \Delta p + Q}_i; x)~
[\delta ^{\beta}_{\alpha} \delta ^{\delta}_{\gamma} -\delta ^{\beta}_{\gamma}
\delta ^{\delta}_{\alpha}] ~ ,
\end{eqnarray}
which is an umklapp process permitted as $2 \bQ_i$ is a commensurate
wavevector.  In both channels we assume the coupling functions to be
momentum-independent and we anti-symmetrize the spin indices.
Now we expand the partition function to second order in $S_R + S_U$,
integrate out the high-energy modes, and study the resulting
effective action generated by this procedure.
In the course of this RG procedure, both $S_R$ and $S_U$ are renormalized
as well as the BCS action, and the resulting flow equations are a
complicated set of coupled differential equations.  As we are only interested in
the initial growth of the BCS channel, we do not need to perform a complete
analysis of the flow equations, we need 
only examine how $S_R + S_U$ renormalizes the BCS channel
in the initial stages of its growth near $s=1$.  The second-order correction 
to the BCS channel is given by:
\begin{eqnarray}
{1\over 2}\langle (S_R + S_U)^2 \rangle
&\approx& \ln(s)~ (f_R^2 + 4 f_U^2)\sum_\bs \sum_{i=1}^4 \sum_{\Delta \bf  p}
\int d\tau~ d^2x~ \psi^{* \alpha}({\bf S}; x)~ \psi^{* \beta}({\bf -S}; x)
\nonumber \\
&&\times
\psi_{\beta}({\bf S + \Delta p + Q}_i; x)~
\psi_{\alpha}({\bf -S - \Delta p - Q}_i; x) \nonumber \\
&&\times {\sqrt{2}\over 2(2\pi )^2}~{1\over \cos \Delta \bp_x/2}
\ln \Big{(} {1+ \cos (\kappa + |\Delta \bp_x|/2)\over
1 - \cos(\kappa +|\Delta \bp_x|/2)} \Big{)}
\nonumber \\
&& + {\ln(s) \over 2}~ f_R^2~ \Big{[} {\sqrt{2}\over (2\pi )^2}
\ln \Big{(} {1 + \cos \kappa \over
1 - \cos \kappa} \Big{)} + {1\over 2\pi} \Big{]}
\nonumber \\
&&\times \sum_{\bs, \bt} \int d\tau~ d^2x~ 
\psi^{* \alpha}({\bf S}; x)~ \psi^{* \beta}({\bf -S}; x)~
\psi_{\beta}({\bf -T}; x)~ \psi_{\alpha}({\bf T}; x) 
\nonumber \\
&&+~ \mbox{non-BCS~terms}~.
\label{crtn2}
\end{eqnarray}
The second term in equation \ref{crtn2} governs the flow of the  
$s$-wave BCS channel but since the $s$-wave channel is repulsive at 
first order it is not a candidate for instability.  However, 
the first term in equation \ref{crtn2}, as it has strong wavevector
dependence, contributes to the RG flow in all of the other BCS channels. 
The flow equations are:
\begin{eqnarray}
{d V_{BCS}({\bf S, T}) \over d \ln(s)} &=&
(f_R^2 + 4f_U^2){\sqrt{2}\over 2 (2\pi )^2}~
{1\over \cos \Delta \bp_x(\bs,\bt)/2}
\ln \Big{(} {1 + \cos (\kappa + |\Delta \bp_x(\bs,\bt)|/2)\over
1 - \cos (\kappa +|\Delta \bp_x(\bs,\bt)|/2)} \Big{)}
\nonumber \\
&& -~{\Lambda \over (2\pi)^2} \sum_\bu {1\over v_F(\bu)}~
V_{BCS}({\bf S, U})~ V_{BCS}({\bf U, T})~,
\label{flowEQ}
\end{eqnarray}
if $\bf S$ and $\bf T$ are in the same quadrant in the $(k_x, k_y)$
plane.  On the other hand, when $\bf S$ and $\bf T$ are in different 
quadrants, 
\begin{equation}
{d V_{BCS}({\bf S, T}) \over d \ln(s)} = -~{\Lambda \over (2\pi)^2}
\sum_\bu {1\over v_F(\bu)}~ V_{BCS}({\bf S, U})~ V_{BCS}({\bf U, T})~.
\end{equation}

Finally, we can determine which BCS channels are attractive
by evaluating the generalized Fourier coefficients of the
double expansion shown in equation \ref{GFE}.
After performing the generalized Fourier transform of the RHS
of equation \ref{flowEQ} we find that, among the fundamental modes, 
there are two diagonal Fourier coefficients which flow to negative values.
Their rate of change, evaluated for the case $\kappa = 0.01$, is:
\begin{eqnarray}
{dC_2^{11}\over d \ln(s)} &=& \Lambda ^2\sum_{\bs,\bt}
{}~{1\over v_F({\bf S})~v_F({\bf T})}~
{dV_{BCS}({\bf S, T})\over d \ln(s)}~a_2(1, {\bf S})~ a_2(1, {\bf T})
\nonumber \\
&\approx &-0.1~(f_R^2+4~f_U^2)-|C_2^{11}|^2/(2\pi)^2~,
\\
{dC_3^{00}\over d \ln(s)} &=& \Lambda ^2 \sum_{\bs,\bt}
{}~{1\over v_F({\bf S})~ v_F({\bf T})}~
{dV_{BCS}({\bf S, T})\over d \ln(s)}~ b_1(0, {\bf S})~ b_1(0, {\bf T})
\nonumber \\
&\approx &-0.23~(f_R^2+4~f_U^2)-|C_3^{00}|^2/(2\pi)^2 ~.
\end{eqnarray}
Here we have set $C_2^{11}(s=1) = C_3^{00}(s=1) = 0$ initially.
From the Landau theorem we infer that there can be 
a BCS instability in either the $B_1$ ($d_{x^2-y^2}$-wave) or $A_2$ 
(extended $s$-wave) channels provided the SDW instability does not preempt it.
Furthermore, as the inequality 
\begin{equation}
\left| {dC_3^{00} \over d \ln(s)} \right| > 
\left| {dC_2^{11} \over d \ln(s)} \right| 
\end{equation}
holds for small values of $\kappa$, 
$d_{x^2-y^2}$-wave superconductivity is preferred.
Superconducting order, if it exists, will most likely possess 
$d_{x^2-y^2}$-wave symmetry.  This conclusion 
is consistent with the expectation that nearest-neighbor
antiferromagnetic exchange favors $d_{x^2-y^2}$-wave superconductivity.
It also agrees qualitatively with a diagrammatic 
approach\cite{Ruvalds} and a parquet solution
obtained for the flat Fermi surface\cite{parquet}.

Of course, the tendency towards spin-density order dominates all other
instabilities close to half-filling.  But slightly below half-filling,
because nesting is not perfect but nonetheless effective, 
there is competition between the SDW and BCS instabilities.
As the energy cutoff $\epsilon_c$ 
around the Fermi surface is reduced via the RG
transformation towards $\epsilon_s$, where $\epsilon_s$
is the energy deviation of the actual
Fermi surface from perfect nesting, the SDW channel stops
flowing but the BCS channel continues to renormalize logarithmically
as its flow is independent of nesting.  
If the initial SDW coupling is small enough 
or if $\epsilon_s$ is large enough (the 
SDW channel does not develop an instability
until $\epsilon_c \rightarrow \epsilon_s$) SDW order will not occur.  
Instead, superconductivity will set in at a sufficiently low temperature.

\section{Conclusion}
\label{conclusion}
\setcounter{equation}{0}
In this review we have shown that multidimensional bosonization, when
combined with the Renormalization Group picture, is a powerful tool for the 
study of strongly interacting fermion systems.  First, 
an effective action for quasiparticles
is derived by integrating out the high energy degrees of freedom
which lie outside of the concentric shells which encircle the Fermi surface. 
The renormalization process is stopped before all of the degrees of freedom
are eliminated; the effective Hamiltonian so obtained describes quasiparticles
confined to a narrow region around the Fermi surface of width 
$\lambda \ll k_F$.  Of course no attempt is made to calculate the values 
of parameters such as the effective mass $m^*$ or the coupling constants
for the short-ranged part of the forward and exchange interactions as this can
only be done reliably in special limits, for example the case of a very weak 
bare interaction.
The Fermi surface is now discretized, tiled into squat boxes with
aspect ratio $\lambda \ll \Lambda$, where $\Lambda$ is the length of the
box along the surface.  In each squat box, currents defined in terms of the 
fermionic quasiparticles also 
have an equivalent representation in terms of bosonic
fields.  When it is legitimate to linearize the quasiparticle spectrum 
about the Fermi momentum by dropping the non-linear terms, the complicated  
many-body interactions reduce to a quadratic form in the boson fields
which possesses the large $U(1)^\infty_k$ symmetry, 
assuming that there is no BCS instability.  Given these caveats, the 
boson propagator can be determined exactly from a generating functional; 
the fermion propagator then follows on using the exponential representation
of the fermions in terms of the bosons.  

As the existence of Landau quasiparticles is not presupposed, multidimensional
bosonization is a useful nonperturbative approach to access non-trivial 
ground states of strongly correlated fermion systems and to study their 
stability.  In one spatial dimension, well-known exact results are recovered.
The ground state is generically a Luttinger liquid, and the low-lying
excitations are collective modes of charge and spin which propagate at 
different velocities.  In higher dimension the Landau Fermi liquid is realized
when fermions interact via a longitudinal interaction of either short-range,
or of the usual Coulomb form.  Unphysically longer ranged interactions,
however, can destroy the Landau quasiparticles.  All of the important 
results of Fermi liquid theory can be recovered within bosonization, the
nonanalytic $T^3 \ln T$ contribution to the specific heat in three dimensions,
the equations of motion of charge and spin collective modes, as well as
screening and
quasiparticle damping in two and three dimensions.  Both long and short
wavelength physics can be described provided the energy is low enough to 
be within a shell surrounding the Fermi surface.  For example the density
response has the usual RPA form at low wavevectors, and at high wavevectors
near $2 k_F$ the Kohn nonanalyticity is found. 

The possibility of non-Fermi liquid ground states has been raised for systems
interacting via transverse gauge fields, as these fields are not screened
at low frequencies.  Such a scenario has been suggested as the mechanism
underlying cuprate superconductivity, but a concrete example is furnished by
quantum Hall systems in two dimensions, specifically a two-dimensional 
electron gas in a perpendicular magnetic field such that only the lowest
Landau level is half-filled.  In the widely used Chern-Simons gauge theory
the physical electrons are mapped by a singular gauge transformation onto
composite fermions which each carry two units of statistical flux and
interact via a transverse gauge field in addition to the usual longitudinal
density-density interaction.  At the mean field level the average flux cancels
the external magnetic field and the ground state is a free Fermi gas.  
The stability of this state against gauge (= charge) fluctuations was 
studied by determining the composite quasiparticle propagator.  
Multidimensional bosonization broke down for longitudinal interactions
shorter-ranged than Coulomb as the transverse gauge fluctuations, which
are moderated by the longitudinal interaction, are too
strong in this instance.  In the case of the Coulomb interaction, two
important results were found.  At equal time and at long distances the
propagator has the form $G_F(\bs; x_\parallel) \sim |x_\parallel|^{-\zeta}
/ x_\parallel$.  In the opposite limit of long temporal separation, 
$|t| \gg |x_\parallel|$, the leading terms can be Fourier transformed and the
self-energy extracted, yielding the marginal Fermi liquid form,
$\Sigma(\omega) \sim \zeta \omega \ln |\omega|$, identical to that found
within RPA.    

This result has been criticized by several authors and remains controversial.
The basis for the criticism is the linearization of 
the quasiparticle dispersion, an approximation which is especially useful
in bosonization as it leads to a Gaussian bosonic action.  Because gauge
fluctuations carry relatively large transverse momentum, it has been 
argued that non-linear 
corrections to the quasiparticle dispersion should be retained, especially
in the case of short-range longitudinal interactions.  Furthermore,
Altshuler, Ioffe, and Millis\cite{Altshuler} argue that, 
in the case of the Coulomb interaction,
when the quadratic $q_\bot^2/(2 m^*)$ term is retained 
the MFL form of the self-energy given
above is the leading term, and all other higher-order terms are subleading.
This claim has been criticized by Kopietz who points out\cite{Kopietz-Vertex} 
that the two-loop
self-energy is more divergent than the single-loop self-energy.  Kopietz and
Castilla\cite{Kopietz-Castilla} 
attempted to incorporate the $q_\bot^2/2 m^*$ term within a bosonized 
treatment and find that Fermi liquid behavior is recovered, even in the 
case of short-range longitudinal interactions where the gauge fluctuations
should be strongest and most disruptive. 
The nature of the composite quasiparticle propagator, and whether or not it
is even physically meaningful, remains the subject of ongoing research.  
Different, direct, schemes for bosonization of Landau levels have been 
proposed\cite{Westfahn,Vignale}, 
and these may prove useful.  Renormalization group
methods, when combined with the Ward identity approach\cite{Italian-RG} are
another route being pursued.   

A particularly important feature of multidimensional bosonization is that it can
be combined with the renormalization group to analyze various instabilities
of the Fermi liquid, all of which are treated on an equal footing.  
Bosonization is most useful in this context as the forward-scattering
parts of the interaction, those which preserve the $U(1)^\infty_k$ symmetry,
can be incorporated exactly at the outset as part of the Gaussian fixed point.  
As a simple example we have studied the BCS instability for a circular Fermi
surface, and have rederived the RG flow equations found by 
Shankar\cite{Shankar} in the
fermion basis.  For a nested Fermi surface close to half-filling, we studied
the competition (for a repulsive bare interaction) between density wave and
superconducting instabilities.  A pleasing conclusion coming from this
calculation is that the $d_{x^2-y^2}$ superconducting channel is 
favored in agreement with experiments\cite{d-wave}.  Above $T_c$ 
the normal state is a marginal Fermi liquid, also in qualitative
agreement with many experiments\cite{MFLphenomenology}.  Bosonization is 
particularly appropriate here as the forward-scattering channel responsible
for the marginal Fermi liquid state is incorporated non-perturbatively.
Unlike other approaches, in multidimensional bosonization Feynman diagrams
receive rather less emphasis than algebraic and geometric ways of understanding
correlated fermions.  When combined with the renormalization group, 
bosonization yields insight into fermion liquids which does not rely on
the resummation techniques of microscopic Fermi liquid theory\cite{AGD}.  
By unifying the microscopic and Landau pictures, 
multidimensional bosonization provides a new way to think about correlated
fermions.

Finally we should mention some outstanding problems which remain unanswered.
First, as we have noted, linearization of the fermion dispersion is a central
approximation, crucial to the success of the approach.  As pointed out 
originally by Haldane\cite{old-Haldane}, 
nonlinear corrections to the dispersion can only 
be included systematically in a perturbative treatment.  When these terms
are important, it is not clear how useful bosonization remains.  The most
important failing of bosonization to date, however, is that it remains 
unable to 
account for the intermingling effects of interaction and disorder leading to
diffusion and the modification of screening\cite{disorder}.  
To understand the role of
disorder it is necessary to confront the effects of interpatch scattering,
which breaks the $U(1)^\infty_k$ symmetry down to just the usual global $U(1)$
charge conservation symmetry.
Some progress has been made in the special case of one spatial 
dimension where the weak-coupling RG equations have been worked 
out\cite{Giamarchi}.  In higher dimension, forward scattering by 
impurities can be included rather easily\cite{Kopietz_book,Kopietz-disorder}, 
but this
has no effect once the $\lambda/k_F \rightarrow 0$ limit is taken, as required
if multidimensional bosonization is to be rigorous.  In the regime of strong
disorder, Belitz and Kirkpatrick have
used a Hubbard-Stratonovich transformation to decouple the fermion
interaction along with replicated disorder by introducing bosonic 
$Q$-fields\cite{BK}.  They emphasize that the soft, low-energy, modes of 
disordered systems differ qualitatively from the low-energy degrees of 
freedom in the clean limit, as they are diffusive rather than ballistic
in character.  It remains to be seen whether or not a theory can be constructed
which interpolates accurately between the limits of weak and strong disorder.   

\section{Acknowledgments}
\label{ack}
We thank P. A. Bares, K. Bedell, D. Belitz, 
C. Castellani, C. Di Castro, S. Das Sarma, J. P. Eisenstein,
B. I. Halperin, T. Kirkpatrick, G. Kotliar, A. Ludwig, A. Maccarone,
W. Metzner, A. Millis, G. Murthy, 
P. Nelson, R. Shankar, A. Stern, and C. M. Varma for
helpful discussions.  This work was initiated at the Institute
for Theoretical Physics in Santa Barbara.  Additional work was carried out
at the Aspen Center for Physics.
The research was supported in part by the National Science Foundation
through grants DMR-9008239 (A.H.) and DMR-9357613 (J.B.M.) and by a grant
from the Alfred P. Sloan Foundation (J.B.M.).

\appendix
\renewcommand{\theequation}{\thesection.\arabic{equation}}

\section{Calculation of the Boson Correlation Function}
\label{boson_correlation}
\setcounter{equation}{0}

Use equation \ref{represent} to find the equal-time correlation function
\begin{eqnarray}
G_\phi(\bs; \bx) &=& \langle \phi(\bs; \bx) \phi(\bs; {\bf 0})
- \phi^2(\bs; {\bf 0}) \rangle
\nonumber \\
&=& \frac{\Omega}{4 \pi}~ \sum_{\bp, \hbn_\bs \cdot \bp > 0}
\theta(\bs; \bp) \frac{(e^{i \bp \cdot \bx} - 1) e^{-a \hbn_\bs \cdot \bp}}
{| \hbn_\bs \cdot \bp |}
\label{A1}
\end{eqnarray}
where we have introduced the convergence factor $e^{-a \hbn_\bs \cdot \bp}$.
To evaluate the integral over the component of momentum parallel to the surface
normal, $p_\parallel \equiv \hbn_\bs \cdot \bp$, consider first the integral
\begin{equation}
I(x_\parallel) \equiv \int_\alpha^{\lambda/2} \frac{d p_\parallel}{4 \pi}~
\frac{e^{i p_\parallel x_\parallel}}{p_\parallel}\ ,
\label{A2}
\end{equation}
here $\alpha$ is an infrared cutoff which drops out of the calculation
at the end.  Then
\begin{equation}
I(0) = \frac{1}{4\pi} \ln \left[\frac{\lambda}{2 \alpha} \right],
\label{A3}
\end{equation}
\begin{equation}
\frac{d I(x_\parallel)}{d x_\parallel} = 
i~ \int^{\lambda/2}_{\alpha} \frac{dp_\parallel}{4\pi} 
e^{i p_\parallel (x_\parallel + i a)} = -\frac{1}{4\pi (x_\parallel +ia)}
\label{A4}
\end{equation}
and 
\begin{equation}
I(x_\parallel) = - \frac{1}{4\pi} \ln (x_\parallel + ia) +C
\label{A5}
\end{equation}
where, from equation \ref{A3},
\begin{equation}
C = \frac{1}{4\pi} \ln \frac{i a \lambda}{2 \alpha}
\label{A6}
\end{equation}
and therefore
\begin{equation}
I(x_\parallel) = -\frac{1}{4\pi} \ln \left[\frac{2\alpha}{\lambda}
\frac{x_\parallel + ia}{ia} \right].
\label{A7}
\end{equation}
Now consider the integrals over the momenta perpendicular to the surface
normal, specializing to the case $D = 2$ for concreteness:
\begin{eqnarray}
\int^{\Lambda/2}_{-\Lambda/2} \frac{dp_\bot}{2\pi}  e^{i p_\bot x_\bot}
&=& \frac{1}{\pi x_\bot} 
\sin \frac{\Lambda x_\bot}{2} 
\nonumber \\
&\cong& \left\{ \begin{array}{ll}
\frac{\Lambda}{2\pi}~, & \Lambda |x_\bot| \ll 1 \\
\\
0~, & \Lambda |x_\bot| \gg 1
\end{array}\right.
\label{A8}
\end{eqnarray}
Combining these results, and using the fact that $\sum_{\bp} = L^D 
\int \frac{d^Dp}{(2 \pi)^D}$
we find that the infrared cutoff $\alpha$ drops out and
\begin{equation}
G_\phi(\bs; \bx) = \left\{ \begin{array}{ll}
-\frac{\Omega^2}{4\pi} \ln \left[\frac{x_\parallel +
ia}{ia}\right], & \Lambda |\bx_\bot| \ll 1 \\
\\
-\infty, & \Lambda |\bx_\bot| \gg 1.
\end{array}\right.
\label{A9}
\end{equation}

\section{Integrating Out Fast Boson Fields}
\label{fast}
\setcounter{equation}{0}
To evaluate the propagator of the $h$ fields, make use of the relationship
between $\phi$ and the canonical bose fields, $a$ and $a^*$, 
equation \ref{represent} to write
\begin{equation} \langle h(\bs, x) h (\bs, 0)\rangle
=\frac{\Omega}{4\pi} \int \frac{d\omega}{2 \pi} {\sum_\bq}^\prime
\frac{\langle a(\bs, q) a^\dag(\bs, q) \rangle}{q_\parallel}
e^{i(\bq \cdot \bx - \omega \tau)}
\label{B1}
\end{equation}
where
\begin{equation}
{\sum_\bq}^\prime = \left( \frac{L}{2\pi}\right)^D \int_{-\Lambda/2}^{\Lambda/2}
d^{D-1}q_\bot \int_{\lambda/2s < |q_\parallel| < \lambda/2} dq_\parallel
\label{B2}
\end{equation}
and 
\begin{equation}
\langle a(\bs, q) a^*(\bs, q) \rangle 
= \frac{-1}{i \omega - v_F^* q_\parallel}
\label{B3}
\end{equation}
The frequency integral is straightforward and equation \ref{B1} reduces to
\begin{equation}
\langle h(\bs, x) h(\bs, 0) \rangle = \frac{\Omega^2}{4\pi}
\int_{\lambda/2s < |q_\parallel| < \lambda/2} 
dq_\parallel \frac{e^{i (q_\parallel 
(x_\parallel + iv_F^* \tau)}}{q_\parallel} 
[\theta(q_\parallel) \theta(\tau) - \theta(-q_\parallel) \theta(-\tau)]
\label{B4}
\end{equation}
where we have used the result (given here for $D = 2$):
\begin{equation}
\int_{-\Lambda/2}^{\Lambda/2} dq_\bot e^{i q_\bot x_\bot} 
= \frac{\sin(\Lambda x_\bot / 2)}{(x_\bot / 2)} \rightarrow 
\left\{ \begin{array}{ll}
\Lambda,  & \Lambda |x_\bot| \ll 1 \\
\\
0, & {\rm otherwise}
\end{array}\right.
\label{B5}
\end{equation}
To evaluate the integral over $q_\parallel$ write it as the difference
$\int_0^{\lambda/2} - \int_0^{\lambda/2s}$ and replace the upper limit by a
smooth cut-off to find
\begin{equation}
\langle h(\bs, x) h(\bs, 0)\rangle = \left\{ \begin{array}{ll} 
\frac{\Omega^2}{4\pi} \ln \left[ \frac{x_\parallel + iv_F^* \tau 
+ (2 is/\lambda) \sgn(\tau)}{x_\parallel + iv_F^*\tau +
(2i/\lambda) \sgn(\tau)}\right], & \Lambda |x_\bot| \ll 1 \\
\\
0, &  {\rm otherwise}\ .
\end{array}\right.
\label{B6}
\end{equation}

\section{Free Fermion Propagator}
\label{fermion_details}
\setcounter{equation}{0}
Expressed in terms of the boson fields the fermion propagator in patch $\bs$ is
given by 
\begin{equation}
G_F(\bs; \bx, \tau) =\frac{-\Lambda^{D-1}}{(2\pi)^D} \frac{e^{i \bk_\bs \cdot
\bx}}{a} \exp \left[\frac{4\pi}{\Omega^2} \langle \phi (\bs; x, \tau)\phi
(\bs, 0) -\phi^2 (\bs, 0)\rangle\right],
\label{C1}
\end{equation}
where the correlation function of the $\phi$ fields expressed in terms of the
canonical boson fields $a$, and $a^*$, is given by
\begin{eqnarray}
G_\phi(\bs; \bx, \tau) &=& \langle \phi(\bs; \bx, \tau) \phi(\bs, {\bf 0}, 0) 
- \phi^2(\bs; {\bf 0}, 0) \rangle\nonumber\\
&=& \frac{\Omega}{4\pi} \sum_\bq \int \frac{d\omega}{2\pi} \left[e^{i 
( \bq \cdot \bx - \omega \tau)} - 1 \right] 
\frac{\langle a(\bs; q) a^\dag(\bs; q) \rangle}{\hbn_\bs \cdot \bq}
\end{eqnarray}
\label{C2}
which determines the fermion Green's function once the Fourier transform of the
boson Green's function
is known.  Now the free boson Green's function is given by equation \ref{B3}: 
\begin{equation}
G_B^0(\bs; \bq, \omega) = \frac{-1}{i\omega - v_F^* \hbn_\bs \cdot \bq}
\end{equation}
\label{C4}
and therefore
\begin{equation}
G_\phi^0(\bs; \bx, \tau) = \frac{i\Omega^2}{4\pi} \int_{-\lambda/2}^{\lambda/2}
dq_\parallel \int_{-\infty}^{\infty} \frac{d\omega}{2\pi} 
\frac{e^{i (q_\parallel x_\parallel - \omega \tau)} - 1}
{q_\parallel (\omega + i v_F^* q_\parallel)}
\end{equation}
\label{C5}
where we have used equation \ref{B5}, valid for $D = 2$, 
to integrate over the perpendicular momentum.  Now consider
\begin{eqnarray}
\frac{\partial}{\partial x_\parallel} G_\phi^0(\bs; \bx, \tau) &=&
i\frac{\Omega^2}{4\pi} \int_{-\lambda/2}^{\lambda/2} dq_\parallel~ 
\big{[} \theta(q_\parallel) \theta(\tau) - \theta(-q_\parallel) \theta(-\tau)
\big{]}~ e^{i q_\parallel (x_\parallel + iv_F^* \tau)}
\nonumber\\
&=& -\frac{\Omega^2}{4\pi} \frac{1}{(x_\parallel + iv_F^* \tau)}
\end{eqnarray}
\label{C6}
and therefore
\begin{equation}
G_\phi(\bs; x, \tau) = -\frac{\Omega^2}{4\pi} \ln(x_\parallel + iv_F^* \tau) 
+ C;
\end{equation}
\label{C7}
but $\lim_{\tau \rightarrow 0^+} G_\phi(\bs; {\bf 0}, \tau) = 0$, which gives
$C = \frac{\Omega^2}{4\pi}~ \ln(i a)$
and
\begin{equation}
G_\phi^0(\bs; \bx, \tau) = -\frac{\Omega^2}{4\pi} 
\ln \left[ \frac{x_\parallel + iv_F^*\tau}{ia}\right]
\end{equation}
\label{C8}
For the case of $D = 2$, it follows from equation \ref{C1} that
\begin{equation}
G_F(\bs; \bx, \tau) = \frac{i\Lambda}{(2\pi)^2} \frac{e^{i\bk_\bs \cdot x}}
{\hbn_\bs \cdot \bx + i v_F^* \tau}
\end{equation}
\label{C9}
The Green's function for real time is obtained by continuing 
$i \tau \rightarrow -t$.

\section{How Kohn's Theorem Is Satisfied}
\label{Kohn_App}
\setcounter{equation}{0}
In this appendix, we derive the result quoted in Section \ref{hfll}, namely
the collective excitations of the interacting quasiparticles of
the half-filled Landau level occur at a frequency $\omega = 2\pi n_f
\tilde{\phi}/m$, the cyclotron frequency of a particle with bare mass
$m$ in a magnetic field $B = 2 \pi n_f \tilde{\phi}/e$.
To prove this result, we add to the action of the Chern-Simons gauge theory a
Fermi liquid interaction in which all the coefficients except $f_1$
have been set equal to zero\cite{Simon}. The contribution to the action,
equation \ref{FLact}, is given in real time by
\begin{equation}
S_{FL}[\psi^*, \psi] = {f_1\over 2k_F^2}\int d^2x~ dt~ \psi^{*}(x) {\bf \nabla}
\psi(x) \cdot \psi^{*}(x) {\bf \nabla} \psi(x)
\end{equation}
which is made gauge covariant by replacing
${\bf \nabla}$ by ${\bf \nabla} - i{\bf A}$. On Fourier transforming
it is given in terms of the currents as
\begin{eqnarray}
S_{FL} &\approx &
-{{f_1}\over{2 V k_F^2}}~ \int {{d\omega}\over{2\pi}}~\big{\{}
\sum_{\bs,\bt; \bq}~ J(\bs; q)~ \bk_\bs \cdot \bk_\bt~ J(\bt; -q)
\nonumber \\
&+& {{f_1}\over{2 V k_F^2}}~
\sum_{\bs, \bq} n_f \big{[} J(\bs; q)~ {\bf k_S \cdot A}(-q)
+ J(\bs; -q)~ {\bf k_S \cdot A}(q) \big{]} 
\nonumber \\
&-& {{f_1}\over{2 V k_F^2}}~ \sum_{\bf q}
n_f^2~ {\bf A}(q) \cdot {\bf A}(-q) \big{\}}
\end{eqnarray}
As in Section \ref{propagator} the term quadratic in $J$ is decoupled by
introducing additional transverse $A_t$ and longitudinal $A_l$
gauge fields.

Including the Fermi liquid interaction the complete action of the
theory is given by
\begin{eqnarray}
S[A^{\mu },a,\xi ,\xi^*] &=& \sum_\bs \sum_{\bf q, \hat{n}_S \cdot q > 0}
\int {d\omega \over 2\pi}~ \Big{\{}
(\omega - v_F^*~ \hbn_\bs \cdot \bq)~ a^{*}(\bs; q)~ a(\bs; q)
\nonumber \\
&+& \xi(\bs; q)~ a^*(\bs; q) + \xi^*(\bs; q)~ a(\bs; q)\Big{\}}
\nonumber \\
&+& {1\over V} \sum_\bs \sum_{\bf q} \int {d\omega \over 2\pi }~
J(\bs;q) \Big{\{} A_0(-q) + (v_F^* + {f_1 n_f \over k_F})
{\bf \hat{n}_S \times q \over |q|}~ A_T(-q)
\nonumber \\
&+& {\bf \hat{n}_S \cdot q \over |q|}~ A_l(-q)+
{\bf \hat{n}_S \times q \over |q|}~A_t(-q)\Big{\}}
\nonumber \\
&+&{1\over 2V}\sum_{\bf q} \int {d\omega \over 2\pi }~\Big{\{}
\big{[}{n_f \over m^*}+
{f_1 n_f^2 \over k_F^2} + {{{\bf q}^2~ V({\bf q})}
\over{(2 \pi \tilde{\phi})^2}} \big{]}~A_T(q)~
A_T(-q)+ {2i|{\bf q}|\over 2\pi \tilde{\phi}}~A_0(q)~A_T(-q)
\nonumber \\
&-&{1\over f_1}~ A_l(q)~A_l(-q)-{1\over f_1}~A_t(q)~A_t(-q) \Big{\}}~ ,
\label{CSFL}
\end{eqnarray}
where $\mu $ runs over $0, T, l$ and $t$.
We have coupled the boson fields $a^*$ and $a$ to $\xi$ and $\xi^*$ 
to construct a generating
functional as in equation \ref{hfllgenfunc}. After integrating out the
boson fields we obtain the effective action:
\begin{eqnarray}
S_{eff}[A^{\mu }, \xi ,\xi^*]  &=&
{1\over 2}\int {d^2q\over (2 \pi )^2} \int {d\omega \over 2\pi }~\Big{\{ }
\big{[} {n_f \over m^*}+{f_1 n_f^2 \over k_F^2}+ {{{\bf q}^2~V({\bf q})}
\over{(2 \pi \tilde{\phi})^2}}
+(v_F^*+{f_1 n_f \over k_F})^2\chi_t(q)\big{]}~A_T(q)~A_T(-q) 
\nonumber \\
&+& {2i|{\bf q}|\over 2\pi \tilde{\phi}}~A_0(q)~A_T(-q)+[-{1\over f_1}
+\chi_l(q)]~A_l(q)~A_l(-q)+[-{1\over f_1}+\chi_t(q)]~
A_t(q)~ A_t(-q)
\nonumber \\
&+&\chi_{l0}(q)~[A_l(q)~A_0(-q)-A_0(q)~A_l(-q)]
+2[v_F^*+{f_1 n_f \over k_F}]~\chi_t(q)~A_T(q)~A_t(-q)
\nonumber \\
&+& \chi_0(q)~ A_0(q)~ A_0(-q) \Big{\} }
- {1\over V}\sum_{\bf q, \hat{n}_S \cdot q > 0} \int {d\omega \over 2\pi }~
\sum_\bs \Big{\{}{\xi(\bs; q) \sqrt{\Omega |\hbn_\bs \cdot \bq|}\over
(\omega -v_F^* \hbn_\bs \cdot \bq)}~
A_{\alpha }(-q)~ b^{\alpha }(\bs; q) 
\nonumber \\
&+& {\xi^*(\bs; q) \sqrt{\Omega |\hbn_\bs \cdot \bq|}\over
(\omega -v_F^* \hbn_\bs \cdot \bq)}~
A_{\alpha }(q)~ b^{\alpha }(\bs; -q)
+{\xi^*(\bs; q)~ \xi(\bs; q) \over \omega -v_F^* \hbn_\bs \cdot \bq}
\Big{\}} ~ ,
\label{AppSeff}
\end{eqnarray}
here
\begin{equation}
b^{\alpha }(\bs; q) \equiv (1, [v_F^* + {f_1 n_f \over k_F}]{\bf \hat{n}_S
\times q \over |q|},{\bf \hat{n}_S \cdot q \over |q|},
{\bf \hat{n}_S \times q \over |q|})
\end{equation}
and the susceptibilities $\chi_{\alpha }(q)$ are defined by:
\begin{eqnarray}
\chi_0(q) &=& -N^*(0)\int {d\theta \over 2\pi } ~{\cos{ \theta }\over x -
\cos{\theta } + i\eta ~{\rm sgn}(x) }
\nonumber \\
&=& N^*(0)\big{[}1-\theta(x^2-1){|x|\over \sqrt{x^2-1} }
+i\theta(1-x^2){|x|\over \sqrt{1-x^2} }\big{]} 
\nonumber \\
\chi_t(q) &=& N^*(0)\int {d\theta \over 2\pi } ~{\cos{\theta }~\sin^2{\theta }
\over x -\cos{\theta } + i\eta ~{\rm sgn}(x)} 
\nonumber \\
&=& N^*(0)\big{[}x^2-{1\over 2}-\theta(x^2-1)|x| \sqrt{x^2-1}
-i\theta(1-x^2)|x| \sqrt{1-x^2} \big{]} 
\nonumber \\
\chi_l(q) &=& N^*(0)\int {d\theta \over 2\pi } ~{\cos^3{\theta }
\over x -\cos{\theta } + i\eta ~{\rm sgn}(x)} 
\nonumber \\
&=& N^*(0)\big{[}-x^2-{1\over 2}+\theta(x^2-1)|x|{x^2\over \sqrt{x^2-1} }
-i\theta(1-x^2)|x|{x^2\over \sqrt{1-x^2} }\big{]} 
\nonumber \\
\chi_{l0}(q)&=&-\chi_{l0}(-q) = N^*(0)\int {d\theta \over 2\pi }~{\cos^2 \theta 
\over x -\cos{\theta } + i\eta ~{\rm sgn}(x)} 
\nonumber \\
&=& N^*(0)\big{[}-x+\theta(x^2-1)|x|{x\over \sqrt{x^2-1} }
-i\theta(1-x^2)|x|{x\over \sqrt{1-x^2} }\big{]} ~ ,
\label{chis}
\end{eqnarray}
and $x=\omega /(v_F^* |{\bf q}|)$.
The boson propagator is obtained in the usual way
\begin{eqnarray}
\langle a(\bs; q)~ a^\dagger(\bs; q)\rangle
&=& {i\over \omega - v_F^* \hbn_\bs \cdot \bq + i \eta~ {\rm sgn}(\omega)}
\nonumber \\
&+& i~ {\Lambda \over (2\pi )^2}~ \hbn_\bs \cdot \bq~
{D_{\alpha \beta }(q)~ b^\alpha (\bs; q)~ b^\beta (\bs; -q)\over
[\omega - v_F^* \hbn_\bs \cdot \bq + i \eta ~{\rm sgn}(\omega)]^2} ~ ,
\label{AppGaa}
\end{eqnarray}
and the collective modes are given by the poles of the gauge propagator.
The inverse gauge propagator can be read off from the action
equation \ref{AppSeff} and in the limit of interest
$\omega \gg  v_F^*|{\bf q}|$ is given by
\begin{equation}
[K(q)]_{\alpha \beta} =
\left( \begin{array}{cccc}
-{N^*(0)\over 2x^2} & {i|{\bf q }|\over 2\pi \tilde{\phi }} & -{N^*(0)
\over 2x} & 0 \\
{i|{\bf q }|\over 2\pi \tilde{\phi }}
& n_f({1\over m^*}+{f_1 n_f\over k_F^2}) 
& 0 & -{N^*(0)\over 8x^2}(1+{m^*f_1 n_f \over k_F^2}) \\
{N^*(0)\over 2x} & 0 & -{1\over f_1} & 0 \\
0 & -{N^*(0)\over 8x^2}(1+{m^*f_1 n_f \over k_F^2}) & 0 & -{1\over f_1}
\end{array} \right)
\end{equation}
The poles of $D_{\alpha \beta}$
are found by setting ${\rm det}|K_{\alpha \beta }| = 0$ which gives
$\omega ^2 = [2\pi \tilde{\phi} n_f ({1 / m^*}+{f_1 / 4\pi })]^2
= (2\pi n_f \tilde{\phi} / m_b)^2$.  The
second equality follows from the application of Galilean invariance to a
Fermi liquid\cite{Tony}.  The energy of the 
collective mode is at the cyclotron frequency determined by the bare
electron band mass in agreement with Kohn's theorem.

\end{document}